\documentclass [12pt]{article}
\usepackage{amsmath,amssymb,cite}
\usepackage{appendix}
\usepackage{feynmp}
\usepackage{float}
\usepackage[vcentermath,enableskew]{youngtab}
\usepackage{tabularx}
\usepackage[english]{babel}
\usepackage{graphicx}
\usepackage{indentfirst}
\usepackage{epsfig}
\usepackage{epstopdf}
\usepackage[]{hyperref}
\usepackage[section]{placeins}
\usepackage[stable]{footmisc}
\usepackage{appendix}
\usepackage{tabularx}
\usepackage[english]{babel}
\usepackage{graphicx}
\usepackage{indentfirst}
\usepackage{epsfig}
\usepackage{slashed}
\usepackage{fancyhdr}
\numberwithin{equation}{section}

\fancypagestyle{plain}{\fancyhead{}
}

\begin{document}

\title{
On the foundations of quantum theory\footnote{This essay is based on invited lectures at Jijel MSB University (29-31 October 2018) and El-oued EHL University (11-15 March 2018).}\\
}

\author{Badis Ydri\\
Department of Physics, Faculty of Sciences, Annaba University,\\
 Annaba, Algeria.
}

\maketitle

\begin{abstract}
We draw systematic parallels between the measurement problem in quantum mechanics and the information loss problem in black holes. Then we proceed to propose a solution of the former along the lines of the solution of the latter which is based on the holographic gauge/gravity duality. The proposed solution is based on 1) the quantum dualism between the local view of reality provided by Copenhagen and the manifold view provided by the many-worlds and on 2) the properties of quantum entanglement in particular its fungibility.
\end{abstract}

\tableofcontents

\section{Introductory remarks}


Quantum mechanics is perhaps the greatest scientific breakthrough ever achieved. It brought with it a seismic paradigm shift in our way of thinking about nature, and at the same time it is underpinning most of the dramatic technological innovations of the modern era, as well as providing a profound lasting impact on our metaphysical conception of reality. So its value is both ontological, epistemic and practical.

But quantum mechanics is only one step (certainly the more colorful, dramatic and puzzling) in a long list of revolutionary paradigm shifts.

The Copernican revolution is the first revolution in physics in which Copernicus moved away from the ptolemaic cosmology towards a heliocentric cosmology. This marks in fact the starting point of the scientific revolution. Then comes Newton and his Newtonian physics which is the first revolution in theoretical physics in which the scientific experimental method was supplemented with the mathematical method. This revolution continued with Euler, Lagrange, Hamilton and others. And this classical revolution continued further with the invention/discovery of thermodynamics, which can certainly be reduced to mechanical motion, and electromagnetism which resisted such a reduction.

This completes (or almost it did) the edifice of classical physics which is truly classical in spirit and methods, i.e. Aristotelian in a precise sense. Einstein revolution (special relativity and general theory of relativity) is a part of this classical realm of physics. The relativistic conception of spacetime due to Einstein is mid-way between Newton's absolutism and Leibniz's relationalism.

After Newton came another revolutionary revolution -little known to us as physicists- which is the Copernican revolution in philosophy due to Kant which is (despite its rather very complicated character) can be stated as the fact that cognition (the observer) determines appearances (the world) and not the other way around which Kant called  transcendental idealism. By hindsight this is very reminiscent of one of the most central conclusion of quantum mechanics regarding the irreducible role of the conscious/zombie observer. Yet, for Kant, metaphysics is not possible.

Then comes the quantum revolution (Bohr, Einstein, Heisenberg, Schrodinger, Dirac, Wigner, Born, Pauli and others in the golden age of modern physics).

To put the problem in a clear perspective we recall first that the {philosophy of classical physics} is based on 1) determinism, 2) locality, 3) the world exists without observation, 4) consciousness is irrelevant to observation, 5) things are knowable, and 6) the world is causally closed. This leads to {physicalism}, i.e. the view that everything including minds and consciousness is reducible to matter.

Quantum mechanics contradicts classical mechanics in every point. So the world which is described faithfully (from experimental evidence) by quantum mechanics is non-deterministic, non-local, it does not seem to exist without observation, the conscious observer on some accounts is implicitly or explicitly crucial to observation, things are not necessarily knowable, and the world is not causally closed because of the irreducible character of the consciousness of the observer.


   This situation is termed the quantum mud by Popper since from one hand there is quantum mechanics  and from the other hand there is its interpretation, i.e.
   the relation between the mathematics of quantum mechanics and the external world is not without a metaphysical burden.

The single most central but controversial aspect of quantum mechanics consists in the so-called measurement problem, i.e. the reduction/collapse of  the state vector when subjected to a quantum measurement. The measurement problem  is a well posed problem mathematically involving the physics of entanglement and decoherence. However,  the reduction/collapse process is a non-unitary, irreversible and stochastic process which can not be accounted for by any known physics. It could be caused by the environment, or by the mind of the observer or by some lesser sort of observer-participancy, or it could be caused by the spacetime structure at the Planck scale. Or perhaps there is no altogether a non-unitary and irreversible reduction of the state vector but there is instead a unitary and reversible many-worlds.

The measurement problem is therefore the most central question in quantum physics since  we are trying to comprehend the world through the lenses of quantum mechanics. Schrodinger summarizes the problem by the intriguing question: {Does quantum mechanics provide a fuzzy picture of a clear reality or is the fuzziness in reality itself and quantum mechanics is only  providing a clear picture of that fuzziness?}. Whereas Wheeler summarizes the situation by noting that the world according to quantum mechanics can be drawn as an eye looking onto itself.

But the quantum revolution underwent really another paradigm shifting revolution from within with Bell and his celebrated Bell's theorem with all its physical consequences and metaphysical ramifications.

In summary, the states of the physical system do not really exist before measurement and either the observer determines the world (the measurement promotes potential existence into real existence) which is very similar to Kant's transcendental idealism or else the world is really a linear superposition of many coherent branches. Thus here, as opposed to Kant's conclusion based on classical physics, quantum metaphysics is a real possibility but tailored  with experimental and theoretical quantum physics.

It is well appreciated by string theorists, quantum information theorists and many other quantum physicists that black holes evaporation and the associated information loss problem provide the ideal laboratory to test the fundamental principles of quantum mechanics and their possible modifications by taking into account the equivalence principle.

The information loss problem arises from the fact that a correlated entangled pure state with zero Killing energy (originating in the gravitational collapse of a black hole) when it reaches near the horizon will give rise to a thermal mixed state outside the horizon. The entangled pure state is formed from particle pairs where one particle of each pair is transmitted through the horizon as information loss whereas the other particle is reflected to infinity as Hawking radiation. The existence of the horizon is what makes this problem such a special and a peculiar problem.

But information is thought not to be lost at the end since it will start coming out with the radiation at the Page time when the entanglement entropy between the interior and the exterior becomes maximal (Page curve). This is the unitarity assumption which will be reinforced in any model based on the holographic gauge/gravity duality.

This paper is organized as follows. In section $2$ we provide a review of the structure of quantum mechanics according to the Copenhagen interpretation and the many-worlds formalism. We also review many physical results/effects and theorems of quantum philosophy with the main emphasis placed on quantum entanglement and Bell's theorem.

In section $3$ we provide a description of Hawking radiation and the corresponding information loss problem. Then we briefly review the holographic gauge/gravity duality and many other related  ideas relevant to the information loss problem and its unitary resolution such as the connection between spacetime geometry and quantum entanglement.

In section $4$ we provide a synthesis based on the quantum dualism, i.e. the fact that the Copenhagen interpretation provides the local view of reality whereas the many-worlds formalism provides the manifold view and the two views are complementary not contradictory.

The information loss problem and the measurement problem share the fundamental characterization that an initial pure state is evolved into a final mixed state since there is in both cases a part of the system which is inaccessible (the environment in the case of quantum mechanics and behind the horizon for the case of black holes). However, the information loss problem admits in principle a solution via the holographic gauge/gravity duality. The goal is to exploit this formal analogy (and the analogy as we will argue is even physical) in order to extend the proposed solution for the information loss problem to the measurement problem.

Section $5$ contains a summary.

\section{The measurement problem and interpretations of quantum mechanics}
\subsection{The wave-particle duality and complementarity principle}


The double-slit interference experiment is "impossible to explain in any classical way" and is "the heart of quantum mechanics" and "contains the only mystery" \cite{Feynman}. Feynman is also reported to have said that quantum interference is the "mother of all quantum phenomena" and that  because of it "nobody understands quantum mechanics".

Quantum interference in the  double-slit interference experiment provides the first direct indication of the wave-particle duality.

  Thus the interference pattern is observed even if we send light through the two slits one single photon at a time. It works for photons, electrons and in principle for all other particles.

  {A wave (interference)-particle (path) duality} is the first duality in quantum physics. The two descriptions are {complementary} (not contradictory as in classical mechanics) to each other since they can not be observed simultaneously.

  Another related complementarity principle in quantum physics is the {position-momentum duality} and the Heisenberg principle. 
    The position $x$ and the momentum $p$ are canonical variables represented by incompatible operators on the Hilbert space satisfying the Dirac commutation relation
    \begin{eqnarray}
      [x,p]=i\hbar.
      \end{eqnarray}
    This leads immediately to the uncertainty relation
    \begin{eqnarray}
      \Delta x\Delta p> \hbar.
      \end{eqnarray}
      In other words, there is a fundamental limitation on the precision of measurements. The quantum phase space becomes discrete, i.e. pointless!!, constituted of elementary Heisenberg cells of volume $\hbar$ containing one state each. The phase space is then {fuzzy} (since we can not discern points) or {noncommutative} (since coordinates are not commuting). This is the prototype for all noncommutative geometry, fuzzy spaces and matrix models which is one proposal for quantum gravity.

      The wave-particle duality is also intimately related to the other complementarity principle: {entanglement-decoherence duality}. The question then arises: Which one is more fundamental?

      The evidence seems to point towards entanglement being the most fundamental quantum effect and even interference can be reduced to entanglement as shown using the ER=EPR conjecture in \cite{Susskind:2014yaa,Susskind:2016jjb} (the two slits are construed as maximally entangled and as such they are connected via a smooth gravitational bridge).
  
      {Wheeler's delayed choice gedanken  experiment} \cite{wheeler78} (which appeared first in his essay "Law without Law") is perhaps the mother of all interference experiments and is one of the greatest quantum effect which  was verified experimentally for example in \cite{Aspect:2007,MKDT}. This which-way experiment threatens time ordering and causality and according to Wheeler this experiment shows that no phenomena is a real phenomena until it is an observed phenomena. 



\begin{figure}[htbp]
\begin{center}
   \includegraphics[width=10cm,angle=-0]{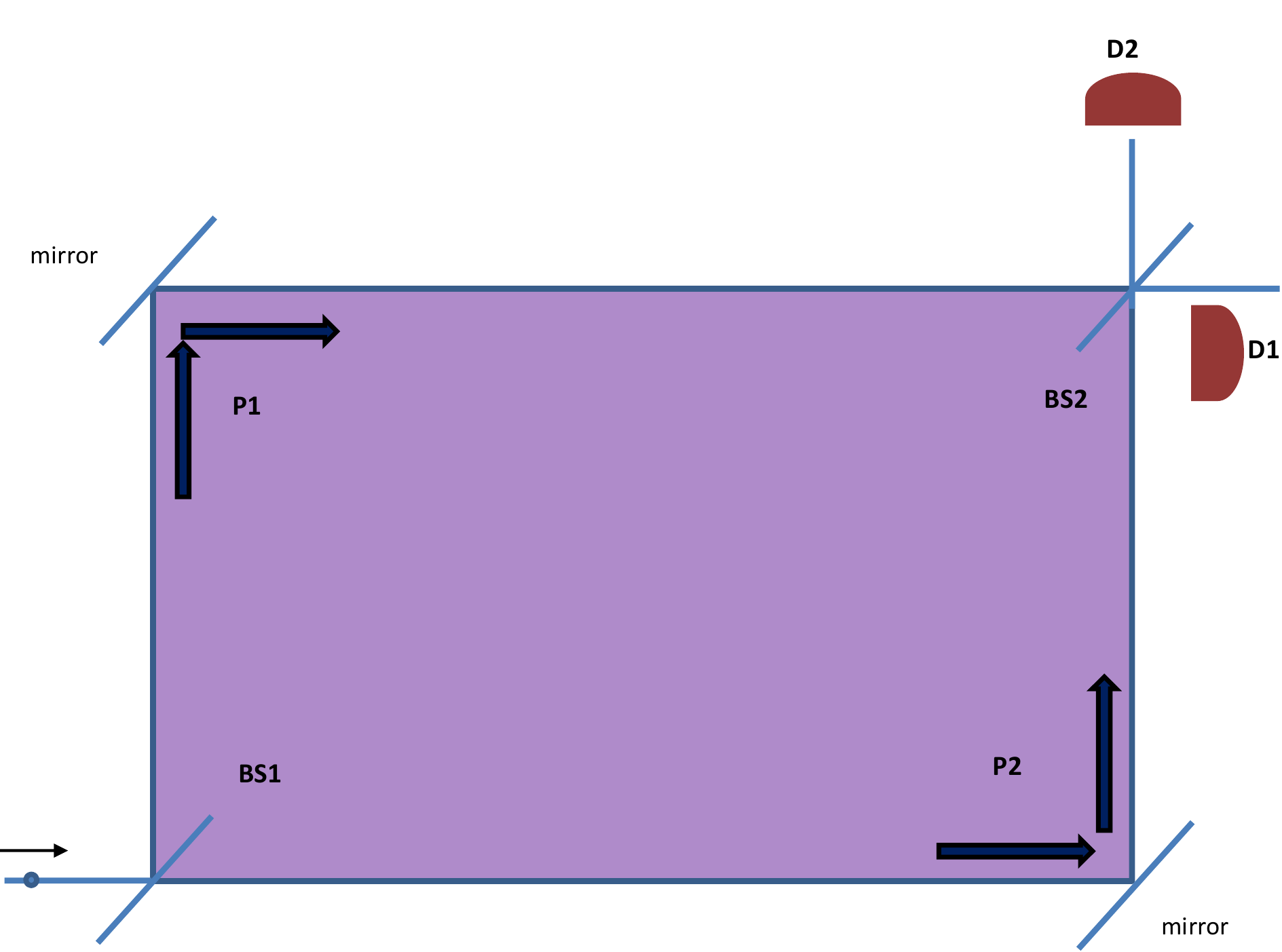}
  \caption{The delayed choice experiment.}\label{dc}
\end{center}
\end{figure}

    A source of light sends photons one by one through the paths shown. The photons pass through a beam splitter BS1. They are either reflected with a probability $1/2$ towards the first mirror (path P1) and then unto the detector D1. 
    Or they can be refracted with a probability $1/2$ towards the second mirror (path P2) and then unto the detector D2. See figure (\ref{dc}).

    The paths are determined: If D1 clicks then the path taken is P1 whereas if D2 clicks then the path taken is P2.

We position another beam splitter BS2 between the detectors D1 and D2. Let $A$ be the amplitude of the emitted light. The reflected wave is $i*A$ whereas the refracted wave is $1*A$. We can reach D1 by two routes: reflection+refraction or refraction+reflection giving the probability amplitude $i*1+1*i=2i$. And we can reach D2 by two routes: reflection+reflection or refraction+refraction giving the probability amplitude $\longrightarrow$ $i*i+1*1=0$.

  In other words, the photons reach D1 (constructive interference) always but never D2 (destructive interference), i.e. {light behaves as waves when paths are not determined}. 

  If only one route is allowed (BS2 removed) light behaves as particles. Whereas if the two routes are allowed (BS2 is not removed) light behaves as waves. {This is complementarity}.

  Wheeler proposes to add BS2 at the last moment after the photons go through BS1 and before they reach the detectors at the intersection points of the two paths P1 and P2. The result does not change: {light behaves as waves if we put BS2 and as particles if we remove BS2}.

  Thus we are deciding {retroactively} whether the photons act as waves (both paths allowed) or as particles (one path allowed) after they complete their journey. Equivalently, the photons decide to behave as particles (one path) or as waves (two paths) at the last moment although the particle behavior requires passing by one path whereas the wave behavior requires passing by two paths.

  Wheeler proposed also a {cosmic delayed choice} experiment. The light source is a {quasar} whereas the beam splitter is a {gravitational lensing}. The emitted photons (since millions of years) act as waves by observing interference patterns or as particles if we employ telescopes to determine their paths.
    We can thus create or alter the distant past by our manner of observing it now as Wheeler puts it. According to this {extreme} view even the {big bang} could have been created by our observation.

\subsection{Copenhagen interpretation}

\subsubsection{The von Neumann processes}
Quantum mechanics according to the standard view (the Copenhagen or Bohr's interpretation \cite{Bohr}) is based on 
two mega-laws (not one) which were mathematically formulated originally by von Neumann in his book \cite{VN55} (see also \cite{D47,LL77,Heisenberg}). These are given by
\begin{itemize}
\item {\bf Process II:} The unitary evolution in time generated by a Hamiltonian $H$ given by the Schrodinger equation, namely
 \begin{eqnarray}
      i\hbar\frac{\partial}{\partial t}|\psi(t)\rangle=H|\psi(t)\rangle.
      \end{eqnarray}
Also one should mention the quantum  superposition principles: If $\psi_1$ and $\psi_2$ are two solutions of the Schrodinger equation then any 
linear combination $\alpha\psi_1+\beta\psi_2$, for any complex numbers $\alpha$ and $\beta$, is also a solution. The superposition principle can be given by the path integral.

The Schrodinger equation allows us to compute the state of the physical system $\psi(t)$ at any given time $t$ 
starting from some initial state $\psi(t_0)$ at the initial time $t_0$.

\item {\bf Process I:} The collapse or reduction postulate termed process I by Von Neumann (process II is the unitary evolution). 
This allows us to compute the state of the system when we subject it to a quantum measurement. Explicitly, it states 
that the state of the system after measurement will collapse to the eigenstate in the Hilbert state corresponding to 
the eigenvalue determined by the outcome of the measurement process. The collapse postulate should be coupled with 
the Born's statistical rule which determines the probability or frequency of finding the various outcomes of the act of 
measurement.
\end{itemize} 
Let us consider now some physical system ${\cal S}$ and let $O_1$ and $O_2$ be two physical oberservers 
(for example position and momentum) associated with ${\cal S}$. The states of the physical system ${\cal S}$ are 
 vectors in a complex vector space ${\cal H}$ with the properties of a Hilbert space whereas the physical observables will be 
 represented by operators denoted by the same symbols which are hermitian, i.e. $O_i^{\dagger}=O_i$.  Alternatively, any pure state  
 of the physical system is given by a corresponding density matrix $\rho$. The state at the initial time $t_0$ is 
 denoted by $\rho_0(t_0)$. We will suppose that the operators $O_1$ and $O_2$ are incompatible operators, i.e. they do not
 commute  under the pointwise multiplication of operators, viz $O_1.O_2\neq O_2.O_1$. 
 
Next, we will measure the observable $O_1$ at the 
 instant $t_1$ to find the value (eigenvalue) $p_1$. This measurement is represented on the Hilbert space ${\cal H}$ by a hermitioan operator
 $P_1$ which is also an idempotent, i.e. $P_1^2=P_1$. The operator $P_1$ is a projection operator on the (subspace) eigenspace of the Hilbert 
 space associated with the eigenvalue $p_1$. The probability of obtaining the eigenvalue $p$ at time $t_1$ if the state 
 of the system is 
 prepared at the time $t_0$ to be in the density matrix $\rho_0(t_0)$ is given by the Born's rule

\begin{eqnarray}
{\cal P}_1=tr(P_1(t_1)\rho_0(t_0)).
\end{eqnarray}
After the first measurement the initial density matrix $\rho_0(t_0)$ collapses to the density matrix $\rho_1(t_1)$ associated 
with the eigenvalue $p_1$ given by the von Neumann's rule

\begin{eqnarray}
\rho_1(t_1)=\frac{P_1(t_1)\rho(t_0)P_1(t_1)}{tr(P_1(t_1)\rho(t_0))}.
\end{eqnarray}
Next, we measure the observable $O_2$ at the instant $t_2$ to find the eigenvalue $p_2$. This second measurement is again 
represented with a projection operator $P_2$ which projects on the eigenspace of the Hilbert space ${\cal H}$ associated 
with the eigenvalue $p_2$. The conditional probability of obtaining the second measurement, provided that the first 
measurement has been performed, is given by the Born's rule

\begin{eqnarray}
{\cal P}_2=tr(P_2(t_2)R_1(t_1)).
\end{eqnarray}
Thus, the probability of obtaining the first measurement $P_1(t_1)$ at time $t_1$ and then the second measurement $P_2(t_2)$
 at time $t_2$ is given by the product of the conditional probability ${\cal P}_2$ and the first probability ${\cal P}_1$, viz

\begin{eqnarray}
{\cal P}=tr(P_2(t_2)P_1(t_1)\rho_0(t_0)P_1(t_1)P_2(t_2)).
\end{eqnarray}
Generalization of this result is straightforward. The probability of obtaining the measurements $P_1(t_1)$, $P_2(t_2)$,...., $P_n(t_n)$ 
at the successive instants of time $t_1$, $t_2$,...,$t_n$, if the state 
 of the system is 
 prepared at the initial instant $t_0$ to be in the density matrix $\rho_0(t_0)$, is given by the generalized Born's rule

\begin{eqnarray}
{\cal P}=tr(P_n(t_n)...P_2(t_2)P_1(t_1)\rho_0(t_0)P_1(t_1)P_2(t_2)...P_n(t_n)).
\end{eqnarray}
The set of projectors $P_1(t_1)$, $P_2(t_2)$,...., $P_n(t_n)$ at the instants $t_1$, $t_2$,...,$t_n$  defines what we call a 
quantum history \cite{RG0}.

\subsubsection{Quantum Zeno effect and the collapse postulate}

        The quantum Zeno effect \cite{Misra:1976by}  is one of the greatest effects in quantum physics due to its intimate connection to time and consciousness.
        {It asserts the cancellation of motion under continuous  quantum measurement} and as a consequence it provides an almost direct test of the reduction/collapse postulate.
        
 It is called the Zeno effect because of the intriguing similarity with the Zeno paradoxes of antiquity (which Russell called  "immeasurably subtle and profound" \cite{Russell}) which are due to Zeno of Elea and his teacher Parmenides. In summary, according to Zeno and Parmenides, there is no motion,  time, change, multitude and infinity in the actual  world and everything of that is illusionary. 

Zeno (as recounted by Aristotle in his physics) provided $4$ paradoxes in defense of his teacher's ideas.

{The arrow paradox for example goes as follows.}  In order for the arrow in flight to move it must change the position it occupies in space. But at any instant of time the arrow is neither moving to where it is nor it is moving to where it is not. It can not move to where it is because it is already there and it can not move to where it is not because there is no elapsed time during the instant of time under consideration. Hence at any instant of time the arrow is not moving, i.e. motion does not occur. And if motion does not exist then time does not exist.



The quantum Zeno effect is effectively saying that there is no unitary evolution in time under a repeated measurement. See the pedagogical presentation \cite{PCV}.

\begin{figure}[htbp]
\begin{center}
  \includegraphics[width=8.0cm,angle=0]{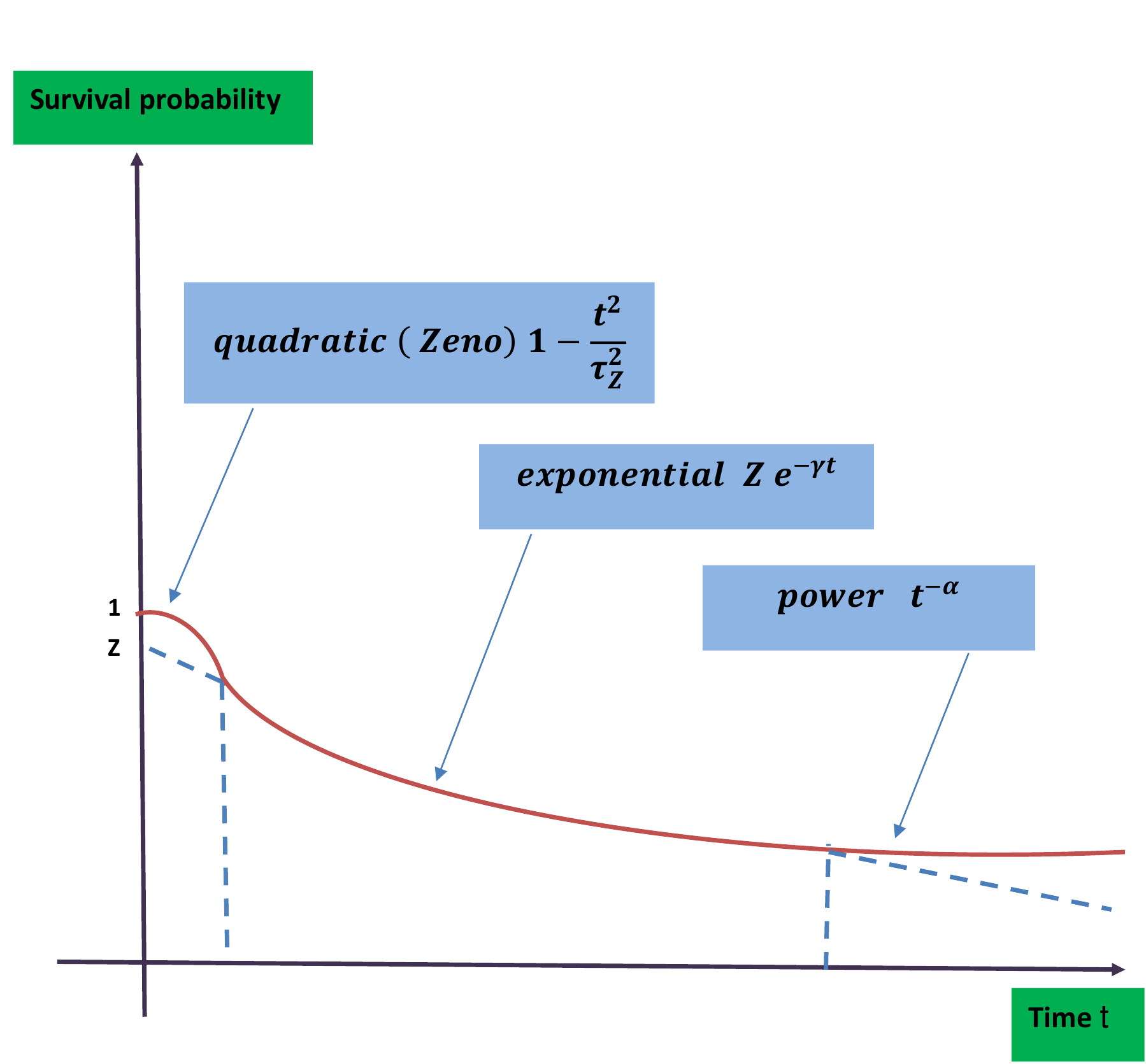}
\end{center}
\caption{The quadratic Zeno region at short times.}\label{QZE}
\end{figure}

Sudarshan and Misra in $1977$ considered  the {survival probability}  which divides into three regimes: { 1) a quadratic behavior,} { 2) an exponential   decay}, then 3) { a power law behavior} as shown on figure (\ref{QZE}).

If the system starts at $t_0=0$ from $\psi_0$ then the probability 
$p(\delta t)$ of finding the system at a time $\delta t$ still in $\psi_0$ is 
\begin{eqnarray}
  p(\delta t)=1-\frac{\delta t^2}{\tau^2_Z}.
  \end{eqnarray}
What is the probability of finding the system in its initial state $\psi_0$ after $N$ continuous measurements?

After the first measurement the unitary evolution of the system starts anew from $\psi_0$ if the system was found in this state (collapse postulate). The probability that the second measurement will reveal the system to be still in $\psi_0$ will be given by $p(\delta t)^2$ and not $p(\delta t)$.
   This second measurement will again collapse the state back onto the initial state if the system was found there and the whole process repeats. 

The probability of finding the system in its initial state $\psi_0$ after $N$ continuous measurements (because of the collapse postulate) is then given by 

\begin{eqnarray}
    p(t)=p(\delta t)^N=\exp(\frac{t^2}{N\tau^2_Z})\longrightarrow 1~,~N\longrightarrow 1.
    \end{eqnarray}
Thus if we perform an increasing number of quantum measurements to  check whether or not the system is still in its initial state the chances of actually finding it there become more certain. 

In some precise sense, the system is frozen in its initial state, i.e. the unitary time evolution is halted 
by the continuous measurement  of its state. 

Among the most recent experimental confirmations of the quantum Zeno effect is by means of a real space measurement
of atomic motion due to Patil, Chakram, Vengalattore \cite{Pascazio:2013dha}.

The atoms in an ultracold gaz will arrange themselves in a lattice and their velocity is vanishingly small. But by the Heisenberg uncertainty principle the position and the velocity  are conjugate variables. Thus the uncertainty in the position of any given atom is very large and as a consequence  the atom can be anywhere in the lattice with equal probability due to quantum tunneling. By subjecting the gaz to a continuous measurement (by illuminating them with imaging laser which causes them to fluoresce) it is observed that quantum tunneling is completely suppressed.

\subsection{Entanglement entropy}

\subsubsection{The reduced density matrix}
       In quantum mechanics it is shown by the EPR experiment for example that entanglement is at odd with locality. The action (due to a measurement) seems to propagate with an infinite velocity and although it can not carry any energy we are left in an uncomfortable position. Entanglement as opposed to energy is not conserved and there are degrees of entanglement. Mathematically, entanglement means that the vector state is not separable, i.e. it can not be written as a tensor product.
       
\begin{figure}[H]
\begin{center}
  \includegraphics[angle=-0,scale=0.6]{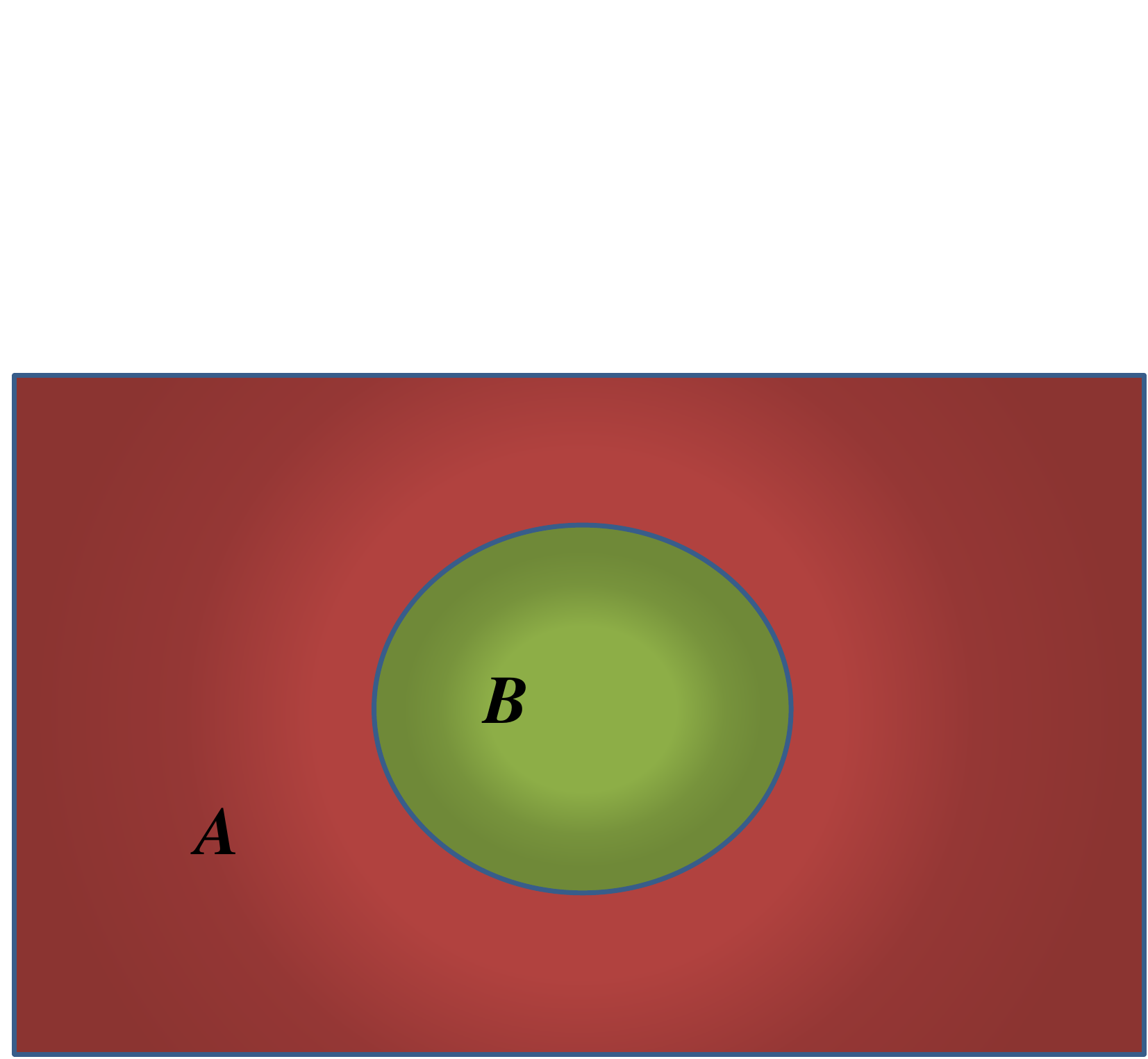}
\end{center}
\caption{The accessible and inaccessible regions.}\label{sketch1}
            \end{figure}

       Quantum entanglement is measured by entropy or more precisely by entanglement entropy. However, entropy has actually two sources: statistical and quantum.
            \begin{enumerate}
\item {\bf The statistical/thermal entropy}: The thermal or Boltzmann entropy of a macroscopic state is the logarithm of the number $n$ of microscopic states consistent with this state. Thus this entropy measures the lack of resolution, i.e. the fact that a large number of microscopic configurations correspond to the same macroscopic thermodynamical state. The thermal entropy is defined in terms of the Blotzmann density matrix $\rho_{\rm ther}=\exp(-\beta E)/Z$ by
\begin{eqnarray}
  S_{\rm ther}&=&-tr \rho_{\rm ther}\log\rho_{\rm ther}\nonumber\\
  &=&\log n.
\end{eqnarray}
The second equality holds if the microstates are equally probable.
\item {\bf Entanglement entropy:}        
\begin{itemize}
          \item {\bf Measurement}: In quantum mechanics, there is another source of entropy associated with the restriction of observers, who are performing the experiments, to finite volume. Indeed, a typical observer performing an experiment on a closed system, which is supposed to be in a pure ground state $|\Psi\rangle$, will only be able to access a particular subsystem, i.e. a partial set of the relevant observables such as those with support in a restricted volume.
            
            We will denote the accessible subsystem by $A$ (where the observers are restricted) and the inaccessible subsystem is $B$. The total system $\Sigma=A\cup B$ is in a pure ground state $|\Psi\rangle$.  See figure (\ref{sketch1}).
            
          \item   {\bf Reduced density matrix:}       The state of the system will be given by a mixed density matrix $\rho$ and the entropy will measure the correlation between the inaccessible subsystem $B$ and the accessible part $A$ of the closed system. The total Hilbert space is  ${\cal H}_{\rm Tot}={\cal H}_A\otimes{\cal H}_B$.

            The observer who can not access the subsystem $B$ will describe the total system by the reduced density matrix (obtained by tracing over the inaccessible degrees of freedom) 
\begin{eqnarray}
  \rho_{\rm Red}\equiv \rho_A=tr_B\rho_{\rm Tot}.
\end{eqnarray}
In other words, we trace (integrate) over the inaccessible subsystem $B$, i.e. we take average over the inaccessible degrees of freedom.

  \item {\bf The mixed versus pure states:} The reduced density matrix is an incoherent (mixed) superposition (statistical ensemble, classical probabilities, no interference terms, random relative phases). It is not an idempotent and it satisfies  $Tr\rho^2<1$.

            In contrast, a pure state is a vector in the Hilbert space which is a coherent superposition (interference terms, coherent relative phases) represented by a projector.

            Mixed states are relevant if the exact initial state vector is unknown.

\item {\bf Entanglement entropy:} The entropy of the subsystem $A$  which measures the correlation between the inaccessible subsystem and the accessible part of the closed system is defined by the von Neumann entropy of this reduced density matrix, viz
\begin{eqnarray}
  S_{\rm Red}\equiv S_A=-tr_A\rho_A\log\rho_A=-\sum_i\rho_i\log \rho_i.
\end{eqnarray}
Thus, entanglement entropy is the logarithm of the number of microscopic states of the inaccessible subsystem $B$ which are consistent with observations restricted to the accessible subsystem $A$, together with the assumption that the total system is in a pure state. It measures the degree of entanglement between $A$ and $B$.  This is different from the thermodynamic Boltzmann entropy.
\item {\bf Properties:} The entanglement entropy satisfies the following properties. For three subsystems $A$, $B$ and $C$ which do not intersect each other we have the so-called strong subadditivity relations
\begin{eqnarray}
  &&S_{A+B+C}+S_B\leq S_{A+B}+S_{B+C}\nonumber\\
  &&S_{A}+S_C\leq S_{A+B}+S_{B+C}.
\end{eqnarray}
By choosing $B$ empty in the above relations we obtain
  \begin{eqnarray}
    &&S_{A+B}\leq S_{A}+S_{B}.
  \end{eqnarray}
  The mutual information is defined by
   \begin{eqnarray}
    && I(A,B)=S_{A}+S_{B}-S_{A+B}\geq 0.
  \end{eqnarray}
  If we choose $B$ to be the complement of $A$ then
\begin{eqnarray}
S_A=S_B\Rightarrow S_{A+B}\leq 2S_A.    
\end{eqnarray}
Hence the entanglement entropy is not an extensive quantity.

\item {\bf Examples:}
For a pure (separable) state, i.e. when all eigenvalues with the exception of one vanish, we get
  $S=0$. For mixed states we have $S>0$.

  In the case of a totally incoherent mixed density matrix in which all the eigenvalues are equal to $1/N$ where $N$ is the dimension of the Hilbert space we get the maximum value of the Von Neumann entropy given by 
\begin{eqnarray}
S_{\rm Red}=S_{\rm max}=\log N.
\end{eqnarray}
In the case that $\rho$ is proportional to a projection operator onto a subspace of dimension $n$ we find 
\begin{eqnarray}
S_{\rm Red}=\log n.
\end{eqnarray}
In other words, the Von Neumann entropy measures the number of important states in the statistical ensemble, i.e. those states which have an appreciable probability. This entropy is also a measure of the degree of entanglement between subsystems $A$ and $B$ and hence its other name entanglement entropy.
\item {\bf Information:} The von Neumann entropy $S_V\equiv S_A$ is not additive as opposed to the thermal entropy $S_B$ defined with respect to Boltzmann distribution. We have $S_B\geq S_V$, i.e. the Boltzmann thermal entropy (coarse grained, macroscopic) is always greater or equal to von Neumann entanglement (fine grained,microscopic) entropy.

  The amount of information is the difference:
\begin{eqnarray}
  I=S_B-S_V.
  \end{eqnarray}
            If $\Sigma=A$ then there is no entanglement and the amount of information is maximal, i.e. $S_V=0\Rightarrow I=S_B$. If $A<<B$ then in this case the amount of information is zero, i.e. $S_V=S_B$. Equivalently, if $A<<B$ then the entanglement entropy becomes maximal equal to the thermal entanglement. 

Remark that the von Neumann entropy of the total system is zero, viz $S_{\rm Tot}=tr\rho_{\rm Tot}\log\rho_{\rm Tot}=0$ since there is no inaccessible part here.
\end{itemize}
\end{enumerate}

\subsubsection{Entanglement entropy in quantum mechanics and quantum field theory}
For detail of the formalism used here we refer to \cite{Bombelli:1986rw}. We will consider a Hamiltonian of the form
\begin{equation}
  \label{1} H=\frac{1}{2}\sum_{A,B} (\delta_{A,B} \pi^A\pi^B+V_{AB}\varphi^A\varphi^B).
\end{equation}
In this equation $V$ is a real symmetric matrix with positive definite eigenvalues. The normalized ground state of this model is given in the Schrodinger representation by
\begin{equation}
  \langle {\varphi^A}|0\rangle = \big[\det \frac{W}{\pi}\big]^{1/4}\exp \big [-\frac{1}{2}W_{AB}\varphi^A\varphi^B\big].
\end{equation}
$W$ is the square root of the matrix $V$. The corresponding density matrix is
\begin{equation}
  \label{3} \rho(\varphi,\varphi')= \big[\det \frac{W}{\pi}\big]^{1/2}\exp\big [-\frac{1}{2}W_{AB}(\varphi^A\varphi^B+\varphi'^A\varphi'^B)\big]
\end{equation}
If we suppose that the field degrees of freedom $ \varphi^\alpha, \alpha=\overline{1,n}$, are inaccessible then the correct description of the state of the system will be given by the reduced density matrix in which we integrate out these inaccessible degrees of freedom, viz
\begin{equation}
  \label{4} \rho_{\rm red}({\varphi}^{n+1},{\varphi}^{n+2},...,{\varphi}^{'n+1},{\varphi}^{'n+2},...)= \int \prod_{\alpha=1}^{n} d{\varphi}^\alpha \rho(\varphi,{\varphi}')
\end{equation}
The entanglement entropy is the associated Von Newman entropy of $ \rho_{\rm red}$ defined by $ S= -{Tr} \rho_{\rm red}\log\rho_{\rm red}$. The entanglement entropy for any Hamiltonian of the form (\ref{1}) can be shown to be given by \cite{Bombelli:1986rw}
\begin{equation}
  \label{5} S_{\rm ent}= \sum_i\bigg[ \log\big(\frac{1}{2}\sqrt{{\lambda}_i}\big)+ \sqrt{1+{\lambda}_i}\log\bigg(\frac{1}{\sqrt{{\lambda}_i}}+ \sqrt{1+\frac{1}{{\lambda}_i}}\bigg)\bigg].
\end{equation}
The ${\lambda}_i$ are the eigenvalues of the following matrix
\begin{equation}
  \label{6} \Lambda_{i,j}= -\sum_{\alpha=1}^{n}W^{-1}_{i\alpha}W_{\alpha j}
\end{equation}
$W_{\alpha j}$ and $ W^{-1}_{i\alpha}$ are elements of $W$ and $W^{-1}$ respectively with $i,j$ running from $n+1$ to ${\cal N}$ and $\alpha$ from $1$ to $n$, i.e. $\Lambda$ is an $({\cal N}-n)\times({\cal N}-n)$ matrix and $i,j$ run from $n+1$ to $N$.

The calculation of entanglement entropy in conformal field theory is more involved but is based on the same formula $ S= -{Tr} \rho_{\rm red}\log\rho_{\rm red}$.  See \cite{Nishioka:2009un} and references therein.

In a QFT on a $(d+1)-$dimensional manifold ${\bf R}\times {\bf N}$ where $d\geq 2$ and ${\bf N}=A\cup B$ it is found that entanglement entropy
\begin{itemize}
\item $1)$ depends only on the geometry of $A$  (this is why entanglement entropy is also called geometric entropy).
\item $2)$ is UV divergent and hence the continuum theory should be regularized by a lattice $a$.
\item $3)$ is proportional to the area of the boundary $\partial A$ of $A$ since the entanglement between $A$ and $B$ occurs strongly obviously on the boundary.
  \end{itemize}
We have explicitly \cite{Bombelli:1986rw,Srednicki:1993im}   
\begin{eqnarray}
S_A=\gamma.\frac{{\rm Area}(\partial A)}{a^{d-1}}+{\rm subleading~terms}.
\end{eqnarray} 
This entanglement entropy formula (which includes UV divergences, proportional to the number of matter fields) is very similar to the Bekenstein-Hawking formula (which does not include UV divergences, is not proportional to the number of matter fields). In fact the quantum corrections to the  Bekenstein-Hawking black hole entropy in the presence of matter fields is given by entanglement entropy \cite{Susskind:1994sm,Fiola:1994ir,Jacobson:1994iw,Solodukhin:1994st}.

\subsection{EPR and Bell's theorem}

  \subsubsection{The Einstein, Podolsky, Rosen (EPR) experiment }
  
  The celebrated Einstein, Podolsky, Rosen (EPR) gedanken experiment \cite{Einstein:1935rr} is based on two assumptions:
  \begin{itemize}
\item {\bf EPR1 or Classical Realism:} 
In other words, the world, or more precisely its properties, really exist independently of any measurement. Thus, a physical quantity is real if its value can be predicted with certainty (hence the need for hidden variables since the Schrodinger equation does not permit this) without disturbing the system being measured (there should be no entanglement which Einstein termed: spooky action at a distance).
\item {\bf EPR2 or Locality:}
  The physical properties of a system A should be independent from the physical properties of a spatially separated system B (no entanglement again). This assumption is closely related to relativity in an almost obvious sense!
  \end{itemize}
These two assumptions led them directly to the conclusion that the quantum wave function given by the solution of the Schrodinger equation is an incomplete description of physical reality and thus hidden variables are needed.

Bohr the father of the orthodox and the Copenhagen interpretations was opposed to EPR1 (realism) more than to EPR2 (locality). Bell then showed that the two EPR assumptions lead directly to what we call now Bell inequality which is badly violated by quantum mechanics  \cite{Bell:1964fg} and nature (the famous Aspect experiment \cite{Aspect:1981zz}).

Putting it differently, one of the two EPR assumptions or both is/are at odd with quantum mechanical predictions.  Bell himself rejected EPR2, i.e. locality or more precisely local causality, which is also the view of the majority of physicists and philosophers with the exception perhaps of consistent (decoherent) histories who reject classical realism in favour of the so-called quantum realism (the single framework rule) \cite{RG0}.

Thus the world according to the views of the majority of physicists and philosophers who understand quantum mechanics in this particular way is certainly not local. And it may even be not classically real. And there is even an implicit danger to free will and/or causality.

\subsubsection{Theorem of quantum philosophy}
The three fundamental theorems of quantum philosophy are:
\begin{enumerate}
\item {\bf The Kochen-Specker theorem (1967):}  The Kochen-Specker theorem  \cite{Kochen:1968zz} states simply that no hidden variable contextual description of quantum mechanics is possible.

  This theorem depends on the no-contextuality requirement: The results of a given measurement which are predicted by the underlying state (wave function and hidden variables) do not depend on what other measurements are being performed on the system.
In the contextual hidden variable theories (such as Bohm's interpretation \cite{Bohm:1951xx0}) the result of a given measurement depends on the state and on the other measurements being performed on the system.
\item {\bf Bell's theorem (1965):} This states that hidden variables theories can only be non-local \cite{Bell:1964kc,Bell:1964fg}. This is the most fundamental of all these theorems.
\item {\bf The Greenberger-Horne-Zeilinger (GHZ) theorem (1989):} The GHZ theorem \cite{GHZ} is a generalization of Bell's theorem which is mid-way between the algebraic no hidden variable theorem (combinatorial considerations) of Kochen and Specker and the statistical hidden variable theorem (multi-particle considerations) of Bell. This situation is termed Bell without statistics by \cite{price}.

  The GHZ theorem involves a maximally entangled tripartite system as opposed to the maximally entangled bipartite system considered in Bell's theorem. As Bell's theorem the GHZ theorem rules out local hidden variable theories. Both Bell and GHZ rely on the absence of advanced action.
\end{enumerate}
These three major theorems are the most difficult objections to the ignorance interpretations of quantum mechanics which assumes that quantum mechanics is incomplete and thus it should be supplemented by hidden variables. These theorems show that any hidden variable description of quantum mechanics must be both contextual and non-local. Only non-local and contextual hidden variable interpretations such as Bohm's interpretation \cite{Bohm:1951xx0} can escape these no-go theorems.

A fourth theorem worth mentioning is the free will theorem of Conway and Kochen \cite{CK,CK2}. This is a mixture of KS and Bell's theorems which leads to some puzzling consequences for the problem of free will (see also \cite{CL,wc}).
\subsubsection{Bell's theorem}

The state vector (or wave function) of a physical system only permits us to calculate the probabilities of various possible outcomes in 
a given measurement. This is the orthodox position. The question one can then ask immediately: did the physical 
system have all along, i.e. before the act of measurement, 
the value found after measurement? 

Einstein, and all those who believe in local realism, would answer this question in the affirmative. This is the 
substance of the famous EPR paradox \cite{Einstein:1935rr} (see also the nice discussion of \cite{RG}). Thus, on this view, the system 
has the measured value long before the act of measurement had took place, and that quantum mechanics is simply an 
incomplete theory, 
as it can only allow us to calculate probabilities. In other words, there must exist extra hidden variables, which together with 
the wave function, will 
allow us to predict precisely the behavior of the physical system exactly and deterministically before the measurement. 

This picture assumes therefore implicitly/explicitly:
1) reality, 2) locality and 3) free choice. Bell's theorem \cite{Bell:1964kc} shows that these extremely reasonable 
assumptions are not compatible. It states precisely 
that no physical theory based on local hidden variables can reproduce all the predictions of quantum mechanics 
\cite{Bell:1987hh,Bell:1964kc}. 

Bell's theorem is justifiably the most profound result in physics and some may even go farther 
and considers it ``the most profound discovery of science`` \cite{HS75}.

Most interpretations of quantum mechanics will relax either the assumption of reality, or the assumption of locality,  but 
rarely the assumption of free choice (Bell's himself discussed super-determinism 
while Price \cite{price} considered backward causation in connection with the transactional interpretation of quantum 
mechanics \cite{cramer}). 

The majority (I guess) view (Bohr and the Copenhagen school \cite{Bohr}) states, on the other hand, that the 
state of the system actually does not exist before measurement (see also \cite{DM82}) which is what seems to be confirmed 
by  Aspect's experiment  \cite{Aspect:1981zz}.

We consider the setup of the EPR (Einstein, Podolsky, Rosen) thought experiment of $1937$ as re-imagined by Bohm \cite{DB51} and Bell \cite{Bell:1987hh} which is given by a 
neutral pion particle decaying into a pair of electron and positron. This is given by the 
decay process

\begin{eqnarray}
\pi_0 \longrightarrow e^{-}+e^{+}.
\end{eqnarray}
The electron and the positron fly away in opposite directions due to the conservation of momentum, i.e. since the pion decays at rest. 
The state vector of the system electron+positron is a maximally entangled spin state (known as Bell's state) given by the singlet state 
\begin{eqnarray}
\frac{1}{\sqrt{2}}(|+\rangle|-\rangle-|-\rangle|+\rangle).
\end{eqnarray}
Thus, if the spin of the electron is up, then the spin of the positron must be down, and if the spin of the electron is down, then the
spin of the electron must be up. The probability for each possibility is given by $1/2$.

We leave now the electron and positron fly away from each other, in this entangled state, until their distance apart 
becomes as large as desired, perhaps of the order of the diameter of the observable universe, i.e. we allow the two systems to become causally disconnected. Then, we (or rather Alice) perform 
a spin measurement on say the electron. 

The idea 
behind this gedanken experiment is that the electron and positron are allowed to separate as far apart as possible, that there 
is strictly no possible causal mutual influence between the two, and hence when Alice performs our measurement on the electron
on this side of the universe, the measurement on the spin of the positron (by Bob) on the other side of the universe can be supposed to be a completely uncorrelated measurement, yet because of entanglement and collapse this measurement is really completely determined.

What does quantum mechanics say precisely about such a situation?

The measurement of the spin of the electron by Alice will yield the values 
$\pm 1/2$ with 
equal probability $1/2$, and similarly the measurement of the spin of the positron by Bob will yield the values 
$\pm 1/2$ with 
equal probability $1/2$ . 

But because the system is found in an entangled state, the measurement of the spin of the electron by Alice 
 will collapse the wave function, and as a consequence the spin of the positron can be determined with certainty, regardless of the measurement of Bob. For example, if when we measure 
the spin of the electron we find spin up, the wave function collapses which means the spin of the positron, which is on
the other side of the universe, 
must be down with certainty. And if we find the spin of the electron to be down, then we know that  
the spin of the positron must be up without any further measurement. Thus, the effect of the collapse propagates 
instantaneously even from one side of the universe to the other side, and this is what Einstein has called 
''spooky action at a distance``, which goes against the spirit of relativity. 

The solution according to Einstein lies 
in the reality of the wave function, i.e. the spin of the electron is well defined even before measurement. In other words, it is really either up or down 
before measurement. And thus quantum mechanics in
allowing us to only calculate probabilities is simply an incomplete  theory. Hence the need for hidden variables, i.e. the wave function ψ must be supplemented by
an extra variable (hidden variable) λ which allows a complete specification of the state of the system. 

As discussed above 
Bell has shown that any local deterministic hidden variable theory can not reproduce all the predictions of quantum mechanics. 

\begin{figure}[htbp]
\begin{center}
  \includegraphics[width=15.0cm,angle=0]{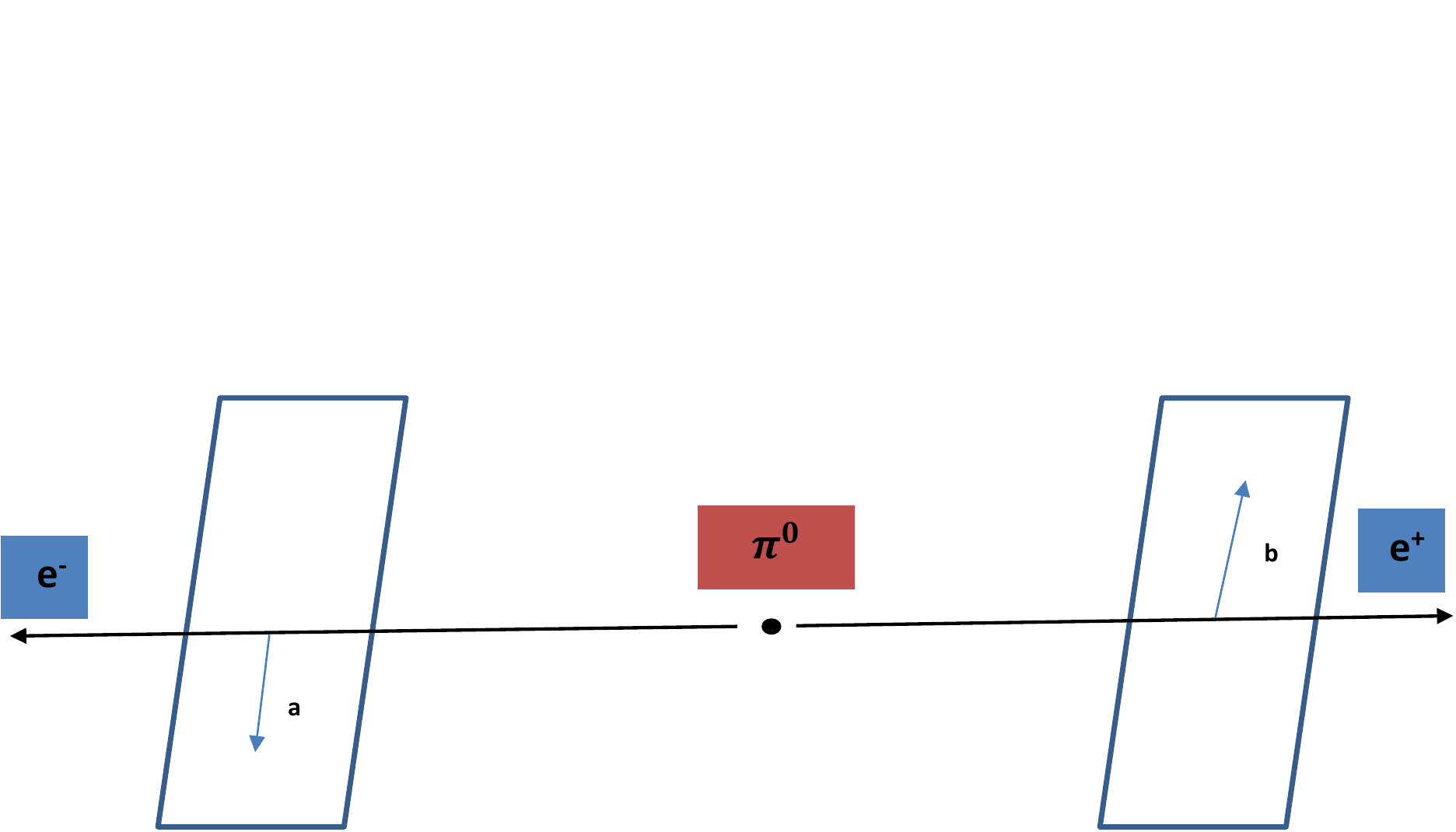}
\end{center}
\caption{The EPRBB experiment.}\label{eprbb}
\end{figure}

This goes as follows (we follow the simplified presentation of \cite{GR94}). We start by following Bell in measuring the spin $S_{1a}$ of the electron in an 
arbitrary direction $\vec{a}$, while  
measuring the spin $S_{2b}$
of the positron in another arbitrary dimension $\vec{b}$. See figure (\ref{eprbb}).

According to the rules of standard quantum mechanics 
outlined above the expectation 
value of the product of the two spins, viz $\langle S_{1a}S_{2b}\rangle $, is given by the scalar product 

 \begin{eqnarray}
P(a,b)=-\vec{a}.\vec{b}.
\end{eqnarray}
But according to Einstein, and all those who have a natural inclination towards local realism and hidden variables, the wave function 
$\psi$ comes with a hidden variable $\lambda$ given by some probability density $\rho(\lambda)$ satisfying as usual
\begin{eqnarray}
\rho(\lambda)>0~,~\int d\lambda \rho(\lambda)=1.
\end{eqnarray}
This is the assumption of realism.

We will further assume locality, which here means the requirement that 
the directions $\vec{a}$ 
and $\vec{b}$ are freely and independently chosen, and which also means in general that physical actions can not 
propagate faster than the speed of light, and thus 
measurements made at places which are space-like separated can not influence each other.

Thus will also assume that the 
measurement $S_{1a}$ and $S_{2b}$ 
of the spins of the electron and positron 
are given by two functions $f$ and $g$ which can only take the two values $\pm 1$, i.e. $f(\vec{a},\lambda)=\pm 1$ and $
g(\vec{b},\lambda)=\pm 1$, such that when the two spins are aligned we get precisely anti-correlated measurements, viz 
\begin{eqnarray}
  f(\vec{a},\lambda)=-g(\vec{a},\lambda).
\end{eqnarray}
The expectation value $P(a,b)$ for the product of two spins  should then be given by the equation 
\begin{eqnarray}
P(a,b)=\int \rho(\lambda)f(\vec{a},\lambda)g(\vec{b},\lambda) d\lambda.
\end{eqnarray}
Any deterministic local hidden variable theory with these general properties
will then give an
expectation value $P(a,b)$ satisfying, for three arbitrary directions $\vec{a}$, 
$\vec{b}$, $\vec{c}$, the inequality 
\begin{eqnarray}
|P(a,b)-P(a,c)|<1+ P(b,c).
\end{eqnarray}
This very simple result is the celebrated Bell's inequality. 

As it turns out, this hidden variable's result is quite incompatible with the 
above quantum 
mechanical prediction, i.e. with $P(a,b)=-\vec{a}.\vec{b}$. For example, if $\vec{a}$ is perpendicular to $\vec{b}$, and 
$\vec{c}$ makes 
a $45$ degree angle with $\vec{a}$ and $\vec{b}$, we obtain  \cite{GR94}

\begin{eqnarray}
P(a,b)=0, P(a,c)=P(b,c)=-0.7.
\end{eqnarray}
This is clearly not satisfied by Bell's inequality.

\begin{figure}[htbp]
\begin{center}
  \includegraphics[width=10cm,angle=-0]{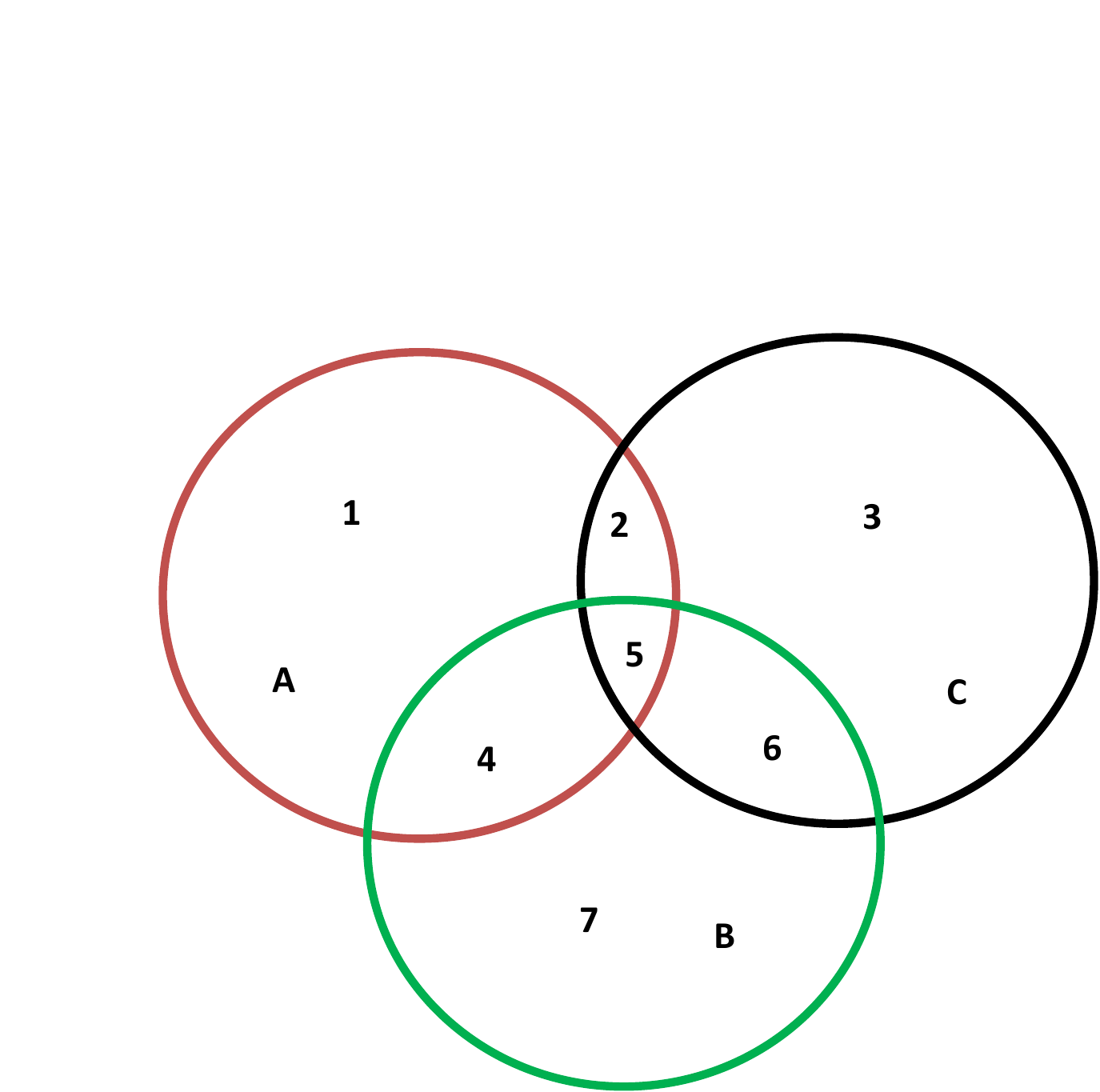}
  \caption{Bell's inequality with Venn sets.}\label{venn}
\end{center}
\end{figure}

To highlight the severity of this violation we consider a simple problem from set theory and logic. Let $A$, $B$ and $C$ three properties with corresponding sets indicated by the Venn diagrams on figure (\ref{venn}).

Let ${\cal N}_1$ be the number of objects which have {property $A$ but not property $B$}, i.e. ${\cal N}_1=N_1+N_2$. And let ${\cal N}_2$ be the number of objects which have {property $B$ but not property $C$}, i.e. $ {\cal N}_2=N_7+N_4$. And let ${\cal N}_3$ be the number of objects which have {property $A$ but not property $C$}, i.e. ${\cal N}_3=N_1+N_4$.

It is then trivial to show (the proof is visual from the Venn diagrams below) that
\begin{eqnarray}
  {\cal N}_1+{\cal N}_2\geq {\cal N}_3.
  \end{eqnarray}
This is Bell's inequality in this context. We are saying that this logical inequality is badly violated by quantum mechanics and nature.

Thus, Bell's has shown in a very simple way that if Einstein's local realism 
is correct, then quantum mechanics is not merely incomplete but it is plainly wrong. On the other hand, if quantum mechanics 
is correct, then no local hidden variable theory can be made consistent with quantum mechanics.

The Aspect, Grangier and Roger 
experiment of $1982$ has decisively shown that Bell's inequalities are violated in reality and quantum mechanics 
predictions are fully vindicated.

For some recent work on the violations of Bell's inequality see \cite{Hensen:2015ccp}.
 Thus, nature at the most fundamental of levels seems to be really not real and 
perhaps also non-local.

\subsection{Decoherence and the measurement problem}
The process of quantum measurement is one of the most fundamental aspects of quantum theory, 
which involves in an essential way many profound quantum effects such as the collapse of the wave function, 
quantum entanglement and decoherence. This makes it a very hard (in fact virtually intractable) problem (so far).  

The phenomena 
of decoherence \cite{zeh,zurek,Tegmark:2001qh} in particular involves the unavoidable interaction between the quantum system and the environment, which looks very much 
like a measurement, and thus the state of the quantum system becomes entangled with the state of the environment, and this 
in turn causes what looks like a collapse of the wave function, i.e. the reduction of a quantum pure state to a 
statistical mixture causing a loss of coherence. In other words, decoherence explains (I think 
very well) how classicality emerges from the underlying quantum nature. What remains debatable, and for some a dubious assertion, is the 
claim that the quantum measurement is, nothing more and nothing less than, the decoherence due to the coupling of 
the quantum system to
the environment (including the measurement devices, brains and minds). 

Thus the physics of decoherence is not controversial, but the claim that the quantum 
measurement is simply decoherence is still very much open to debate. For one, the Copenhagen school still maintains that the state of the 
system does not exist before measurement, while the many-worlds interpretation maintains that it exists in various branches of the many-worlds, 
and consistent histories maintains that the state is simply indeterminate. As it seems decoherence does not favor any of these positions over the 
others.

In summary, we have:
\begin{itemize}
\item{} Decoherence is the unavoidable interaction between the quantum system (open system, i.e. Schrodinger equation is inapplicable) and its environment. Coherence of the wave function is destroyed and {quantum entangled pure states} are turned into {classical statistical mixtures}. 

\item{} The {boundary} between the quantum (linear combination) and the classical (determinism) is therefore dynamically determined by decoherence and not by the act of measurement. This in fact is an {interpretation} of the collapse of the wave function!

\item{} The {measurement} yields an {entangled state} of the quantum system and the measuring device (detector). This entangled state obeys Schrodinger equation and is a correlated and non-separable state (Aspect experiment) which violates Bell inequalities.
  
\item{} Conclusion: {The states of the quantum system in the entangled state do not, and in fact can not, exist before measurement}.
 \end{itemize}
However, an observer who did not inspect the detector will describe the system by a {density matrix}. The density matrix $\rho_e$ associated with the entangled state is pure (interference terms or coherences). {Measurement} will take this pure density matrix to a reduced density matrix $\rho_r$ which is mixed (non-unitary, collapse, no coherences).

To illustrate this fundamental point, we consider as a system S a spin one-half particle with two states $|+\rangle$ and $|-\rangle$. The detector D clicks if it measures spin up and does nothing otherwise.

The interaction between the system S and the detector D produces then an entangled pure state as follows
  \begin{eqnarray}
 |\Phi_i\rangle=(\alpha|+\rangle+\beta|-\rangle)|D-\rangle\Rightarrow\alpha |+\rangle|D+\rangle+\beta|-\rangle|D-\rangle=|\Phi_e\rangle.
\end{eqnarray}
This pure state defines a pre-measurement or an incompleted  measurement and it is alternatively described by the pure density matrix
  \begin{eqnarray}
 \rho_e&=&|\Phi_e\rangle\langle\Phi_e|\nonumber\\
 &=&|\alpha|^2|+\rangle|D+\rangle\langle+|\langle D+|+|\beta|^2|-\rangle|D-\rangle\langle-|\langle D-|\nonumber\\
 &+&\alpha\beta^*|+\rangle|D+\rangle\langle-|\langle D-|+\beta\alpha^*|-\rangle|D-\rangle \langle +|\langle D+|.\label{rhoc}
 \end{eqnarray}
The completed measurement is described by the mixed density matrix (off-diagonal/interference terms are canceled)
  \begin{eqnarray}
 \rho_r
 &=&|\alpha|^2|+\rangle |D+\rangle \langle +|\langle D+|+|\beta|^2|-\rangle |D-\rangle \langle-|\langle D-|.
  \end{eqnarray}
The (discontinuous, irreversible, instantaneous, non-deterministic and non-unitary) transition $\rho_e\longrightarrow \rho_r$ is the collapse postulate in Copenhagen. But decoherence claim that it is a dynamical process obtained by taking into account the {environment}.

  Yet, there is no known process which effectuates the transition $\rho_e\longrightarrow \rho_r$. This is the measurement problem.

  We should then consider the system consisting of the quantum system S, the detector D and the environment E. The interaction of the environment  with $S+D$ is also described by an entangled state and a pure density matrix as follows
 \begin{eqnarray}
&&|\Phi_e\rangle|E_0\rangle =(\alpha |+\rangle |D+\rangle +\beta|-\rangle |D-\rangle)|E_0\rangle \Rightarrow \alpha |+\rangle |D+\rangle |E+\rangle +\beta|-\rangle |D-\rangle |E-\rangle=|\Psi\rangle.\nonumber\\
  \end{eqnarray}
     {The density matrix of the system $S+D$ is then obtained by tracing over the degrees of freedom in the environment E (which are inaccessible) to obtain the reduced density matrix}, viz
     \begin{eqnarray}
      \rho_{S+D}&=&Tr_{E}|\Psi\rangle \langle \Psi|\nonumber\\
      &=&|\alpha|^2|+\rangle |D+\rangle |+\rangle \langle D+|+|\beta|^2|-\rangle |D-\rangle |-\rangle \langle D-|\nonumber\\
&=&\rho_r.
    \end{eqnarray}
  This is the claim of decoherence! or more precisely the claim of those who use decoherence to interpret the collapse or reduction of the state vector.   

\subsection{The many-worlds formalism}

	\subsubsection{The many-worlds and coherent branching}
The many-worlds formalism was introduced in 1957 by Hugh Everett III in his doctoral dissertation under Wheeler \cite{mw1,mw2}. He actually dubbed it the "{relative state formulation}" and it is reported that he was in fact dismissive of the term "many-worlds" introduced by DeWitt and Graham \cite{DG} when they revived this formalism in the 1970's.

The only postulate of the many-worlds formalism is the unitary evolution in time given by the Schrodinger equation which is the only admitted process (as opposed to Copenhagen's two processes). The wave function is of a real ontology not of a merely descriptive value and the collapse of the wave function never occurs.

In the measurement process the collapse of the wave function is replaced by the splitting or branching (which is a fully reversible and unitary process) of the world which is a literal and direct reflection of the linear superposition principle. The frequency of branching is given precisely by Born's rule.

The many-world formalism is complementary to the Copenhagen not contradictory (this is this author's view). This is much stronger than the view (Susskind) that the Copenhagen is a very good approximation of the many-worlds. In fact the Copenhagen is thought of as an exact statement with respect to the conscious/zmobie observer whereas the many-worlds is an exact statement with respect to a super-observer who does not interfere with the world.

An analogy due to Tegmark \cite{Tegmark:1997me} is to think of  the many-worlds as playing the role of the manifold structure of spacetime in general relativity whereas the Copenhagen plays the role of the local flatness observed by every observer around each point in spacetime.

Thus the non-unitary observer-participancy (as Wheeler puts it \cite{wheeler78,Wheeler}) or consciousness-causes-collapse (von Neumann-Wigner interpretation \cite{Wigner1,Wigner2})  found in the single-world of the Copenhagen interpretation is replaced by a unitary many-worlds formalism (or a many-minds formalism \cite{AL88,Lock} in which the unitary branching occurs in the mind and not in the world). In some sense the extreme view of a non-unitary efficacious role of consciousness in quantum mechanics (quantum dualism in a single-world  \cite{stapp}) is dual/complementary to a no less extreme view of a unitary reality with many coherent and parallel worlds (physicalism in a many-worlds).


	\subsubsection{The Schrodinger's cat and quantum immortality}

        The Schrodinger's cat experiment is definitely among the greatest quantum experiments ever devised. It was introduced by Schrodinger in 1935 to highlight the conceptual problems with the Copenhagen interpretation \cite{Schrodinger0,Schrodinger}. The physical system here is a {conscious~cat}. Thus, there is an object (the cat), a subject (the observer, i.e. mind or the detector) and the inaccessible and unavoidable environment.

        The object and the subject are related through the quantum measurement.
   If no measurement is made on the cat then the state of the cat is a {linear superposition} given by
  \begin{eqnarray}
    |{\rm cat}\rangle=\frac{1}{\sqrt{2}}(|{\rm alive}\rangle+|{\rm dead}\rangle).
  \end{eqnarray}
  \begin{itemize}
  \item Question 1:Is the cat dead and alive in the same time or is she neither dead nor alive?
  
  \item Question 2: When a measurement is performed what do we find?
  \end{itemize}
  The many-worlds answer (no collapse, branching, wave function has an objective reality) is given by the state vector
  \begin{eqnarray}
    \frac{1}{\sqrt{2}}(|{\rm alive}\rangle|{\rm happy}\rangle+|{\rm dead}\rangle|{\rm dead}\rangle).
  \end{eqnarray}
  Thus, there is a branch of the many-worlds in which the cat is alive and another branch in which the cat is dead and the two branches are coherent. It is {decoherence} that destroys this linear superposition between the two branches and turn them into parallel (independent) worlds. Thus decoherence is precisely the relation between object and environment from one hand and subject and environment from the other hand. In some sense decoherence acts as a measurement.

   So the cat is alive in one world and is dead in another world and the two worlds are both genuinely real. This is to be contrasted with Copenhagen which states that these states do not exist before measurement.
  
  In summary, the Copenhagen destroys the {objectivity of classical reality} by giving the subject a special role (but reality is really not classical but quantum and the quantum dualism describes it perfectly as such!).
  
  On the other hand, the many-worlds maintains the objectivity of the classical reality which is formed (we have to accept it) of an infinite number of coherent branches and decohered parallel worlds. No special role is given to the subject.
  
In some sense the many-worlds is an external view in which the mathematics has an objective reality whereas the Copenhagen is an internal view in which the mathematics is a representation or approximation of reality \cite{Tegmark:1997me}.

Another related gedanken experiment which is profoundly puzzling is quantum immortality proposed by Tegmark \cite{Tegmark:1997me}. It is claimed to be the only  experiment which can
discriminate between Copenhagen and many-worlds.

In the current case the Schrodinger's cat is replaced by the Schrodinger's experimenter. A quantum gun  is prepared in the state
  \begin{eqnarray}
    \frac{1}{\sqrt{2}}(|{\rm up}\rangle+|{\rm down}\rangle).
  \end{eqnarray}
The trigger will be pulled and the gun fires if the measurement of a qubit yields the value $-1$ otherwise nothing happens. We repeat $n$ times. {According to the Copenhagen the probability of survival after $n$ steps is obviously $1/2^n$.

  The state after the first measurement according to the many-worlds is
 \begin{eqnarray}
    \frac{1}{\sqrt{2}}(|{\rm up}\rangle|{\rm alive}\rangle+|{\rm down}\rangle|{\rm dead}\rangle).
 \end{eqnarray}
 We suppose $1)$ {oblivion} (no physical consciousness after death), and $2)$ {continuity of identity} (time between measurements is much smaller than the time of human consciousness).

 We immediately conclude that the experimenter will find herself alive in each step, i.e. probability of survival is $1$. This is because  there is one conscious person after and before the experimenter, her identity is continuous, and the other persons in the other branches have all suffered oblivion.

 In fact, the experimenter in most branches is dead but there exists one branch where the experimenter survives, and because the assumptions of oblivion and continuity of identity hold, the experimenter never dies and she acts as if she is immortal.  However, this experimenter will be the only person who knows this and thus she can objectively
 discriminate between Copenhagen and many-worlds  favoring the many-worlds.

\subsubsection{The many-minds interpretation}

The many-mind interpretation is a very close relative of the many-world interpretation which involves the following crucial modification/twist: The split or branch of the world into parallel branches when a quantum measurement is performed is shifted to a split or branch of the mind into parallel minds. 
In both interpretations, it is assumed that quantum mechanics as it stands is a complete theory of nature and that there is no collapse of the wave function under measurement which is what is accounted for by the splitting into branches. 

This means in particular that the fundamental law is given by the Schrodinger equation alone, and the relationship between branching and relative frequencies, which is ill defined  a priori in both pictures, should be given for consistency by Born's rule.

The most important versions of the many-mind interpretations are:
\begin{itemize}
\item Albert and Loewer theory \cite{AL88}. This is an in intrinsically dualistic theory which assumes in the words of Albert "that every sentient physical system there is is associated not with a single mind but rather with a continuous infinity of minds". However, in this theory, there is no supervenience of brane states and mind states.
\item Lockwood theory \cite{Lock}: In this theory there is a complete supervenience of the physical and mental.
\end{itemize}
In the following we will follow \cite{HP}.

Before we begin we mention few other implications of the Albert-Loewer theory which is the most important one for us here because of its dualistic character.

Firstly, an epistemological implication of the Albert-Loewer theory is the observation that our current experience could be fully compatible with the fact that the universe has always been in the vacuum state. This seem to me to be also an ontological implication. Another implication of the Albert-Loewer theory is that quantum non-locality is removed completely from the physical and delegated to the mental world.
In fact all many-worlds and many-minds interpretations are no-collapse models and they avoid non-locality by claiming that Bell correlations (predicted by Bell's theorem) are not fully objective correlations but they are observer-dependent. Price critique of these claims reach the conclusion that these no-collapse models do not really eliminate non-locality but they simply explain it.

The system $S$ under consideration is assumed to be composed of a single electron. We are interested in the measurement of the $z$ component of the spin. The measurement apparatus is denoted $M$ and the observer is denoted $O$. The total system $S+M+O$ is initially prepared in the state
\begin{eqnarray}
  |\Psi_0\rangle=(\alpha|-\rangle+\beta|+\rangle)\otimes|\psi_0\rangle\otimes|\phi_0\rangle.
  \end{eqnarray}
The state $|\psi_0\rangle$ is the initial state of the apparatus and $|\phi_0\rangle$ is the initial state of the observer's brain. The complete state $|\Psi_0\rangle$ is supposed to obey the Schrodinger equation only. In other words, we assume that there is no collapse.

The measurement interaction between the system $S$ and the measurement apparatus $M$ creates a one-to-one correlation between the states of up and down spins of the electron and the pointer states $|\psi_{\pm}\rangle$ of the apparatus. Hence the system $S$ and the measurement apparatus $M$ become entangled, i.e. the measurement interactions results in taking the above state $|\Psi_0\rangle$ to the combination
\begin{eqnarray}
  |\Psi_1\rangle=(\alpha|-\rangle\otimes|\psi_+\rangle+\beta|+\rangle\otimes|\psi_-\rangle)\otimes|\phi_0\rangle.
  \end{eqnarray}
Next we assume that the brain states corresponding to all possible outcomes of all possible experiments form a preferred basis in the brain's Hilbert space. These states correspond to those mental states associated with the conscious perception of the outcomes of the experiments. Let us denote the two brain states associated with the conscious perception of the states of up and down spins of the electron by $|\phi_{\pm}\rangle$. Then the interaction between the measurement apparatus $M$ and the observer $O$ will take the state $|\Psi_1\rangle$ to the final state
\begin{eqnarray}
  |\Psi_f\rangle=\alpha|-\rangle\otimes|\psi_+\rangle\otimes |\phi_+\rangle+\beta|+\rangle\otimes|\psi_-\rangle\otimes|\phi_-\rangle.
  \end{eqnarray}
The measurement has no definite result and thus this theory (called the bare theory by Albert) is not complete and it should then be supplemented by extra ingredients.

The many-mind interpretation of Albert and Loewer is a no-collapse interpretation in which we suppose that the bare theory is complete with respect to the physics including the brain. However, regarding the relationship of the brain states $|\phi_{\pm}\rangle$ to the mental states of the observer $O$ we also assume the following two postulates:

\begin{itemize}
\item Each brain state $|\phi\rangle$ is associated at all times with an infinity of non-physical minds.
\item The minds do not obey the Schrodinger equation but evolve in time in a stochastic way with a probability given by the Born rule.
\end{itemize}
Thus we start with an infinity of minds associated with the initial brain state $|\phi_0\rangle$. In some sense the minds are degenerate described all by the single brain state $|\phi_0\rangle$. Each mind then evolves in a stochastic way to either the state $|\phi_+\rangle$ with the Born probability $|\alpha|^2$ or to the state $|\phi_-\rangle$ with the Born probability $|\beta|^2$. Thus the state $|\Psi_f\rangle$ should be replaced by
\begin{eqnarray}
  |\Psi_f(m,n)\rangle=\alpha|-\rangle\otimes|\psi_+\rangle\otimes |\phi_+(m)\rangle+\beta|+\rangle\otimes|\psi_-(n\rangle\otimes|\phi_-\rangle.
  \end{eqnarray}
The notation $|\phi(m)\rangle$ means that the brain state $|\phi\rangle$ corresponds to and is indexed by the subset $m$ of the set of minds. In other words, the description of the post measurement state includes the quantum states of the system $S$ and of the apparatus $M$ and the subsets of the set of minds in the $+$ and $-$ branches of the superposition.

This interpretation is truly probabilistic since before the divergence of the minds into the branches of the state $|\Psi_1\rangle$ it is fully random which branch each mind will actually follow. The probability is given by the quantum mechanical Born rule. The requirement of an infinite number of minds is put forward in order to avoid 1) the so-called mindless hulk problem and also in order to avoid 2) Bell's non-locality.

More importantly is the fact that this interpretation is dualistic in the sense that only subsets of the set of minds (and not individual minds) supervene on the brain states. Thus any $m-$mind can be exchanged  with any $n-$mind leaving the physics invariant.

The other issue concerns the relationship between branching and relative probability which is a major problem in the many-world interpretation as well. This is solved by simply assuming the Born rule as shown originally by Everett in $1957$. It can then be proved that the probability of each branch on a given tree is given by the quantum mechanical Born rule and that each individual mind performs a classical random walk on this tree with this probability. This does not mean that there exists a non-contextual classical probability distribution which can assign the correct probability to all branches at once in accordance with the violation of Bell's inequality.

Indeed, the probability of the branching must be conditional on the measurement performed. If the probability were pre-determined then the minds will act as hidden variables and they will necessarily violate Bell's inequality, i.e. we have a non-local hidden variables theory. Hence in order to avoid this non-locality we must assume a random distribution of the minds which is conditional on the given measurement.

In Lockwood interpretation there is a complete supervenience of the (continuous infinity of) minds on the brain states. Also the minds are supposed to be not stochastic. Thus the final post measurement state is given by $|\Psi_f\rangle$ and not $|\Psi_f(m,n)\rangle$. In other words, subsets of minds are indexed by brain states as opposed to brain states being indexed by subsets of minds. Thus the fraction of minds in the branch $+$ is proportional to $|\alpha|^2$ whereas the fraction of minds in the branch $-$ is proportional to $|\beta|^2$. On the other hand, the dynamics of minds, which is not random in this case, is not clear and some possibilities are discussed for example in \cite{HP}.

\subsection{Bohmian mechanics}
\subsubsection{A deterministic non-local theory}
Bohmian mechanics is the only deterministic, and thus causal, hidden variable interpretation theory of quantum mechanics which was conceived originally by de Broglie \cite{debroglie} and then really constructed by Bohm \cite{Bohm:1951xx0}. It is the only explicit non-local formulation of quantum mechanics, through the introduction of the so-called quantum potential, which thus attempts to reflect, consciously or otherwise,  the non-local, i.e. action at a distance, character of physical reality as described by quantum mechanics.

Being non-local means in particular that Bohm evades the constraints imposed by Bell's theorem, which was confirmed experimentally for example by Aspect et al.,  by considering a non-local hidden variable extension of quantum mechanics.

The state of the system in this interpretation is given by the usual wave function $\psi$ in the Hilbert space together with the usual generalized coordinates $q_i$ of classical mechanics. As usual, the set of generalized coordinates $q_i$ defines a point $\vec{q}$ called a configuration in configuration space. It can be argued that the wave function is in fact the hidden variable here since it is not measurable as opposed to the measurable positions $q_i$ \cite{DGZ}. A more serious discrepancy is the fact that the $q_i$ are actually the classical positions not the actual quantum positions $\langle\hat{x}_i\rangle$.  This discrepancy can be alleviated somewhat if we keep in mind this difference and translate back to the actual quantum positions whenever is needed.

The evolution of the positions $q_i$ in time is  given in terms of the wave function $\psi$ itself by means of the so-called guiding equation. Hence Bohmian mechanics contains besides the usual Schrodinger equation, which governs the evolution of the wave function $\psi$, the guiding equation which governs the evolution of the  configuration $\vec{q}$ in terms of the wave function. The evolution of positions $q_i$ is then in a clear sense guided by the wave function. This is clearly a deterministic system. Thus according to Bohm quantum mechanics is as deterministic as classical mechanics.

As pointed out originally by Bohm himself   the predictions of Bohmian mechanics and quantum mechanics should fully coincide. He says in \cite{Bohm:1951xx0}: "as long as the present general form of Schrodinger's equation is retained the physical results obtained with this suggested alternative interpretation are precisely the same as those obtained with the usual interpretation".

This is true almost by construction as we will see. The only possible source of confusion is the existing difference between the generalized coordinates of the system ${q}_i$ and the quantum positions $\langle \hat{x}_i\rangle$.

We start then with the wave function $\psi=\psi(t,\vec{x})$ which obeys as usual the Schrodinger equation

\begin{eqnarray}
i\hbar\frac{\partial\psi}{\partial t}=H\psi~,~H=-\frac{\hbar^2}{2m}\vec{\nabla}^2+V. \label{B1}
\end{eqnarray}
Following Bohm's original derivation we polar decompose the wave function as

\begin{eqnarray}
\psi=R\exp(i S/\hbar).
\end{eqnarray}
We compute immediately

\begin{eqnarray}
i\hbar\frac{\partial \psi}{\partial t}=(i\hbar\frac{\partial\ln R}{\partial t}-\frac{\partial S}{\partial t})\psi.
\end{eqnarray}
And
\begin{eqnarray}
H\psi=-\frac{\hbar^2}{2m}\bigg(\frac{1}{R}\vec{\nabla}^2R+\frac{2i}{\hbar}\vec{\nabla}S\vec{\nabla}\ln R+\frac{i}{\hbar}\vec{\nabla}^2S-\frac{1}{\hbar^2}(\vec{\nabla}S)^2\bigg)\psi+V\psi.
\end{eqnarray}
By equating the two terms we get

\begin{eqnarray}
\frac{\partial S}{\partial t}+\frac{1}{2m}(\vec{\nabla}S)^2+V-\frac{\hbar^2}{2m}\frac{\vec{\nabla}^2R}{R}=i\hbar\bigg(\frac{\partial\ln R}{\partial t}+\frac{\vec{\nabla}S}{m}\vec{\nabla}\ln R+\frac{1}{2m}\vec{\nabla}^2S\bigg).\label{5}
\end{eqnarray}
Now we introduce the velocity operator $\vec{\hat{v}}$ acting on the Hilbert space ${\cal H}$ through the correspondence principle, viz

\begin{eqnarray}
\vec{\hat{v}}\psi=\frac{\vec{\hat{p}}}{m}\psi=\frac{1}{m}\frac{\hbar}{i}\vec{\nabla}\psi=\frac{1}{m}\bigg[\frac{\hbar}{i}\vec{\nabla}\ln R+\vec{\nabla}S\bigg]\psi.
\end{eqnarray}
In the classical limit $\hbar\longrightarrow 0$ the action of this velocity operator becomes simply

\begin{eqnarray}
\vec{{v}}\psi=\frac{\vec{{p}}}{m}\psi=\frac{1}{m}\vec{\nabla}S\psi.
\end{eqnarray}
Thus, $S$ is Hamilton's principal function (effectively what we call the action).

Bohm apparently defines the velocity not as an operator on the Hilbert space but as the rate of change $\vec{v}_{\psi}(t,\vec{x})$ of the so-called configuration $\vec{Q}(t,\vec{x})$ of the system by the relation 

\begin{eqnarray}
\vec{v}_{\psi}=\frac{d\vec{Q}}{dt}=\frac{1}{m}\vec{\nabla}S.
\end{eqnarray}
This rate of change is equal to the classical velocity and not to the quantum velocity and $\vec{Q}$ is the hidden variable we need to adjoin to the wave function $\psi(t,\vec{x})$ in order to obtain a complete description of the system.  This definition is also motivated by the definition of the probability current density (see below). In terms of the wave function, Bohm's velocity can then be rewritten as

\begin{eqnarray}
\vec{v}_{\psi}=\frac{d\vec{Q}}{dt}=\frac{\hbar}{m}{\rm Im}\frac{\psi^*\vec{\nabla}\psi}{\psi^*\psi}.\label{B2}
\end{eqnarray}
This form can also be deduced on general grounds by employing symmetry considerations: Galilean invariance (normalization), time reversal (complex conjugation) and rotational invariance (derivation), etc \cite{DGZ}.

Thus, the wave function provides the source for Bohm's velocity $\vec{v}_{\psi}$ which means in particular that Bohm's position of the particle, which is given by $\vec{Q}$,  is guided by the wave function and hence the name "pilot wave" of this interpretation. In the words of \cite{DGZ}: "the wave function governs the evolution of the position of the particle".

Equations  (\ref{B1}) and (\ref{B2}) where the state of the system is given by the pair $(\psi,\vec{Q})$ define Bohmian (Bohemian) quantum mechanics.

The configuration $\vec{Q}$ lives in a configuration space similar to the configuration space of generalized coordinates found in classical mechanics. Thus, $\vec{Q}$ which is a point in a configuration space is not the same as $\vec{\hat{x}}$ which is an operator on the Hilbert space. However, in Bohemian mechanics what is interpreted as the actual vector position of the quantum particle in ordinary space is in fact $\vec{Q}$ and not the eigenvalue $\vec{x}$ of the vector position operator $\vec{\hat{x}}$. The difference between the two velocities is also exhibited by the fact that since $\vec{Q}$ and $\vec{v}_{\psi}$ are sourced by the wave functions they must depend on $\vec{x}$ and $t$. Only in the classical limit the two of course coincides.

We go back now to equation (\ref{5}) and substitute Bohm's velocity. We get

\begin{eqnarray}
\frac{\partial S}{\partial t}+\frac{1}{2}m\vec{v}_{\psi}^2+V+U=\frac{i\hbar}{2\rho}\bigg(\frac{\partial\rho}{\partial t}+\vec{v}_{\psi}\vec{\nabla}\rho+\rho\vec{\nabla}\vec{v}_{\psi}\bigg),\label{10}
\end{eqnarray}
where we have introduced the probability density in the usual way

\begin{eqnarray}
\rho=R^2=\psi^{*}\psi,
\end{eqnarray}
and $U$ is the so-called quantum potential defined by

\begin{eqnarray}
U=-\frac{\hbar^2}{2m}\frac{\vec{\nabla}^2R}{R}.
\end{eqnarray}
Let us also recall the continuity equation (by using the Schrodinger equation)

\begin{eqnarray}
\frac{\partial}{\partial t}(\psi^*\psi)=-\frac{\hbar}{2im}\vec{\nabla}(\psi^*\vec{\nabla}\psi-\psi\vec{\nabla}\psi^*).
\end{eqnarray}
The current is then defined by

\begin{eqnarray}
\vec{J}=\frac{\hbar}{2im}(\psi^*\vec{\nabla}\psi-\psi\vec{\nabla}\psi^*)=\rho\vec{v}_{\psi}.
\end{eqnarray}
The continuity equation becomes

\begin{eqnarray}
\frac{\partial}{\partial t}\rho=-\vec{\nabla}(\rho\vec{v}_{\psi}).
\end{eqnarray}
The right hand side of equation (\ref{10}) is thus equal to the continuity equation and by substitution we get also the modified (by the quantum potential) Hamilton-Jacobi equation

\begin{eqnarray}
\frac{\partial S}{\partial t}+\frac{1}{2}m\vec{v}_{\psi}^2+V+U=0\Rightarrow -\frac{\partial S}{\partial t}=\frac{1}{2m}(\vec{\nabla}S)^2+V+U.
\end{eqnarray}
This is truly a deterministic formalism since the configuration  (the quantum vector position) $\vec{Q}=\int dt \vec{\nabla}S/m$ obeys classical dynamics with an extra piece (the quantum potential $U$) added to the potential. But it is a non-local formulation since the evolution in time of $\vec{Q}$ is sourced by the wave function $\psi$ which can exist everywhere. The Born rule is imposed here as an initial condition on the wave function.


\subsubsection{Beables} 
Bell is without doubt the most profound thinker about quantum mechanics of all time. He is the originator of Bell's theorem which remains one of the most fundamental concrete results in the foundation of quantum mechanics which has also been confirmed experimentally. In this section, we will discuss one of his ingenious interpretation of quantum mechanics \cite{Bell:1982xg,Bell:1984bf} which is based on Bohm's interpretation of non-relativistic many-particle quantum mechanics \cite{Bohm:1951xx0}. See also \cite{sud1,sud2}.

In Bohm's deterministic theory, particles have always definite positions and their motion is fully deterministic guided by the wave function (a pilot wave as Bell called it) which acts as a quantum force rather than as a description of the state of the system.

By analogy, in Bell's indeterministic theory we consider a set of commuting observables (operators or variables) called the beables which then can be diagonalized simultaneously with simultaneous eigenspaces denoted by $S_i$ called the viable subspaces.  In other words, the commuting observables have definite eigenvalues on these eigenspaces so that the actual state of the system is a state vector in one of the viable subspaces $S_i$.

The evolution of this state is however governed by the pilot wave which is a state vector $|\psi(t)\rangle$ obeying the Schrodinger equation with a Hamiltonian given by the physics of the system. This is in direct analogy with the fact that the position of the particles in Bohm's theory (here played by the eigenvalues of the commuting observables) are guided by the Schrodinger wave function. The real state of the system at any given time $t$ is given by one of the components $|\psi_i(t)\rangle$ of the pilot wave, viz
\begin{eqnarray}
  |\psi(t)\rangle=\sum_i |\psi_i(t)\rangle.
  \end{eqnarray}
Now the real state changes in time stochastically (this is where the indeterministic component enters the formalism) by making transitions between the viable subspaces with transition probabilities given by Bell's postulate which I will not state explicitly here.

The end result is that the probability $p_i(t)$ that the real state at any time $t$ is $|\psi_i(t)\rangle\in S_i$, if the probability at the initial time $t=0$ is given by the Born rule,  is also given by the quantum mechanical Born rule

 \[p_i(t)=\langle\psi_i(t)|\psi_i(t)\rangle.\]
It can also be shown that the above Bell's indeterministic theory reduce in the continuum limit (to be defined) to Bohm's deterministic theory.

Thus we can have a theory in which any chosen set of commuting observables have a definite value yet the results of measurements are given by the probabilities of quantum mechanics.

\subsection{On observer-participancy or consciousness}

\subsubsection{Wigner's friend experiment}

The Wigner's friend experiment is one of the most profound gedanken experiments ever devised. It is an extension of the Schrodinger's cat experiment in which the cat is replaced by Wigner's friend. It shows among other things that the collapse of the wave function is a fundamentally different process than unitarity, and in fact collapse can not be reduced to unitarity, and furthermore it shows that the conscious mind seems indeed to play a genuine real role in measurement. 

In some sense collapse is an entirely different interaction in the universe, a sort of a fifth force so to speak, which occurs  after the interaction between the conscious observer and the physical system during the process of quantum measurement.

Wigner's friend experiment can be described as follows. We consider an experimenter $F$ (Wigner's friend) performing an experiment on a two-state quantum system: perhaps a coin $C$ with orthonormal basis states $|{\rm head}\rangle_C$ and $|{\rm tail}\rangle_C$. This coin can be replaced by Schrodinger's cat with orthonormal states $|{\rm alive}\rangle_C$ and $|{\rm dead}\rangle_C$. In Wigner 's original version \cite{Wigner1}  of this experiment this two-state quantum system is given by an object with the states $|\psi\rangle_1$, if a flash emitted by the object has been seen by the friend $F$, and $|\psi\rangle_2$ if no flash was seen.

The Wigner's friend experiment contains also Wigner $W$ who performs measurement on the joint system of friend $F$ plus the two-state system. The initial state of the two-state system is assumed to be a linear combination of $|\psi\rangle_1$ and $|\psi\rangle_2$  given by the state vector (assuming the original language of Wigner)
\begin{eqnarray}
  |\psi\rangle=\alpha|\psi\rangle_1+\beta|\psi\rangle_2.
  \end{eqnarray}
The complex numbers $\alpha$ and $\beta$ are the probability amplitudes corresponding to the pure states $|\psi\rangle_1$ and $|\psi\rangle_2$  and their modulus square $|\alpha|^2$ and $|\beta|^2$ give precisely the probabilities of seeing a flash (alive cat, head) and not seeing a flash (dead cat, tail). 

If the state of the object is $|\psi\rangle_1$ then after the interaction between the object and Wigner's friend, which occur during the measurement performed by the friend on the object, the state of their joint system becomes $|\psi\rangle_1\otimes|F\rangle_1$ where $|F\rangle_1$ is the state of Wigner's friend in which he responds to the question: have you seen a flash (dead cat, head)? with the answer "yes". Similarly, if the state of the object is $|\psi\rangle_2$ then after the measurement performed by the friend on the object the state of their joint system becomes $|\psi\rangle_2\otimes|F\rangle_2$ where $|F\rangle_2$ is the state of Wigner's friend in which he responds to the above question: have you seen a flash  (dead cat, head)? with the answer "no". By the linearity of the Schrodinger equation the joint system friend+object is described by the state vector
\begin{eqnarray}
  |\psi\rangle_{\rm joint}=\alpha|\psi\rangle_1\times|F\rangle_1+\beta|\psi\rangle_2\times|F\rangle_2.
  \end{eqnarray}
This is a maximally entangled Bell state \cite{Bell:1964kc}. 

Now, Wigner will perform his measurement on the joint system friend+object. He will ask his friend whether or not he saw a flash (dead cat, head) and inspect the object. The probabilities according to the Born rule are as follows:
\begin{itemize}
\item There is a probability $|\alpha|^2$ that the friend says "yes" and the object from then on behaves as if it is in the state $|\psi\rangle_1$ of a flash being emitted (or alive cat or head for the coin). 
\item There is a probability $|\beta|^2$ that the friend says "no" and the object from then on behaves as if it is in the state $|\psi\rangle_2$ of a flash not being emitted (or dead cat or tail for the coin). 
\item There is a probability zero that the friend says "yes" but the object from then on behaves as if it is in the state $|\psi\rangle_2$ of a flash not being emitted (or dead cat or tail for the coin). 
\item There is a probability zero that the friend says "no" but the object from then on behaves as if it is in the state $|\psi\rangle_1$ of a flash being emitted (or alive cat or head for the coin). 
\end{itemize}
If the corresponding vector states of Wigner in the cases where there is a non-zero probability are denoted by $|F\rangle_{1,2}$ the total state of the joint system friend+object+Wigner is given by the maximally entangled tripartite Greenberger-Horne-Zeilinger or GHZ state \cite{GHZ}
\begin{eqnarray}
  |\psi\rangle_{\rm total}=\alpha|\psi\rangle_1\times|F\rangle_1\times|W\rangle_1+\beta|\psi\rangle_2\times|F\rangle_2\times|W\rangle_2.
  \end{eqnarray}
Everything seems good, but is it really?

If we substitute for Wigner's friend a device, i.e. a measurement apparatus taken to be just an atom in Wigner's original description, and then repeat the experiment, everything will go through as described above, and indeed nothing can be discerned to be especially wrong about the above picture. 

However, with Wigner's friend instead of the atom performing the measurement on the object, Wigner can simply and surely ask his friend, after completing the experiment, whether or not he saw a flash before he actually had asked him. 

It is for certain that the friend will say that he saw the flash or that he did not see the flash, as the case may be, before Wigner asked him. This means in particular that in the reference frame (so to speak) of Wigner's friend the state vector, even before Wigner's measurement, was already either $|\psi\rangle_1\times|F\rangle_1$ or $|\psi\rangle_2\times|F\rangle_2$ and not their linear combination, which is in gross contradiction to the above quantum mechanical rules verified experimentally for the atom to a great accuracy. 

This is not to say that the friend's position is less reasonable  since quantum mechanics assumes him (in the reference frame of Wigner) to occupy the linear combination $|\psi\rangle_{\rm joint}$ which  implies in a clear sense as Wigner puts it: "that my friend was in a state of suspended animation before he answered my question" \cite{Wigner1}.

Wigner concludes this experiment by saying: "It follows that the being with a consciousness must have a different role in quantum mechanics than the inanimate measuring device: the atom considered above. In particular, the quantum mechanical equations of motion cannot be linear if the preceding argument is accepted."

But are we really confident that the description of the Wigner's friend experiment given above is correct. Another assumption entertained by Wigner himself is "to assume that the joint system of friend plus object cannot be described by a wave function after the interaction". And that the correct description is given in terms of a density matrix. In other words, we should describe the system by a mixed state instead of a pure state.  This corresponds also to the statement that the equation of motion becomes highly non-linear when a measurement by a conscious being is performed. Since as we have already said that the measurement (if it can be called so)  carried out by the atom is certainly described by a vector in the Hilbert space, i.e. a pure state.  The density matrix can be given by
\begin{eqnarray}
\left( \begin{array}{cc}
|\alpha|^2 & \alpha\beta^*\cos\delta  \\
\alpha^*\beta\cos\delta &  |\beta|^2 \end{array} \right).
\end{eqnarray}
Only the case $\delta=0$ corresponds to orthodox quantum mechanics, i.e. to a pure state, whereas all other states are statistical mixtures with all the properties required by the theory of measurement. The above density matrix defines a continuous transition from a pure state $|\psi\rangle_{\rm joint}$ to the mixtures $|\psi\rangle_1|F\rangle_1$ and $|\psi\rangle_2|F\rangle_2$. 

In summary, this is an objective-collapse model. In general we have \cite{BHW}

\begin{itemize}
\item i) No-collapse models such as many-worlds and Bohm. 
\item ii) Objective-collapse model which is a possible view of Wigner himself. Thus, every measurement will produce a collapse for everybody and hence in this case even for Wigner the joint total system is not described by a wave function.
\item iii) Subjective-collapse models in which every observer is assigned a collapse in her own measurement only. This is the standard view of Wigner and most of the Copenhagen school. Thus,  every measurement will produce a collapse only with respect to the observer performing the measurement. 
\end{itemize}
But does assuming the existence of consciousness and collapse imply any contradictions with physical laws. In other words, do we really have "a violation of physical laws where consciousness plays a role" as Wigner himself puts it in his article \cite{Wigner1}. We do not think this to be the case although Wigner himself has since wavered from his position (see \cite{esfeld} for a brief review). On the contrary we think that the objective-collapse model can provide a powerful physical handle on consciousness via the interaction between the universe and the mind. In other words, the Cartesian mind/body problem is not just another metaphysical theory but it can be turned by means of the collapse into a full physical theory. This picture is further strengthen with the established duality between Copenhagen and many-worlds interpretations as we will see. 

\subsubsection{The von Neumann-Wigner interpretation}

        The {Heisenberg cut} is a concept introduced by von Neumann to delineate the boundary between the observer and the observed. In classical mechanics the Heisenberg cut is at $\infty$ placing thus the observed effectively outside the influence of the observer. But in quantum mechanics its placement is arbitrary.

      The von Neumann-Wigner interpretation is an extreme limit of Copenhagen in which the Heisenberg cut is placed between the physical brain and the non-physical conscious mind. Hence, the mind is a fundamental entity not reducible to matter, i.e. it is an independent substance, and consciousness is thus a fundamental aspect of nature on equal footing with atoms and elementary particles (as advocated by Chalmers on purely philosophical grounds \cite{chalmers}).

      The physical universe is the only true quantum system and the non-physical mind is the only true measuring device. 
           It seems then that the physical world is quantum and the non-physical mind is classical  (similar to the many-minds interpretation where the minds are classical as we will discuss shortly) which creates the dichotomy.
            
  The property of unitarity holds true until information enters the conscious mind, i.e. the collapse is caused by the mind performing measurement on brain.

  In some sense the physical world does not exist without observing it  
  and the collapse is a real causal interaction between the physical world and the mind. It is in fact a {fifth force}.  The collapse in this interpretation is objective but it is not Penrose's orchestrated objective reduction Orch OR. 
           However, it is the mind that causes the collapse of the wave function (and not the other way around as in Orch OR). Thus, the collapse is the causal link between matter and mind.

Also {Cartesian dualism}  is seen as an intrinsic property of von Neumann-Wigner interpretation and viewed as a shortcoming. 
However, {quantum dualism} is fundamentally different from Cartesian dualism  as we will try to argue (see also Stapp, Albert and Lockwood).  

Another objection against the von Neumann-Wigner interpretation is due to Bohm and Hiley 
who say that "it is difficult to believe that the evolution of the universe before the appearance
of human beings depended fundamentally on the human mind" \cite{BH}. 

Bohm and Hiley themselves provided then, although in a cynical tone, an answer to this same objection by positing a universal mind.

Yet, another answer, is to assume that the mind is formed exactly of the kind of dark matter and/or dark energy which permeates the universe with real measurable effects without being directly or easily observable. In this way, the mind formed out of this dark stuff is interacting with the brain formed from the luminous stuff, in the same way that dark energy and dark matter found in the universe interact with ordinary matter. In other words, this dark energy/dark matter is in some sense the universal mind that Bohm and Hiley are positing, and hence the dark stuff in the universe acts as a universal mind which causes the collapse, upon measurement, to a particular history which we find ourselves observing from the inside.

Another speculative answer to the above objection is to simply deny the existence of the big bang (and the corresponding evolution of the universe) and to say that all history is actually fake which can be shown  by employing the argument of Price against Boltzman view concerning the emergence of time from the second law of thermodynamic \cite{price}. 


It was noted by \cite{Schreiber:1994st}  that the von Neumann-Wigner interpretation can be tested experimentally since if the mind performs a measurement on the brain at time $t$ for the position of a particle, then we can observe the effect of the collapse by measuring the momentum of the same particle at times $t−a$ and $t+a$, i.e. we can measure that the particle does not obey the Schrodinger equation!. 

\subsubsection{Extended Wigner's friend experiment}

We consider now the extension of the Wigner's friend experiment outlined recently by Frauchiger and Renner \cite{FR}.
See also Baumann, Hansen, Wolf \cite{BHW} and Sudbery\cite{S}.

In this experiment we consider Wigner $W$ and his assistant $A$ which perform measurements on Wigner's two friends $F_2$ and $F_1$ respectively which in turn performs their measurements on two two-state quantum systems: an electron spin $S$ in the states $|+\rangle $, $|-\rangle $ and a quantum coin $C$ in the states $|{\rm head}\rangle$ , $|{\rm tail}\rangle$. The experiment consists in the following 
\begin{itemize}
\item We prepare the quantum coin in the state
  \begin{eqnarray}
    |\psi_0\rangle_C=\frac{1}{\sqrt{3}}|{\rm head}\rangle+\sqrt{\frac{2}{3}}|{\rm tail}\rangle.
     \end{eqnarray}
\item At time $t_1$ the friend $F_1$ measures the face of the coin with memory states denoted by $|H\rangle_1$ if she finds head and $|T\rangle_1$ if she finds tail. The state of the joint system becomes
  \begin{eqnarray}
    |r\rangle=\frac{1}{\sqrt{3}}|{\rm head}\rangle|H\rangle_1+\sqrt{\frac{2}{3}}|{\rm tail}\rangle|T\rangle_1.
     \end{eqnarray}
The friend $F_1$ sets then the spin of the electron to $|-\rangle$ if she gets head and to $|+\rangle_x=(|+\rangle+|-\rangle)/\sqrt{2}$ if shes gets tail. The joint state of $F_1+S+C$ is
\begin{eqnarray}
  |F_1SC\rangle=\frac{1}{\sqrt{3}}|{\rm head}\rangle|H\rangle_1|-\rangle+\sqrt{\frac{2}{3}}|{\rm tail}\rangle|T\rangle_1|+\rangle_x.
  \end{eqnarray}
This can be put into the form
\begin{eqnarray}
  |F_1SC\rangle=\frac{1}{\sqrt{3}}|{\rm tail}\rangle|T\rangle_1|+\rangle+\sqrt{\frac{2}{3}}|{\rm fail}\rangle_{1C}|-\rangle,
   \end{eqnarray}
where the new state is defined by
\begin{eqnarray}
  |{\rm fail}\rangle_{F_1C}=\frac{1}{\sqrt{2}}(|{\rm head}\rangle|H\rangle_1+|{\rm tail}\rangle|T\rangle_1).
   \end{eqnarray}
The orthogonal state is naturally defined by
\begin{eqnarray}
  |{\rm ok}\rangle_{F_1C}=\frac{1}{\sqrt{2}}(|{\rm head}\rangle|H\rangle_1-|{\rm tail}\rangle|T\rangle_1).
   \end{eqnarray}
\item  At time $t_2$ the friend $F_2$ measures the spin of the electron in the basis $\{|+\rangle, |-\rangle\}$ with corresponding memory states $|U\rangle$ and $|D\rangle$. The system of the joint system $F_2+F_1+S+C$ is immediately given by
  \begin{eqnarray}
    |F_2F_1SC\rangle=\frac{1}{\sqrt{3}}|{\rm tail}\rangle|T\rangle_1|+\rangle|{U}\rangle_2+\sqrt{\frac{2}{3}}|{\rm fail}\rangle_{1C}|-\rangle|D\rangle_2.
     \end{eqnarray}
Similarly, we introduce the two states
\begin{eqnarray}
  |{\rm fail}\rangle_{F_2S}=\frac{1}{\sqrt{2}}(|-\rangle|D\rangle_2+|+\rangle|U\rangle_2).
   \end{eqnarray}
  \begin{eqnarray}
    |{\rm ok}\rangle_{F_2S}=\frac{1}{\sqrt{2}}(|-\rangle|D\rangle_2-|+\rangle|U\rangle_2).
     \end{eqnarray}
Then the state of the joint system $F_2+F_1+S+C$ can be put into the form
\begin{eqnarray}
  |F_2F_1SC\rangle=\frac{1}{2\sqrt{3}}\big(|{\rm ok}\rangle_1|{\rm ok}\rangle_2-|{\rm ok}\rangle_1|{\rm fail}\rangle_2+|{\rm fail}\rangle_1|{\rm ok}\rangle_2+3|{\rm fail}\rangle_1|{\rm fail}\rangle_2\big).
   \end{eqnarray}
\item At time $t_3$ Wigner's assistant $A$ measures $F_1$ with the coin in the basis $\{|{\rm fail}\rangle_{F_1C},|{\rm ok}\rangle_{F_1C}\}$. The relevant state of the joint system is
  \begin{eqnarray}
    |a\rangle=\frac{1}{2\sqrt{3}}|{\rm ok}\rangle_1(|{\rm ok}\rangle_2-|{\rm fail}\rangle_2)|{\rm ok}\rangle_a+\frac{1}{2\sqrt{3}}|{\rm fail}\rangle_1(|{\rm ok}\rangle_2+3|{\rm fail}\rangle_2)|{\rm fail}\rangle_a,
     \end{eqnarray}
where $|{\rm ok}\rangle_a$ and $|{\rm fail}\rangle_a$ are the corresponding memory states. Remark that this measurement does not affect the measurement of $F_2$ in the previous moment since the entangled state of $F_2$ and $S$ is unchanged.
\item At time $t_4$ Wigner $W$ himself measures $F_2$ with the spin in the basis $\{|{\rm fail}\rangle_{F_2S},|{\rm ok}\rangle_{F_2S}\}$.  We get the state
  \begin{eqnarray}
    |w\rangle&=&\frac{1}{2\sqrt{3}}\bigg(|{\rm ok}\rangle_1|{\rm ok}\rangle_a+|{\rm fail}\rangle_1|{\rm fail}\rangle_x\bigg)|{\rm ok}\rangle_2|{\rm ok}\rangle_w\nonumber\\
    &+&\frac{1}{2\sqrt{3}}\bigg(-|{\rm ok}\rangle_1|{\rm ok}\rangle_a+3|{\rm fail}\rangle_1|{\rm fail}\rangle_a\bigg)|{\rm fail}\rangle_2|{\rm fail}\rangle_w,
    \end{eqnarray}
where $|{\rm ok}\rangle_w$ and $|{\rm fail}\rangle_w$ are the corresponding memory states. Again we notice that the entangled state of $F_1$ and $C$ is unchanged and thus the previous measurement of $F_1$ is not affected. In some limited sense, the measurements of Wigner and his assistant are independent of each other.
\item Wigner and his assistant compare their results. If $a=w={\rm ok}$ the experiment stops otherwise the experiment repeats.
\end{itemize}
In the above equations we have implicitly assumed standard quantum mechanics, i.e. Copenhagen rules including
the projection postulate, which the authors Frauchiger and Renner demanded of any quantum theory to satisfy as a 
fundamental postulate which they called  "compliance with the quantum theory" (QT). This is the most important among 
their three postulates. The second postulate which they called "single-world" (SW) is also very natural even
in many-worlds. The third postulate they called "self-consistency" (SC) in which they required logical consistency
between the theory's statements about measurement. We have then
\begin{enumerate}
\item QT: compliance with the quantum theory.
\item SW: single-world.
  \item SC: self-consistency.
  \end{enumerate}
Their main result is the following theorem: "No physical theory T can satisfy (QT), (SW), and (SC)".

It is almost obvious that it is the third postulate that can break down and the physics of the relativity 
of simultaneity provides a very good example. We have no (experimental) doubt that standard quantum theory 
applies through and through even if the collapse is not a fundamental law but an approximation.

We have now the following results (following the presentation of \cite{S})

\begin{itemize}
\item If the measurement of $F_1$ at time $t_1$ gives $r_1=$"tail", she will prepare the spin in the state $|+\rangle_x$. The measurement of $F_2$ at time $t_2$ puts the system $F_2+F_1+S+C$ in the state
  \begin{eqnarray}
    |F_2F_1SC\rangle=\frac{1}{\sqrt{2}}(|{\rm fail}\rangle_1-|{\rm ok}\rangle_1)|{\rm  fail}\rangle_2.
    \end{eqnarray}
  The state of $F_2$ and $S$ is not affected with the measurement of $A$ at time $t_3$ of $F_1+C$. The measurement of $W$ at time $t_4$ is then given immediately by the result $w_4=$"fail". We have then
  \begin{eqnarray}
    r_1="{\rm tail}"\Rightarrow w_4="{\rm fail}"\leftrightarrow  w_4="{\rm ok}"\Rightarrow r_1="{\rm head}".
    \end{eqnarray}
The last statement is a consequence of the SW postulate. 
\item If the measurement of $F_1$ at time $t_1$ gives $r_1=$"head", she will prepare the spin in the state $|-\rangle$. The measurement of $F_2$ at time $t_2$ gives then the spin $s_2=-1/2$ with certainty (in the reference frame of $F_1$).  We have then
  \begin{eqnarray}
    r_1="{\rm head}"\Rightarrow s_2=-1/2.
    \end{eqnarray}
\item However, in the reference frame of $F_2$ the full state of the system $F_1+S+C$ after the $F_1$ measurement is actually 
\begin{eqnarray}
  |F_1SC\rangle=\frac{1}{\sqrt{3}}|{\rm tail}\rangle|T\rangle_1|+\rangle+\sqrt{\frac{2}{3}}|{\rm fail}\rangle_{1C}|-\rangle.
  \end{eqnarray}
Thus, the spin in not $s_2=-1/2$ with certainty. In the case the measurement of $F_2$ at time $t_2$ gives the spin $s_2=-1/2$, the measurement of $A$ at time $t_3$ of the joint system $F_1+C$ will return $a_3=$"fail". We have then
\begin{eqnarray}
  s_2=-1/2 \Rightarrow a_3="{\rm fail}".
  \end{eqnarray}
\item
  From the above three results we have immediately
  \begin{eqnarray}
    w_4="{\rm ok}"\Rightarrow r_1="{\rm head}"\Rightarrow s_2=-1/2\Rightarrow a_3="{\rm fail}".
    \end{eqnarray}
In other words, the two super-observers Wigner and his assistant can never agree and the procedure will never halt.
\item
  On the other hand, from the above state $|F_2F_1SC\rangle$ it is obvious that the coefficient of $|{\rm ok}\rangle_1|{\rm ok}\rangle_2$ is non-zero. Hence
  \begin{eqnarray}
    a_4=w_4={\rm ok}~,~{\rm probability}=\frac{1}{12}.
    \end{eqnarray}
  The notation $a_4$ means that if the assistant $A$ performs his experiment at $t_4$ he will find $a_4=$"ok". But $A$ performs his measurement at $t_3$. Fortunately, the measurement of Wigner $W$ at $t_4$ of $F_2+S$ does not affect the entangled state of $F_1$ and $C$ and also it does not affect the entangled state of $A$ and $F_1+C$. Hence $a_3=a_4$, i.e. the measurement of $A$ is the same whether he performs it at $t_3$ or $t_4$. We have then
  \begin{eqnarray}
    a_3=w_4={\rm ok}~,~{\rm probability}=\frac{1}{12}.
    \end{eqnarray}
This is simply in gross contradiction to the previous result.
\end{itemize}
As pointed out in \cite{BHW} to make sense of the above contradiction we should differentiate between three models: 
\begin{itemize}
\item i) No-collapse models.
\item ii) Objective-collapse models.
\item iii) Subjective-collapse models.
\end{itemize}
The Wigner-Frauchiger-Renner experiment shows in a drastic way that subjective-collapse models are highly untenable in the case of encapsulated observers. The problem lies, as noted by \cite{BHW}, in the questionable possibility of communication between Wigner $W$ and his assistant $A$. In fact Wigner $W$ and his assistant $A$ play similar roles to the Schwarzschild (external) and infalling (internal) observers in black holes dynamics. Further discussion of this issue, with connection with the ER=EPR conjecture, is delegated until we discuss the ER=EPR conjecture.

\begin{figure}[htbp]
\begin{center}
\includegraphics[width=12.0cm,angle=0]{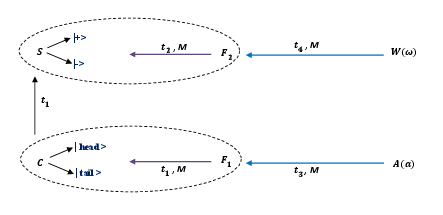}
\end{center}
\caption{The extended Wigner's friend experiment.}\label{EWE}
\end{figure}

\subsubsection{The mind/body problem and quantum Zeno effect: Stapp's theory}

The quantum Zeno effect plays an essential role in the mind/body theory of Stapp \cite{stapp} (see also Eccles \cite{eccles}) which is claimed to avoid the usual problems of classical physicalism and the philosophy of mind given in general by the two problems \cite{kim}
\begin{itemize}
\item The problem of mental causation: how could the mental cause anything material especially in a world which is conceived in a fundamental way as physical substance.
\item The hard problem of consciousness: how could subjective conscious experience even exist in a physical world dominated by matter and governed by physical laws.
\end{itemize}
These two problems are solved at once according to Stapp by orthodox quantum mechanics in which 1) the world is not purely physical and perhaps it does not exist independently of conscious observation, 2) conscious experience is an objective feature of reality and 3) consciousness is causally efficacious via the collapse postulate.  

Stapp's asserts rightly that orthodox quantum mechanics is not Cartesian dualism but its intrinsic dualism is physicallly motivated since the conscious experiences dealt with in quantum mechanics are the core realities of science which consist of "what we have done and what we have learned" \cite{Bohr}. However, in quantum mechanics and as opposed to physicalism conscious experience is much more than a mere physical activity. Indeed, events in quantum mechanics are psychophysical events where the state vector describes only their potentialities. This psychophysical event has two aspects: the psychological conscious experience and the physical aspect given by the reduction of the state vector under observation. Thus, consciousness according to quantum mechanics can certainly exist in a physical world which is dynamically and logically complete and also causally closed.

The mind/body problem is then resolved by means of the quantum zeno effect as follows. Each event, as we have said, is a psychophysical event whose physical aspect is the von Neumann process I, whereas whose associated psychophysical aspect is the conscious experience of intending or choosing to do some physical or mental action. The so-called "template of action" for some action $X$ is an actualization, via the  quantum collapse, of a particular brain pattern and which if held in place for sufficiently long time will cause the physical action $X$ to occur. Thus, if a sequence of similar process I actions, corresponding to some measurement actions, is repeated consciously in a sufficiently rapid succession, the physical brain state corresponding to that particular template of action will be forced with high probability, and the corresponding physical action will follow. According to Stapp the process I probing actions are freely chosen since they are not determined by the state of the universe.   In summary,   mental effort and conscious choice are causally efficacious in the physical world since they can influence and affect the person's physical brain processes in the required way via quantum Zeno effect.

A specific theory of the brain/mind connection which relies heavily on quantum Zeno effect in the brain is also proposed by Stapp. In this theory two coupled harmonic oscillators (HO's) are described semi-classically via coherent states. These HO's are modeling the $40$Hz synchronous oscillations of the electromagnetic field at various brain sites which are thought to accompany conscious experience. The first HO describes the brain state of the conscioussness of the observer whereas the second HO describes the environment.

If no process I action is initiated then energy is conserved and oscillates back and forth between these two HO's with a period inversely proportional to their coupling. However, if a sequence of process I probing actions is performed in a rapid sucession faster then the period then the state of the first HO describing brain correlates of the consciousness of the observer will tend to the uncoupled solution.  This is the quantum Zeno effect in this setting. Hence, the brain/mind connection in this theory relies on process I dynamics where the quantum Zeno effect causes the observer's brain to behave in the prescribed way that causes the body to act in accordance with the observer's conscious intent.

\subsection{On observer-determinacy or Orch OR}
\subsubsection{Orchestrated objective reduction}

The Schrodinger's cat can be thought of as the most fundamental encapsulation of the so-called measurement problem which is arguably the most central paradox in quantum mechanics. Stated differently, this paradox lies in the profound conflict between von Neumann's processes I and II.

Thus, the state of the cat at time $t_0$, where $t_0$ is the time required by an atom before it can decay with a probability amplitude $a$ or not decay with a probability amplitude $b$, must be given by
\begin{eqnarray}
  |\psi(t_0)\rangle=a|{\rm decayed}\rangle|{\rm dead}\rangle+b|{\rm undecayed}\rangle| {\rm alive}\rangle.
  \end{eqnarray}
In our case $t_0=1$ hour, $a=b={1}/\sqrt{2}$. The state of the cat for later times $t>t_0$ is then given by means of Process II as follows
\begin{eqnarray}
  |\psi(t)\rangle=U(t,t_0)|\psi(t_0)\rangle=a(t)|{\rm decayed}\rangle|{\rm dead}\rangle+b(t)|{\rm undecayed}\rangle|{\rm alive}\rangle.
  \end{eqnarray}
In other words, the unitary evolution of the quantum state vector $|\psi(t)$ given by Schrodinger equation 
preservers the coherent superposition of the live and dead cats, and in fact it gives the live and dead cats with the same probability since $|a(t)|^2=|a|^2$ and $|b(t)|^2=|b|^2$.

But was the cat live or dead in reality before measurement?

According to  the Diosi-Penrose-Hameroff interpretation known also as Orch OR (see \cite{hp} and references therein) the superposition of the decayed-nucleus/dead-cat and undecayed-nucleus/living-cat will decay/collapse objectively, i.e. dynamically, at a random instant to either being decayed-nucleus/dead-cat or undecayed-nucleus/alive-cat.

This random instant is on average given by the lifetime $\tau=\hbar/E_G$ where $E_G$ is the gravitational self-energy of the difference between the two stationary mass distributions involved in the superposition, namely the decayed-nucleus/dead-cat and undecayed-nucleus/living-cat.

The resulting density matrix after time $\tau$ is therefore given by the reduced density matrix $\rho_r$, i.e. if there are $N$ worlds for examples then in $N|a|^2$ of them the cat will be indeed alive and in $N|b|^2$ the cat will be dead after $\tau$ ($a$ and $b$ do not decrease from their initial values but they simply oscillate in time). 

In contrast in the Copenhagen and von the Neumann-Wigner interpretations the unitary evolution given by the Schrodinger equation will continue indefinitely until and unless an observation is made on the system. Putting it differently, the Diosi-Penrose-Hameroff interpretation assumes that in the unknown theory of quantum gravity (in which space-time, gravity and matter are unified) a fine-grained unitary evolution must replace the gross-grained Schrodinger evolution which would terminate dynamically.

Also the envisaged quantum structure of spacetime seems to act as a hidden variable in the sense that if the laws of quantum gravity were known to us we would have predicted when the $\rho_c$ (the pre-measurement entangled state) collapses to $\rho_r$ at least on average.

Thus, in Orch OR the system is collapsed long before any measurement has been performed. For the Schrodinger's cat the threshold for objective collapse is reached at a time of the order of $\tau=10^{-43}$ seconds. This seems to contradict Bell's theorem which shows (or seems to show) that the quantum state does not exist before measurement. Here clearly the state of the cat exists at time $\tau=10^{-43}$ seconds as either being dead or being alive long before any measurement has taken place.

In some sense then the envisaged new physics underlying Orch OR acts like hidden variables which converts the quantum mechanical predictions into a classical probability theory.

Thus, the photons emerging from the system (the Schrodinger's cat) towards the retina of the observer carry each a classical bit (the cat is either dead or alive in the Orch OR scheme since the superposition has already collapsed).

The brain in Orch OR is assumed to be a quantum system where the basic units of information or qubits are the tubulins with their two conformational states (open and close conformations). The photons arriving at the retina cause a neuronal activity in the brain which constitutes the initial state of this quantum system. The classical computations performed at the level of the neuronal network in the brain and their underlying quantum computations performed by the vast number of tubulins in the vast number of microtubules constituting the cytoskeletons of the vast numbers of neurons provide most of the unconsciousness,  pre-consciousness and functional consciousness exhibited by the brain.

The quantum computation proceeds from the initial state via the unitary evolution of the Schrodinger equation and it involves a highly complex superposition of tubulins quantum states which will be resolved by OR. However, the quantum computations at the level of the microtubules performed by the vast number tubulins, which are isolated against environmental decoherence  in the warm, wet and noisy brain (Tegmark), are orchestrated by the entanglement between the tubulins.

These quantum computations must terminate objectively at some time $\tau$ by an orchestrated objective reduction (an Ocrh OR) event which marks one moment of qualitative consciousness (we are seeing, i.e. we are understanding that we are looking at a dead/alive cat).

The neural correlate of the conscious perception emerges then at a time scale of the same order of the lifetime of the coherent superposition before the onset of objective collapse. The continuous flow or stream of consciousness emerges from repeated measurement and OR which hold the conscious perception in place according to the quantum Zeno effect (Stapp). 

\subsubsection{Time and free will}
The proposal that quantum effects do actually occur in microtubules in the brain seems to have been confirmed by the experiments Bandyopadhyay et al. \cite{bandy}.

Equally important is the claim that Orch OR solves many outstanding problems in physics and neuroscience. As a very important example we consider briefly the problem of temporal non-locality and free will discovered by Libet \cite{libet}.

In 1965 Kornhuber (physician) and Deecke (his student) discovered the famous so-called readiness potential (RP) which is a large and solw potential in the brain (motor cortex) which precedes voluntary movement. Thus it is an event-related potential in the brain. Then in 1983 Libet (neuroscientist) discovered that the RP potential begins in the brain almost 350 micro seconds before the conscious decision to move.

Libet asked his experimental subjects to report exactly the time W when they decided to initiate a particular voluntary movement. Then he compared this time W with the time of the onset of the readiness potential RP and he found that the readiness potential precedes the conscious decision to move with at least 350 micro seconds.

What does this mean?

On the face of it, this simply means that Libet's subjects have consciously and freely decided to initiate a movement only a long time after the preparatory action in the brain has already started. Simply put: the decision to move is not the cause of the motion!

A

As a consequence, if conscious decisions are not the cause of actions then our conscious free will is not free at all, and it might even be an illusion. This counter-intuitive conclusion led Libet himself (who rejected the obvious and possibly the inevitable consequences for free will) to propose the so-called the veto response hypothesis which states that although our conscious decision is not the cause of the action it can still veto the action before it occurrence.

However, this is not satisfactory at all, and a theory more consistent with this experimental finding is the hypothesis of compatibilism (as opposed to libertarian free will and determinism). This is a very respected and powerful position in the metaphysics of free will (it states that free will and determinism are compatible which is a position adopted by Hume and Leibniz and many other great philosophers).

The veto response hypothesis is effectively a dualistic theory of a libertarian free will executed by a non-deterministic agent. Whereas in compatibilists view, free will is not absolute freedom but the unfettered ability to act when constraints are absent. In other words, freedom really rely on the absence of external constraints rather than requiring being outside the causal chain.

The experiments of Libet were repeated by many other groups with the same results. And several systematic critiques of this sort of experiments are also put forward. In conclusion the results stood up to scrutiny.


According to Penrose and Hameroff this effect can naturally be explained by means of Orch OR. Indeed, since consciousness is a quantum effect the correlation between free will and physical action is not clear-cut but it is in principle fuzzy. And furthermore quantum entanglement in the tubulins by its nature is not expected to be causal in a normal way. Indeed, Orch OR should allow for temporal non-locality and anomalies in the time ordering of consciousness is expected.

\section{The information loss problem in quantum black holes}

\subsection{Schwarzschild black hole}

	\subsubsection{Schwarzschild black hole and Rindler spacetime}

The {Minkowski spacetime} is arguably the most important metric in physics. It is given in spherical coordinates by 
 \begin{eqnarray}
ds^2=-dt^2+dr^2+r^2d\Omega^2.
 \end{eqnarray}
The Schwarzschild eternel black hole  is the second most important solution of Einstein's equation given by the metric 
 \begin{eqnarray}
ds^2=-(1-\frac{2GM}{r})dt^2+(1-\frac{2GM}{r})^{-1}dr^2+r^2d\Omega^2.
 \end{eqnarray}
 This is the geometry generated by a point mass $M$ placed at the center of spacetime as seen by an observer (Schwarzschild observer) position at infinity. It is unique in the sense that any
 spherically solution of  Einstein's equation is necessarily static and thus it must reduce to  the Schwarzschild metric (Birkhoff's theorem \cite{Birkhoff}). The horizon where the coordinates system terminates (the time-like Killing vector vanishes) is located at
 \begin{eqnarray}
r_s=2GM.
 \end{eqnarray}
 Another important spacetime is {Rindler spacetime} which is a uniformly accelerating reference frame with respect to Minkowski spacetime (acceleration $a$). The near-horizon geometry of a Schwarzchild black hole is in fact a Rindler spacetime with acceleration
\begin{eqnarray}
  a=\frac{1}{2r_s}.
\end{eqnarray}
In fact Rindler spacetime plays with respect to Minkowski spacetime the same role that the Schwarzschild observer plays with respect to the so-called Kruskal-Szekers spacetime which is the maximally extended Schwarzschild solution.
 	\subsubsection{Hawking temperature from Unruh effect}
 The Unruh Effect is the statement that a Rindler observer sees the Minkowski vacuum as a thermal canonical ensemble with temperature \cite{Unruh:1976db}
\begin{eqnarray}
  T=\frac{1}{2\pi}.
\end{eqnarray}
The celebrated {Hawking temperature} which is seen by the Schwarzschild observer is thus due to gravitational redhift plus Unruh effect. 
    Indeed, the Rindler time $\omega$ is related to the Schwarzschild time $t$ by the relation 
\begin{eqnarray}
\omega=\frac{t}{4GM}.
\end{eqnarray}
This leads immediately to the fact that frequency as measured  by the Schwarzschild observer $\nu$ is redshifted compared to the frequency $\nu_R$ measured by the Rindler observer given by
\begin{eqnarray}
\nu_R=4GM.\nu\Rightarrow \nu=\frac{\nu_R}{4GM}.
\end{eqnarray}
Hence the temperature as measured by the Schwarzschild observer is also redshifted as 
\begin{eqnarray}
T_R=4GM.T_H\Rightarrow T_H=\frac{T_R}{4GM}=\frac{1}{8\pi GM}.
\end{eqnarray}
This is precisely Hawking temperature.

	\subsubsection{Particle motion in Schwarzschild spacetime}
       The motion of a scalar particle of energy $\nu$ and angular momentum $l$ in the background gravitational 
field of the Schwarzschild black hole is exactly equivalent to the motion of a quantum particle, i.e. a particle obeying the 
Schrodinger equation, with energy $E=\nu^2$ in a scattering potential given in the tortoise coordinate $r_*$ with the expression \cite{Susskind:2005js}
          \begin{eqnarray}
V(r_*)=\frac{r-r_s}{r}\big(\frac{r_s}{r^3}+\frac{l(l+1)}{r^2}\big)~,~r_*=r+r_s\ln(\frac{r}{r_s}-1). 
          \end{eqnarray}
          Particles are free both at infinity and at horizon. There is a barrier at $r\sim 3r_s/2$ where the potential reaches its maximum and the height of this barrier is proportional
to the square of the angular momentum, viz
          \begin{eqnarray}
V_{\rm max}(r_*)\sim \frac{l^2+1}{G^2M^2}\sim (l^2+1)T_H^2~,~T_H=\frac{1}{8\pi GM}.
          \end{eqnarray}
          But particles are in thermal equilibrium at the Hawking temperature $T_H$ and thus their energy $\nu$ is proportional to $T_H$. Hence only $l=0$ particles can escape the potential to infinity (Hawking radiation).


	\subsubsection{Kruskal-Szekers metric and Penrose diagrams}
        
          The Kruskal-Szekers metric which defines the maximal extension of the Schwarzschild
              solution is given by the metric (for $T$ and $R$ given in terms of $t$ and $r$ by some coordinate transformations)
            \begin{eqnarray}
ds^2=\frac{32G^3M^3}{r}\exp(-\frac{r}{2GM})(-dT^2+dR^2)+r^2d\Omega^2.
            \end{eqnarray}
            This spacetime consists of four regions separated by future and past horizons.

            Regions II is the interior of the black hole. Any
future directed path in this region will hit the singularity.
              While region IV  is the interior of the so-called white hole. These are called
              called future and past interiors respectively.

              The regions I and III are precisely the exterior regions (asymptotically flat universes) in which we live in one of them.

              A white hole is the time reverse of the black hole. This corresponds to a singularity
in the past at which the Universe originated. This is a part of spacetime from
 which observers can escape to reach us while we cannot go there.
        
%



  
We consider now the formation of a black hole from gravitational collapse of a thin spherical shell of massless matter (red line in figure (\ref{PD3})).
 The geometry inside the spherical shell is flat Minkowski spacetime. But by Birkoff’s theorem the geometry outside the spherical shell is nothing other than Schwarzschild geometry.

The Penrose diagram of Minkowski spacetime is a triangle bounded by five infinities (two light-like infinities, two time-like infinities and one space-like infinity). Whereas the Penrose diagram of  Schwarzschild geometry consists of the four regions exhibited in the  Kruskal-Szekers metric which are separated by horizons and ending at singularities.
  
  Thus the Penrose diagram of the formation of a black hole from gravitational collapse is obtained by gluing the Penrose diagrams of flat Minkowski spacetime (inside shell) and Schwarzschild black hole (outside shell).
\begin{figure}[htbp]
\begin{center}
\includegraphics[width=5.0cm,angle=0]{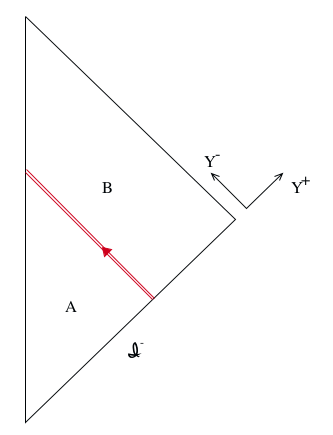}
\includegraphics[width=7.0cm,angle=0]{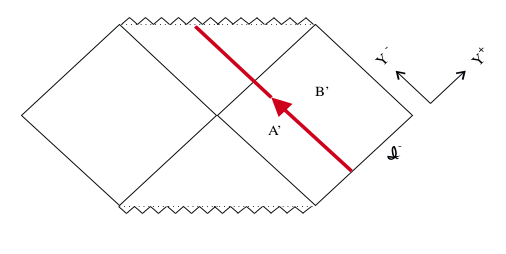}
\includegraphics[width=7.0cm,angle=0]{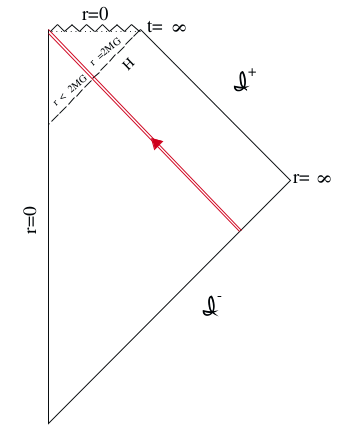}
\end{center}
\caption{The formation of a black hole from gravitational collapse of a thin spherical shell of massless matter (red line)}\label{PD3}
\end{figure}

\subsection{Hawking radiation and information loss problem}

	\subsubsection{Hawking temperature}
        
        The Kruskal vacuum state  $|0_K\rangle$ plays for Schwarzschild observer exactly the same role played by the Minkowski state for the Rindler observer while the Schwarzschild vacuum (also called Boulware vacuum) plays the same role played by the Rinlder vacuum.

The Schwarzschild observer sees (by computing the Bogolubov coefficients as done for example in \cite{Mukhanov:2007zz}) the Kruskal vacuum as a heath bath containing $n_{\omega}$ particles given by
\begin{eqnarray}
    n_{\omega}=\langle 0_k |N_{\omega}|0_K\rangle=\frac{\delta(0)}{\exp(2\pi\omega/a)-1}~,~N_{\omega}=b_{\omega}^{\dagger}b_{\omega}.
    \end{eqnarray}
This a blackbody Planck spectrum with the temperature 
  \begin{eqnarray}
T_H=\frac{a}{2\pi}=\frac{1}{4\pi r_s}=\frac{1}{8\pi GM}.
\end{eqnarray}
By inserting SI units we obtain 
\begin{eqnarray}
T_H=\frac{\hbar c^3}{8\pi GM k_B}.
\end{eqnarray}
The black hole as seen by a distant observer is radiating energy, thus its mass decreases, and as a consequence its temperature increases, i.e. the black hole becomes hotter, which indicates a negative specific heat.

       \subsubsection{Thermodynamics}

In summary, to a distant observer the Schwarzschild black hole appears as a body with 
energy given by its mass $M$ and a temperature $T$ given by Hawking temperature 
\begin{eqnarray}
T=\frac{1}{8\pi GM}.\nonumber
\end{eqnarray}
The thermodynamical entropy $S$ is related to the energy and the temperature by the formula $dU=T dS$. Thus we obtain for the black hole the entropy 
\begin{eqnarray}
dS=\frac{dM}{T}=8\pi GM dM\Rightarrow S=4\pi GM^2.
\end{eqnarray}
However, the radius of the event horizon of the Schwarzschild black hole is $r_s=2MG$, and thus the area of the event horizon (which is a sphere) is 
\begin{eqnarray}
A=4\pi (2MG)^2.
\end{eqnarray}
By dividing the above two equations we get 
\begin{eqnarray}
S=\frac{A}{4G}.
\end{eqnarray}
The entropy of the black hole is proportional to its area. This is the famous Bekenstein-Hawking entropy formula \cite{Bekenstein:1973ur,Bekenstein:1974ax}.

The last point of primacy importance concerns black hole thermodynamics. The thermal entropy is the maximum amount of
information contained in the black hole. The entropy is mostly localized near the horizon, but quantum field theory 
(QFT) gives a divergent value, instead of the above Bekenstein-Hawking value. 
The number of accessible quantum microscopic states is determined by this
entropy via the formula 
\begin{eqnarray} 
n=\exp(S). 
\end{eqnarray}
Since QFT gives a divergent entropy instead of the Bekenstein-Hawking value it must be replaced by quantum 
gravity (QG) near the horizon and this separation of the QFT and QG degrees of freedom can be implemented by the 
stretched horizon which is a time like membrane, at a distance of one Planck length $l_P=\sqrt{G\hbar}$ from the actual
horizon, and where the proper temperature gets very large and most of the black hole entropy accumulates.

	\subsubsection{Gravitational collapse}

           We consider  following  \cite{Jacobson:2003vx} the more realistic situation in which 
a Schwarzschild black hole is formed by gravitational collapse of a thin mass shell as in the Penrose diagram shown in figure (\ref{PD3}). The vacuum state before collapse is Minkowski state whereas after collapse it is Kruskal state.


        The initial/{incoming} state of the black hole is a pure state $|\psi\rangle$ (with incoming negative frequency modes). The final/{outgoing state} corresponds to a positive frequency mode $P$, i.e. centered around some positive frequency $\omega$, with support only at large radii $r$ at late times $t\longrightarrow +\infty$.

In order to find the initial state  we run $P$ backwards in time towards the black hole. A reflected part $R$ will scatter off the black hole and return to large radii and a transmitted part $T$ with support only immediately outside the event horizon, i.e.  $P=R+T$. At the event horizon both positive and negative frequency modes
will be seen in $T$ by a freely falling observer.  
Since the reflected wave packet $R$ has only support in the asymptotic flat region and since $|\psi\rangle$ contains no positive frequency incoming excitations we have
\begin{eqnarray}
a(R)|\psi\rangle =0.
\end{eqnarray}
 Also the black
    hole state $|\psi\rangle$ does not contain positive high frequency modes throughout (adiabatic principle). We have then
    \begin{eqnarray}
a(T^+)|\psi\rangle=a(\bar{T}^-)|\psi\rangle =0.\label{unruh}
    \end{eqnarray}
    These equations define the so-called Unruh vacuum $|U\rangle$.

    If the state $|\psi\rangle $ were annihilated exactly by $T$ it would have been identical with the Schwarzschild or Boulware state $|B\rangle$. Recall that the Schwarzschild vacuum $|B\rangle$ is annihilated by $a(R)$, $a(T)$ and $a(\bar{\tilde{T}})$ where ${\tilde{T}}$ is defined inside horizon.
 
%

The solution of equations  (\ref{unruh}) gives a maximally entangled state describing pairs of particles with zero Killing energy each, viz

    \begin{eqnarray} 
    |U\rangle\sim \sum_n \exp(-\frac{n\pi\omega}{a})|n_R>|n_L>. 
    \end{eqnarray}
The states $|n_R\rangle$ and $|n_L\rangle$ are the level n-excitations of the exterior modes $T$ and the interior
  modes $\bar{\tilde{T}}$.
    Thus one of the pair $|n_R\rangle$ goes outside
the horizon (Hawking radiation) whereas the other pair  $|n_L\rangle$  falls behind the horizon (information lost).


However, the asymptotic Schwarzschild observer registers a thermal mixed state with Hawking temperature, viz
\begin{eqnarray} 
\rho_R={\rm Tr}_L|U\rangle\langle U|\sim \sum_n\exp(-\frac{2n\pi\omega}{a})|n_R\rangle\langle n_R|. 
\end{eqnarray}
 In summary, a correlated entangled pure state near the horizon gives rise to a thermal mixed state outside the horizon. This is the information loss problem.


	\subsubsection{Information loss problem and laws of physics}

In gravitational collapse an entangled pair of particles is created on the past null infinity ${\cal J}^-$ at $r=\infty$ then scatter off the black hole. Then one of the particles escapes to the future null infinity ${\cal J}^+$ at $r=\infty$ whereas the other one passes through the horizon and reaches the singularity ${\cal S}$ at $r=0$.

From the assumption of {locality} we have
 \begin{eqnarray}
H_{\rm in}=H_-~,~H_{\rm out}=H_+\otimes H_S.
 \end{eqnarray}
From the assumption of {unitarity} we have 
\begin{eqnarray}
|\psi_{\rm out}\rangle=S|\psi_{\rm in}\rangle.
\end{eqnarray}
However, with respect to the outside observer at the infinity ${\cal J}^+$ the final state can only be given by a reduced density matrix since she can not access the states on the singularity ${\cal S}$. We have then
\begin{eqnarray}
  |\psi_{\rm in}\rangle\langle\psi_{\rm in}|\longrightarrow \rho_{\rm out}={\rm Tr}_{\cal S}|\psi_{\rm out}\rangle\langle \psi_{\rm out}|.
  \end{eqnarray}
  This is the information loss paradox  \cite{Hawking:1974sw,Hawking:1976ra}.

Thus black hole dynamics contradicts one of the following laws of physics (called laws of nature by Susskind):

\begin{itemize}
    \item Information conservation (quantum mechanics):  The information is defined as the difference between the coarse-grained Boltzmann thermodynamic entropy and the fine-grained von Neumann entanglement entropy. This quantity is conserved due to the unitarity of quantum mechanics.

    \item Equivalence principle (general relativity): Spacetime is a manifold which is locally Minkowski flat (special relativity) everywhere.

    \item Quantum zerox principle (information theory): The linearity of quantum mechanics forbids the duplication of quantum information.
    \item The Bekenstein-Hawking entropy formula \cite{Bekenstein:1973ur,Bekenstein:1974ax}
      \begin{eqnarray}
S=\frac{A}{4G}=\ln n.\nonumber
\end{eqnarray}
\end{itemize}

\subsubsection{Unitarity and Page curve}


The black hole starts in a pure state and after its complete evaporation the Hawking radiation is also in a pure state.  This is the assumption of unitarity. The information is given by the difference between the thermal 
entropy of Boltzmann and the entanglement entropy of Von Neumann whereas the entanglement entropy is given by the Von Neumann entropy.

The so-called Page curve \cite{Page:1993wv,Page:2013dx} is expected to give an entanglement entropy which starts at zero value then it reaches a maximum value 
  at the so-called Page time then drops to zero again. The Page time is the time at which the black hole evaporates around one-half of its mass and the information starts to get out with the  radiation. Before the Page time only energy gets out with the radiation, while only at the Page time the information starts to get out, and it gets out completely at the moment of evaporation. This is guaranteed to happen because of the second principle of thermodynamics and the assumption of unitarity.

     The computation of the Page curve starting from first principles will provide, in some precise sense, the mathematical solution of the black hole information loss problem (Strominger as recounted by Harlow).

\subsection{Black hole complementarity}

        
          Two properties are said to be dual or complementarity if they can not be observed simultaneously. Black hole complementarity due to 't Hooft  \cite{tHooft:1984kcu,tHooft:1990fkf} and Susskind \cite{Susskind:1993if} is a prime example.
 It states simply that no single observer will ever witness a violation of a law of physics.

Thus, the {external observer} sees the infalling matter heating up the stretched horizon (not the event horizon but a physical membrane above it) which then reradiates back the infalling information in the form of Hawking radiation.
  The information is uniformly spread out over the stretched horizon.

Whereas the {infalling observer} sees (because of the equivalence principle) that there is nothing special happening at the horizon and the infalling information or matter will actually pass through and reach the singularity. The information is seen by this observed as localized on the stretched horizon.

A better statement is to say that the information is both reflected at (with respect to the external external observer) and passed through (with respect to the infalling observer) the event horizon. These two stories are in fact complementary not contradictory. The two observes can not communicate (infinite time dilation at the horizon, no signal can be sent from behind the horizon).

In some sense the complementarity principle resolves the violation of quantum xerox or no cloning theorem. Thus, information is either inside the black hole (with respect to infalling observer) or outside the black hole (with respect to external observer).

The postulates of black hole complementarity are (external observer):
\begin{itemize}
\item $1)$ Unitarity (quantum mechanics). 
\item $2)$ Semi-classical field approximation (stretched horizon).
\item $3)$ Hawking-Bekenstein law: The number of microstates of a black hole of mass $M$ is equal to the exponential of the entropy $S(M)=A(M)/4G$ where $A(M)$ is the surface area of the black hole. 
\end{itemize}
The conservation of information (no-cloning or xerox theorem) and the equivalence principle are also implicitly assumed.


\begin{figure}[H]
\begin{center}
\includegraphics[width=10.0cm,angle=0]{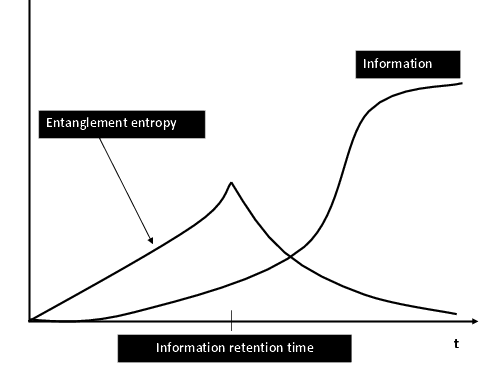}
\end{center}
\caption{Page time, the entanglement entropy and the information.}\label{information}
\end{figure}

\subsection{The firewall and monogamy}

	\subsubsection{Monogamy}
 



Hawking radiation is formed of pairs of entangled particles created around the horizon with one of each pair passing through the horizon whereas the other one goes to infinity. The final state of the particle outside after complete evaporation of the black hole is mixed (information is lost).  The final state must become purified (unitarity).

By using {the black hole complementarity we know that the infalling observer sees the bits of information going inside the black hole whereas the external observer sees them in the late Hawking radiation. 

  However, the infalling observer sees the horizon as flat spacetime whereas the external observer sees it as a hot membrane (
  strong 
  non-locality).

In conclusion, a particle $B$ from the late Hawking radiation must be entangled with the earlier part $R_B$ of Hawking radiation (unitarity). But this particle $B$ must also be entangled with a particle $A$ inside the horizon (locality). And as it turns out, this contradicts another cherished principle which is {the principle of  monogamy of quantum entanglement}: The same system can not be entangled with two different other systems at the same time. 

	\subsubsection{Firewall}

          In summary, we have to give either:
\begin{enumerate}
\item    {Unitarity (purity):} Black hole information is carried out with the late Hawking radiation. This effect is measured by an external observer.
\item    {Effective Field Theory (EFT):} Quantum field theory in curved backgrounds and locality must hold. In other words, we suppose that semi-classical gravity is valid outside the horizon, i.e. small curvature near and outside the horizon. Thus, the external observer will see the horizon as a physical membrane (stretched horizon).
\item    Equivalence Principle (no-drama): Nothing special happening as we fall across the horizon.
   This effect is measured by an infalling observer.
\end{enumerate}
The {AMPS firewall paradox \cite{Almheiri:2012rt} in our view is the most precise formulation of the information loss paradox. It states that black hole complementarity (first and second postulates) are inconsistent with the equivalence principle (third postulate). It was expected (according to the AMPS) that  EFT (locality) breaks down and complementarity is then sufficient. But the situation seems more complicated.


  The firewall is a release of high energy at the horizon arising from breaking the entanglement between the inside and outside modes in accordance with monogamy. This results from the following contradictory statements:
\begin{itemize}
\item    First, from (1), any particle of the late Hawking radiation (after page time) is maximally entangled with some subsystem of the early radiation because the early particles have much more states. It is an eigenstate of $bb^{\dagger}$.

\item    However, by assumption (2), we can propagate (essentially freely) any late Hawking mode back from infinity (where the unitarity assumption should hold) to the horizon (where the no-drama assumption is supposed to hold). The mode becomes highly blue shifted, i.e. it has a very high energy. In other words, we obtain an early mode of the Hawking radiation with very high energy which is encountered by the infalling observer.

\item    But from (3) we conclude that this same Hawking mode is maximally entangled with a modes inside the horizon. The field is in the ground state $a^{\dagger}_{\omega}a_{\omega}=0$ and the mode is not an eigenstate of $bb^{\dagger}$.
\end{itemize}


\subsection{ER=EPR and modification of quantum mechanics}

  \subsubsection{ER=EPR}

The ER=EPR proposal \cite{Maldacena:2013xja} (see also \cite{Susskind:2014yaa,Susskind:2016jjb}) is a much more fundamental conjecture which states that general relativity (Einstein-Rosen or ER wormhole  \cite{Einstein:1935tc}) is equivalent to quantum mechanics (Einstein-Podolsky-Rosen or EPR entanglement \cite{Einstein:1935rr}). 

This profound idea is inspired by van Raamsdonk observation \cite{VanRaamsdonk:2010pw} (see also \cite{Israel:1976ur}) that a maximally extended AdS black hole is dual via AdS/CFT correspondence  to a pair of maximally entangled thermal conformal field theories living at the boundaries \cite{Maldacena:2001kr}.

Thus according to the ER=EPR conjecture there is no need for the firewall, which results from the violation of the equivalence principle through the breaking of the quantum entanglement between the interior and the exterior of the black hole, since the spacetime manifold is genuinely smooth across the horizon. Indeed, it is argued that maximal entanglement between different spacetime regions (such as the interior and the exterior of the black hole) acts exactly as a smooth physical bridge between those regions and thus no contradiction results with the monogamy principle.

The ER=EPR conjectures entails therefore some substantial modification of quantum mechanics by re-interpreting quantum entanglement as physical gravitational bridges.

    The ER  bridge or wormehole is a solution of Einstein's equations which connects spacetime points which are very far away. For example in the Schwarzchild spacetime (Kruskal-Szekeres coordinates) the two asymptotically flat universes (regions I and III) are connected by a non-traversable Einstein-Rosen bridge (a whormhole). This ER bridge is however non-traversable (locality is maintained) and the two observers can only meet inside the horizon rather than communicate.

      The ER bridge generates two maximally entangled regions by Hawking particles.  For example the Hartle-Hawking $AdS$ ER bridge yields the maximally entangled state \cite{Hartle:1976tp,Maldacena:2001kr}
      \begin{eqnarray}
        |\psi\rangle\propto \sum_i\exp(-\beta E_i/2)|i^*\rangle_L|i\rangle_R.
      \end{eqnarray}
      We recall that maximal entanglement is described by Bell states such as EPR  pairs and thus  locality is maintained since EPR correlations can not be used to send information faster than the speed of light corresponding to the fact that the ER bridge is generally a non-traversable Lorentzian wormehole.
      
      Conversely,  two  distant  black holes connected through the interior via an ER bridge can  be  interpreted  as  maximally  entangled states of two black holes that form an EPR pair.

\begin{figure}[htbp]
\begin{center}
\includegraphics[width=7.0cm,angle=0]{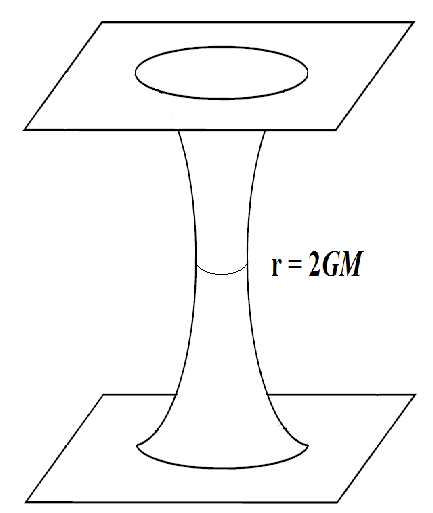}
\end{center}
\caption{Two asymptotically flat universes are connected by a non-traversable ER bridge, i.e. the two observers can only meet inside the horizon. }
\end{figure}

%


In summary, every EPR correlated state is connected by an ER bridge (not necessarily the one given by general relativity). 
The {ER=EPR resolves the information loss problem} 
  by stating that the Hawking radiation is  actually connected by ER bridges to itself and also to the black hole horizon. 


A straightforward generalization of the ER=EPR conjecture, based on the holographic gauge/gravity duality or the AdS/CFT correspondence \cite{Maldacena:1997re} and the remarkable Ryu-Takayanagi formula \cite{Ryu:2006bv,Ryu:2006ef} which generalizes the  Bekenstein-Hawking entropy formula, is the idea that spacetime geometry is equivalent to quantum entanglement \cite{Lashkari:2013koa,Faulkner:2013ica,VanRaamsdonk:2016exw,Casini:2011kv}.

\subsubsection{Interference as entanglement}
  As an application of the ER=EPR conjecture we can argue (following Susskind) that entanglement is more fundamental than interference by showing how to reduce the latter to the former.  

  Every single electron emerges from the two slits in a linear superposition of two non-overlapping wave packets with some relative phase. The wave packets will in time overlap 
    giving rise to the interference pattern. 
   We treat the wave function as a field (second quantization) with the main degrees of freedom of the field are found inside the boxes centered around the peaks of the two emerging wave packets. 
   The state of the system is a maximally entangled Bell pair, i.e.
   \begin{eqnarray}
     |\psi\rangle=\frac{1}{\sqrt{2}}(|10\rangle-|01\rangle).
     \end{eqnarray}
The state $|10\rangle$ corresponds to a particle in the left box and no particle in the right box whereas $|01\rangle$ corresponds to no particle in the left box a particle in the right box. 


Acoording to the {ER=EPR the two wave packets are connected by an ER bridge or a wormhole which is the one that reminds each electron that the other electron is there.}  That is how the  electron going through one slit knows that the other slit is open or closed.

\subsubsection{Extended Wigner's friend experiment revisited}

The Wigner's friend experiment can also be discussed within the context of the ER=EPR conjecture \cite{Susskind:2016jjb,Susskind:2014yaa}. Here we propose a resolution of the apparent contradiction found earlier in the context of the extended Wigner's friend experiment by using the ER=EPR conjecture.

The first set of encapsulated observers consists of the assistant $A$, the friend $F_1$ and the coin $C$ which can all be collapsed into entangled black holes (fungibility  of entanglement). Communication between various observers is described by entanglement. The friend $F_1$ and the coin $C$ are described by a Bell state which is equivalent to an EPR bridge or a wormhole by the ER=EPR conjecture.

This bridge is snipped when the assistant makes his measurement and thus no messages can be sent in between the friend $F_1$ and the coin to meet inside the wormhole. However, with respect to a many-worlds description, the assistant $A$, the friend $F_1$ and the coin $C$ are described by a tripartite GHZ state which is geometrically equivalent to something called the GHZ brane by the the ER=EPR conjecture.  By the separability condition satisfied by GHZ states no entanglement exists between any two parties and thus no messages can be sent between any two parties to meet inside the GHZ brane. But, many-worlds preserves entanglement and thus reversibility. In particular, the union of any two parties in the GHZ state is still entangled with the third party, and hence messages can be sent between any two cooperating parties ($A$ and $F_1$) and the third one ($C$) to meet inside the GHZ brane. 

The same picture holds for the second set of encapsulated observers consisting of Wigner himself $W$, the friend $F_2$ and the spin $S$, with messages between any two cooperating parties (Wigner and $F_2$) and the third one ($S$) can be sent to meet inside their corresponding GHZ brane. There is no direct entanglement between Wigner and his assistant and hence their only route of exchanging messages is through the entanglement between the coin  $C$ and the spin $S$, which in turn entangle the two GHZ branes corresponding to the two sets of encapsulated observers.

With respect to Wigner (or his assistant $A$), his own measurement has already severed the above structure at the level of the EPR bridge relating the friend $F_1$ and the coin $C$ (or the friend $F_2$ and the spin $S$). So, indeed the possibility of communication is quite questionable in the case of encapsulated observers.

\subsection{The holographic gauge/gravity duality}

\subsubsection{Holography}
The number of degrees of freedom, or equivalently the amount of information, contained in a quantum system is measured as we know by thermodynamic entropy. In quantum mechanics the entropy is an extensive quantity and thus the entropy of a $d-$dimensional spatial region ${\bf R}^d$ is proportional to its volume  $V_d$. 

However, in quantum gravity the entropy is sub-extensive, i.e. the entropy of a $d-$dimensional spatial region ${\bf R}^d$ is actually proportional to the surface area $S_{d-1}=\partial V_d$ which bounds its volume $V_d$ and not proportional to the volume $V_d$ itself.

In other words, the entropy of a $d-$dimensional spatial region, in a gravitational theory, is bounded by the entropy of the black hole which fits inside that spatial region. This is essentially what is called the holographic principle introduced first by 't Hooft \cite{tHooft:1993dmi} and then extended to string theory by Susskind \cite{Susskind:1994vu} (see also \cite{Bousso:2002ju}). As we can see this principle is largely inspired by the Bekenstein-Hawking formula which  states that the entropy of a black hole ${\cal S}_{BH}$ is proportional to the surface area $A_H$ of the black hole horizon with the constant of proportionality equal $1/4G_N$ where $G_N$ is Newton's constant, viz 
\begin{eqnarray}
{\cal S}_{BH}=\frac{A_H}{4G_N}.
\end{eqnarray} 
The holographic principle provides therefore a partial answer to the question of how could a higher dimensional gravity theory (${\rm AdS}_{d+1}$) contain the same number of degrees of freedom, the same amount information, and have the same entropy as a lower dimensional quantum field theory (${\rm CFT}_d$), i.e. it lies at the heart of the celebrated AdS/CFT correspondence \cite{Maldacena:1997re}.  

\subsubsection{The black p-branes and supersymmetric Yang-Mills gauge theory}
Maldacena in \cite{Maldacena:1997re} (see also \cite{Gubser:1998bc,Witten:1998qj}) produced perhaps the greatest feat in theoretical physics since the days of the quantum revolution of Bohr and the relativity revolution of Einstein. Indeed, one can argue that this achievement is even more astounding and even more impressive than the achievement of the standard model of particle physics and the standard model of cosmology. 


But what is the gauge/gravity duality?

This can be explained in two axes:
\begin{enumerate}
\item Firstly, we consider maximally supersymmetric $U(N)$ gauge field theory in $p+1$ dimensions. The action principle of this theory is invariant under the maximum number of supersymmetric transformations in the given dimension (recall that supersymmetry is a generalization of translations and rotations by adding odd Grassmannian coordinates to spacetime). The gauge theory $U(N)$ is a generalization of electromagnetism, i.e. of U(1), to the case of $N$ charges and $N$ anti-charges called colors.

We consider this theory in t' Hooft limit given by taking the gauge coupling constant $g^2$ to zero and $N$ to infinity in such a way that we keep the so-called 't Hooft coupling $\lambda$ constant where \cite{tHooft:1973alw}
\begin{eqnarray}
\lambda=g^2.N.
\end{eqnarray}
It is well known that the above gauge theory in t' Hooft limit describes N coincident Dirichlet(p)-branes (or Dp-branes for short) which are (in the words of Polchinski): ".. extended objects defined by mixed Dirichlet-Neumann boundary conditions in string theory (which) break half of the supersymmetries of the type II superstring and carry a complete set of electric and magnetic Ramond-Ramond charges.." \cite{Polchinski:1995mt}.

\item Secondly, the system of $N$ coincident Dp-branes form something we call the black p-brane \cite{Gibbons:1987ps,Horowitz:1991cd,Itzhaki:1998dd} which is a generalization of black holes. This black p-brane has mass and as such it curves the spacetime around it with corresponding metric given by a type II supergravity solution. This near extremal p-brane solution is given explicitly by (with $\alpha^{\prime}=l^2_s\longrightarrow 0$ where $l_s$ is the string fundamental length)
\begin{eqnarray}
ds^2&=&\alpha^{\prime}\bigg[\frac{U^{\frac{7-p}{2}}}{g_{\rm YM}\sqrt{d_pN}}\bigg(-(1-\frac{U_0^{7-p}}{U^{7-p}})dt^2+dy_{||}^2\bigg)+\frac{g_{\rm YM}\sqrt{d_pN}}{U^{\frac{7-p}{2}}}\frac{dU^2}{(1-\frac{U_0^{7-p}}{U^{7-p}})}\nonumber\\
&+&g_{\rm YM}\sqrt{d_pN}U^{\frac{p-3}{2}}d\Omega^2_{8-p}\bigg].
\end{eqnarray}
The $(t,y_{\parallel})$ are the coordinates along the p-brane world volume, $\Omega_{8-p}$ is the transverse solid angle associated with the transverse radius $U$, whereas $U_0$ corresponds to the radius of the horizon,  and it is given in terms of the energy density of the brane $E$ by 
\begin{eqnarray}
U_0^{7-p}=a_pg_{\rm YM}^4E.
\end{eqnarray}
The string coupling constant $g_s$ in this limit is given by 
\begin{eqnarray}
g_s=(2\pi)^{2-p}g_{\rm YM}^2(\frac{d_pg_{\rm YM}^2N}{U^{7-p}})^{\frac{3-p}{4}}.
\end{eqnarray}
The factors $d_p$ and $a_p$ in the above equations are some numerical constants which depend only on the dimension $p$. The corresponding Hawking temperature can be determined from the conical singularity of the Euclidean metric to be given by 
\begin{eqnarray}
T=\frac{(7-p)U_0^{\frac{5-p}{2}}}{4\pi g_{\rm YM}\sqrt{d_pN}}.
\end{eqnarray}
The  Bekenstein-Hawking formula ${\cal S}=A/4$ where $A$ is the area density of the horizon can also be calculated for this metric from the first law of thermodynamics  $d{\cal S}=dE/T$.
\end{enumerate}

The gauge/gravity duality states simply that the above maximally supersymmetric $U(N)$ gauge theory in $p+1$ dimensions in the t' Hooft limit is exactly equivalent to type II supergravity in $10$ dimensions around the above black p-brane. Thus, "hidden within every non-abelian gauge theory, even within the weak and strong nuclear interactions, is a theory of quantum gravity"  \cite{Horowitz:2006ct}.

In other words, a higher dimensional curved spacetime manifold emerges in this duality from a lower dimensional gauge theory in a flat spacetime manifold. The emerging extra spatial dimensions are described in the gauge theory by adjoint scalar fields given by $N\times N$ matrices. The extra dimensions emerges in the gauge theory precisely in the limit $N\longrightarrow \infty$ whereas strongly quantum gauge fields give rise to effective classical gravitational fields in the limit $\lambda\longrightarrow \infty$. 

This duality provides therefore a very  concrete non-perturbative definition of quantum gravity and its quantum geometry in terms of gauge theory.
Indeed, since the gauge theory is fully defined quantum mechanically, and in fact it is fully defined non-perturbatively on the lattice a la Wilson\cite{Wilson:1974sk},  the gauge/gravity duality gives then a full non-perturbative definition of quantized gravity which is the holly grail of theoretical physics.

Quantum corrections to the gravity side are given by string theory loop expansion, i.e. in the string coupling constant $g_s$. This corresponds on the gauge side to $1/N^2$ corrections. Whereas stringy corrections to the gravity side are given by the fundamental string length $l_s$ and they correspond on the gauge side to $1/\lambda$ corrections where $\lambda$ is the above 't Hooft coupling.

The most important examples of the gauge/gravity duality are:
\begin{enumerate}
\item The celebrated $AdS_5/CFT_4$ on $AdS_5\times S^5$ in $(3+1)-$dimension which is relevant to the D3-brane \cite{Maldacena:1997re}.
\item The ABJM theory in $(2+1)-$dimension on $AdS_4\times S^7$ relevant to the M2-brane  \cite{Aharony:2008ug}. The dual gauge theory relevant to the M5-brane (which is the magnetic dual of the M2-brane) on $AdS_7\times S^4$ is still largely unknown.
\item Matrix String Theory  in $(1+1)-$dimension \cite{Dijkgraaf:1997vv} relevant to the Gregory-Laflamme instability \cite{Gregory:1993vy} (for example a black string breaking into black holes).
\item The BFSS model or M-(atrix) theory in $(0+1)-$dimension \cite{Banks:1996vh} which is perhaps the most important case. Indeed, the DLCQ (discrete light cone quantization) of M-theory should be described by the BFSS model. The compactification of the BFSS model on a circle (high temperature limit) gives a $U(N)$ gauge theory in $(0 + 0)-$dimension known as type IIB matrix
model (also known as the IKKT model) which provides a non-perturbative
regularization of type IIB superstring theory in the Schild gauge \cite{Ishibashi:1996xs}.
\item The BMN model in $(0+1)-$dimension \cite{Berenstein:2002jq} which is a simple mass deformation of the BFSS model describing the M2-brane and the M5-brane in the Penrose limit of the $AdS_4\times S^7$ and $AdS_7\times S^4$.
\end{enumerate} 
An overview of these theories from the lattice point of view is given in \cite{Hanada:2016jok}.

In summary, we have 
\begin{itemize}
\item The gauge theory in the limit $N \longrightarrow \infty$ (where extra dimensions will emerge) and $\lambda\longrightarrow\infty$ (where strongly quantum gauge fields give rise to effective classical gravitational fields) should be equivalent to classical type II supergravity around the p-brane spacetime.
\item The gauge theory with $1/N^2$ corrections should correspond to quantum loop corrections, i.e. corrections in $g_s$ , in the gravity/string side.
\item The gauge theory with $1/\lambda$ corrections should correspond to stringy corrections, i.e. corrections in $l_s$ , corresponding to the fact that degrees of freedom in the gravity/string side are really strings and not point particles.
\end{itemize}

\subsubsection{AdS/CFT correspondence}

The near-horizon geometry of the black p-branes involves always the product of spheres and AdS spaces which are maximally symmetric spaces given by the coset spaces $S^{d+1} = SO(d + 2)/SO(d + 1)$ and $AdS_{d+1} =SO(d, 2)/SO(d, 1)$ respectively. The dual gauge theory is always a conformal field theory ${\rm CFT}_d$ living on the boundary of $AdS_{d+1}$.

The Klein-Gordon equation in $AdS_{d+1}$ reads explicitly 
\begin{eqnarray}
z^{d+1}\partial_z(z^{1-d}\partial_z\phi)+z^2\partial^2\phi-m^2L^2\phi=0.
\end{eqnarray}
Let $f_k(z)$ be the Fourier transform in the $x-$space.
Near the conformal boundary $z=0$ we have the behavior 
\begin{eqnarray}
f_k(z)\longrightarrow A(k)z^{d-\Delta}+B(k)z^{\Delta}~,~z\longrightarrow 0.\label{KGF1}
\end{eqnarray}
The exponent $\Delta$ is the scaling dimension of the field given by
\begin{eqnarray}
\Delta=\frac{d}{2}+\sqrt{\frac{d^2}{4}+m^2L^2}.
\end{eqnarray}
For $m^2>-{d^2}/{4L^2}$ we have $d-\Delta \leq \Delta$ and hence $z^{d-\Delta}$ is dominant as $z\longrightarrow 0$.  The behavior on the boundary is then 
\begin{eqnarray}
\phi(z=\epsilon,x)= A(x)\epsilon^{d-\Delta}~,~\epsilon\longrightarrow 0.
\end{eqnarray}
This is divergent for $m^2>0$ 
  and hence the scalar field living on the boundary is identified with $A(x)$, viz
\begin{eqnarray}
  \varphi(x)={\rm lim}_{\epsilon\longrightarrow 0}\epsilon^{\Delta-d}\phi(\epsilon,x).
\end{eqnarray}
This is the scalar field representing (or dual to) the anti-de Sitter scalar field $\phi(z,x)$ at the boundary $z=0$. 

Let ${\cal O}(z,x)$ and ${\cal O}(x)$ be the dual operators of the scalar fields $\phi(z,x)$ and $\varphi(x)$ respectively, i.e.
\begin{eqnarray}
  S_{\rm bound}=\int d^dx\sqrt{\gamma}\phi(\epsilon,x){\cal O}(\epsilon,x)
  =L^d \int d^dx\varphi(x){\cal O}(x).
\end{eqnarray}
We get 
\begin{eqnarray}
   {\cal O}(\epsilon,x)=\epsilon^{\Delta}{\cal O}(x).
\end{eqnarray}
This shows explicitly that $\Delta$ is the scaling dimension of the dual operator ${\cal O}$.
 
We are interested in computing  the boundary CFT correlation functions
 \begin{eqnarray}
\langle{\cal O}(x_1)...{\cal O}(x_n)\rangle.
 \end{eqnarray}
In the CFT this is done via the formula 
 \begin{eqnarray}
   \langle{\cal O}(x_1)...{\cal O}(x_n)\rangle=\frac{\delta^n\log Z_{CFT}[J]}{\delta J(x_1)...\delta J(x_n)}|_{J=0}.
 \end{eqnarray}
 Where
 \begin{eqnarray}
   Z_{\rm CFT}[J]=\langle\exp(\int J(x){\cal O}(x))\rangle.
 \end{eqnarray}
The AdS/CFT correspondence states then that the CFT generating functional with source $J=\phi_0=\phi(0,x)$ is equal to the path integral on the gravity side evaluated over a bulk field which has the value $\phi_0$ at the boundary of AdS  \cite{Gubser:1998bc,Witten:1998qj}. We write 
  \begin{eqnarray}
    Z_{\rm CFT}[\phi_0]=\langle\exp(\int \phi_0(x){\cal O}(x))\rangle=\int_{\phi\longrightarrow \phi_0} {\cal D}\phi\exp(S_{\rm grav}).
 \end{eqnarray}
In the limit in which classical gravity is a good approximation we use the classical on-shell gravity action, i.e.
 
  \begin{eqnarray}
    Z_{\rm CFT}[\phi_0]=\exp(S_{\rm grav}^{\rm on-shell}[\phi\longrightarrow\phi_0]). 
 \end{eqnarray}
After holographic renormalization 
  the on-shell action is renormalized and the AdS/CFT prescription becomes
  \begin{eqnarray}
    Z_{\rm CFT}[\phi_0]=\exp(S_{\rm grav}^{\rm renor}[\phi\longrightarrow\phi_0]). 
  \end{eqnarray}
The correlation functions are then renormalized as
  \begin{eqnarray}
\langle{\cal O}(x_1)...{\cal O}(x_n)\rangle=\frac{\delta^n S_{\rm grav}^{\rm renor}[\phi\longrightarrow\phi_0]}{\delta\varphi(x_1)...\delta\varphi(x_n)}|_{\varphi=0}.
  \end{eqnarray}
See for example \cite{Ramallo:2013bua}.
    
\subsection{The geometry/entanglement connection}

\subsubsection{Ryu-Takayanagi formula}
 The Ryu-Takayanagi formula \cite{Ryu:2006bv,Ryu:2006ef} is a generalization of the  Bekenstein-Hawking formula, based on the ${\rm AdS}/{\rm CFT}$ correspondence, in which we identify the entanglement entropy in $(d+1)-$dimensional QFT with a geometric quantity in $(d+2)-$dimensional gravity.

We consider the metric in $AdS_{d+2}$ given by
\begin{eqnarray}
ds^2=g^{\mu\nu}dx_{\mu}dx_{\nu}=\frac{R^2}{z^2}(-dt^2+\sum_{i=1}^ddx_i^2+dz^2).
\end{eqnarray}
The dual $(d+1)-$dimensional CFT lives on the boundary located at $z=0$. The radial coordinate $z$ is a lattice spacing and the theory on the boundary should be properly understood as the continuum limit (in the sense of RG) of a regularize CFT, i.e. with a cutoff $\Lambda$.  This regularized CFT lives on a surface $z=a$ where $a=1/\Lambda$ and $a\longrightarrow 0$.

Our observers live on the boundary $z=a\longrightarrow 0$ of $AdS_{d+2}$. The accessible region $A$ and the inaccessible region $B$ are both on this boundary $z=a$.

The entanglement entropy in the   ${\rm CFT}_{d+1}$ which lives on this boundary can be compute from the  gravity theory which lives in the bulk ${\rm AdS}_{d+2}$ by the formula


\begin{eqnarray}
S_A=\frac{{\rm Area}(\Gamma_A)}{4G_N^{(d+2)}}.
\end{eqnarray}
The $\Gamma_A$ is a minimal area surface in the  time slice $M$ which is the extension of  ${\bf N}=A\cup B$ with boundary  $\partial\Gamma_A=\partial A$. See figure (\ref{sketch2}).

\begin{figure}[H]
\begin{center}
  \includegraphics[angle=-0,scale=0.4]{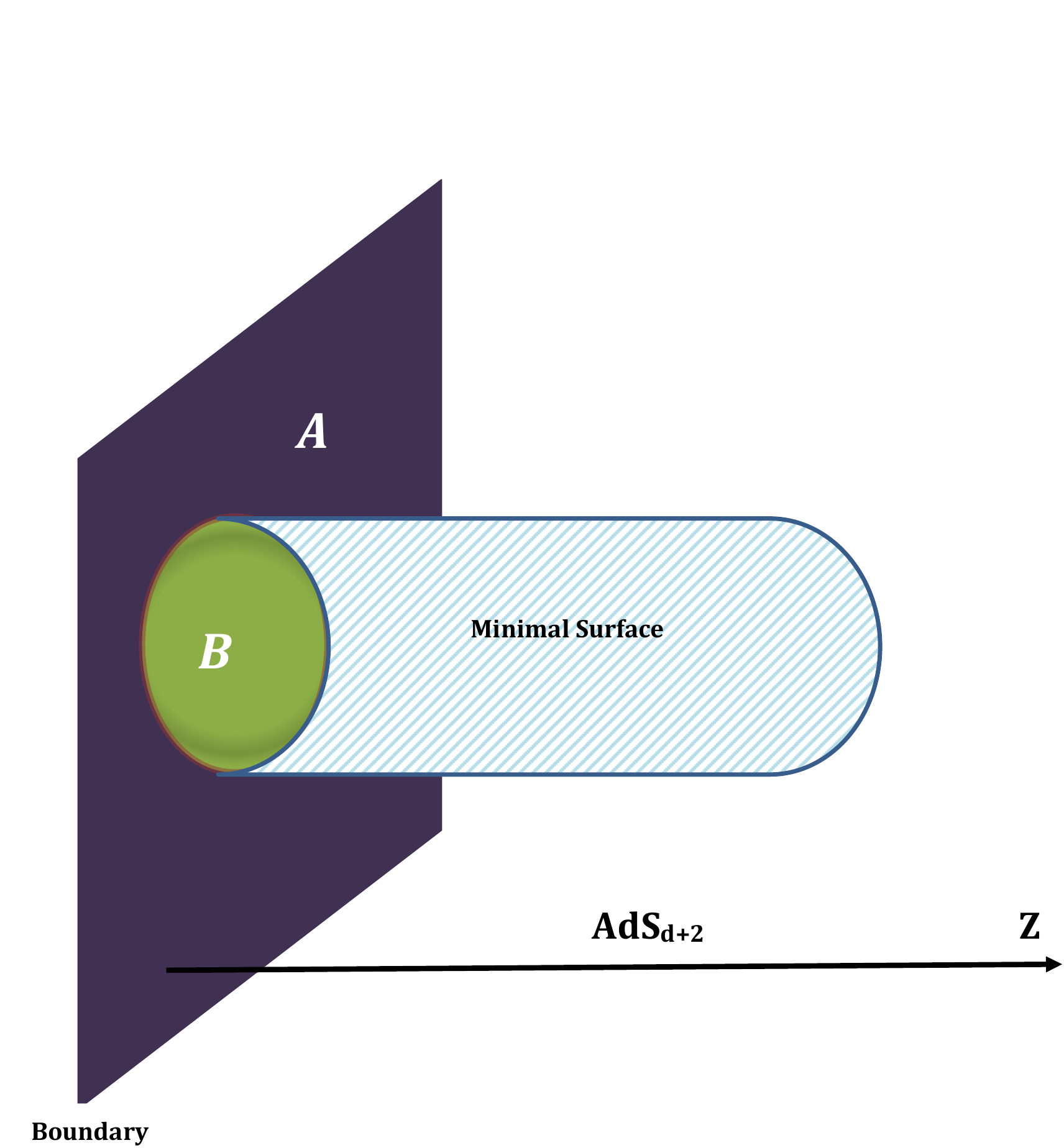}
\end{center}
\caption{The accessible and inaccessible regions in AdS.}\label{sketch2}
            \end{figure}

As an example we consider ${\rm AdS}_3/{\rm CFT}_2$. In this case we are interested in the line segment $x\in [-l/2,l/2]$ on $z=a$. This segment is extended in the bulk to a circle (which we can check is the line of minimal length) given by

\begin{eqnarray}
  z=\frac{l}{2}\sin s~,~x=\frac{l}{2}\cos s~,~\epsilon \leq s\leq \pi-\epsilon~,~\epsilon=\frac{2a}{l}\longrightarrow 0.
\end{eqnarray}
The length of the circle is given by $L=2R \ln l/a$.
The entanglement entropy becomes
\begin{eqnarray}
S_A=\frac{L}{4G_N^{(3)}}= \frac{c}{3}\ln \frac{l}{a}.
\end{eqnarray}
The central charge $c$ of ${\rm CFT}_2$ is related to the radius $R$ of ${ AdS}_3$ by the relation  \cite{Brown:1986nw}
\begin{eqnarray}
c=\frac{3R}/{2G_N^{(3)}}.
  \end{eqnarray}
The calculation on the field theory side is more involved. For a conformal field theory in two dimensions with central charge $c$ on the torus we find explicitly \cite{Holzhey:1994we}
  \begin{eqnarray}
    S_A
  &=&\frac{c}{3}\ln \frac{\Sigma}{\epsilon}.
\end{eqnarray}

 \subsubsection{The CFT/BH correspondence}

 The CFT/BH correspondence between conformal field theories and black holes  is the most understood case of the holographic gauge/gravity duality.

 Let $|\psi(0)\rangle$ be the vacuum state of the ${\rm CFT}_d$. This state is dual to pure AdS metric given by (Poincare coordinates)
\begin{eqnarray}
  ds^2= \frac{L^2}{z^2}(dz^2+dx_{\mu}dx^{\mu}).
\end{eqnarray}
Let $|\psi(\zeta)\rangle$ be a one-parameter family of ${\rm CFT}_d$ excited states which are dual to the perturbed metrics (Fefferman-Graham coordinates)
  \begin{eqnarray}
ds^2= \frac{L^2}{z^2}(dz^2+dx_{\mu}dx^{\mu}+z^d\bar{h}_{\mu\nu}(z,x)dx^{\mu}dx^{\nu}).
\end{eqnarray}
This corresponds to a spacetime $M_{\zeta}$ with boundary at $z\longrightarrow 0$ denoted by $\partial M_{\zeta}$ where the state  $|\psi(\zeta)\rangle$ is living.

This metric can also be understood as corresponding to a spacetime  $M_{\zeta}$, which is a perturbation of pure AdS, dual  to a small perturbation  $|\psi(\zeta)\rangle_{\zeta\longrightarrow 0}$ of the ${\rm CFT}_d$ vacuum  $|\psi(0)\rangle$. This is  an asymptotically AdS spacetime.

For higher excited states   $|\psi(\zeta)\rangle$ we can not assume classical supergravity solution since $l_s$ is no longer much less than $L$ and as a consequence stringy corrections of the order $l_s^2$ and higher become important. The geometry (and even the topology) of $AdS_{d+1}$ becomes therefore very different.

As an example we consider the Schwarzschild-AdS black hole in $d+1$ dimensions given by the metric 

\begin{eqnarray}
 ds^2= -f_Mdt^2+\frac{dr^2}{f_M}+r^2d\Omega_{d-2}~,~ f_M=1-\frac{2\mu}{r^{d-3}}+\frac{r^2}{L^2}.
\end{eqnarray}
The Penrose diagram of the   Schwarzschild-AdS spacetime is shown on figure (\ref{sch1p}). The two light-like infinities ${\cal J}^{\pm}$ and the space-like infinity $i^0$ are replaced with the universal covering of global AdS spacetime in both regions I and III.
The asymptotic regions are denoted $A$ and $B$ and they are the conformal boundary of AdS, i.e. ${\bf R}\times {\bf S}^{d-1}$. 
The regions I and III are causally disconnected but signals from region I can intersect with signals from region III behind the horizon.
  This is then a two-sided black hole, i.e. with two different exterior regions, which can also be viewed as a wormhole.

The Schwarzschild-AdS black hole is conjectured  to be dual to the thermofield double state $|\Psi\rangle$ given by \cite{Maldacena:2001kr}
\begin{eqnarray}
  |\Psi\rangle=\frac{1}{\sqrt{Z(\beta)}}\sum_i\exp(-\beta E_i/2)|E_i^A\rangle\otimes|E_i^B\rangle.
\end{eqnarray}
This is the vacuum state  of two identical non-interacting copies of the conformal field theory living on the boundaries $A$ and $B$ (which can only interact via entanglement).

The Schwarzschild-AdS black hole is a connected spacetime in which light signals traveling from $A$ and $B$ can intersect. This connectedness is equivalent to the entanglement between the two subsystems $Q_A$ and $Q_B$, which contain the degrees of freedom of the local ${\rm CFT}$ living on $A$ and $B$, as expressed by the thermofield double state $|\Psi\rangle$.




The reduced density matrix for the subsystem $Q_A$ is immediately given by integrating out the degrees of freedom associated with the subsystem $Q_B$ as $\rho_A=tr_B |\Psi\rangle\langle\Psi|$ with a non-zero entanglement entropy $S_A=-tr_A \rho_A\log \rho_A$.

  In the limit of zero temperature the reduced density matrix $\rho_A$ approaches $\rho_A=|E_0^A\rangle\langle E_0^A|$ and as a consequence the entanglement entropy vanishes, i.e. 
\begin{eqnarray}
  |\Psi\rangle\longrightarrow |\Phi\rangle=|E_0^A\rangle\otimes|E_0^B\rangle~,~T\longrightarrow 0.
\end{eqnarray}
This is dual to a spacetime consisting of the product of two pieces corresponding to the two disconnected regions $A$ and $B$, i.e. light signals traveling from $A$ and $B$ can not intersect. Thus in the limit $T\longrightarrow 0$ entanglement between $Q_A$ and $Q_B$ is removed and correspondingly connectivity between regions $A$ and $B$ is broken. This shows clearly that entanglement between $Q_A$ and $Q_B$ is a necessary condition for classical connectivity between $A$ and $B$ \cite{VanRaamsdonk:2010pw} .

  \begin{figure}[H]
\begin{center}
\includegraphics[angle=-0,scale=0.5]{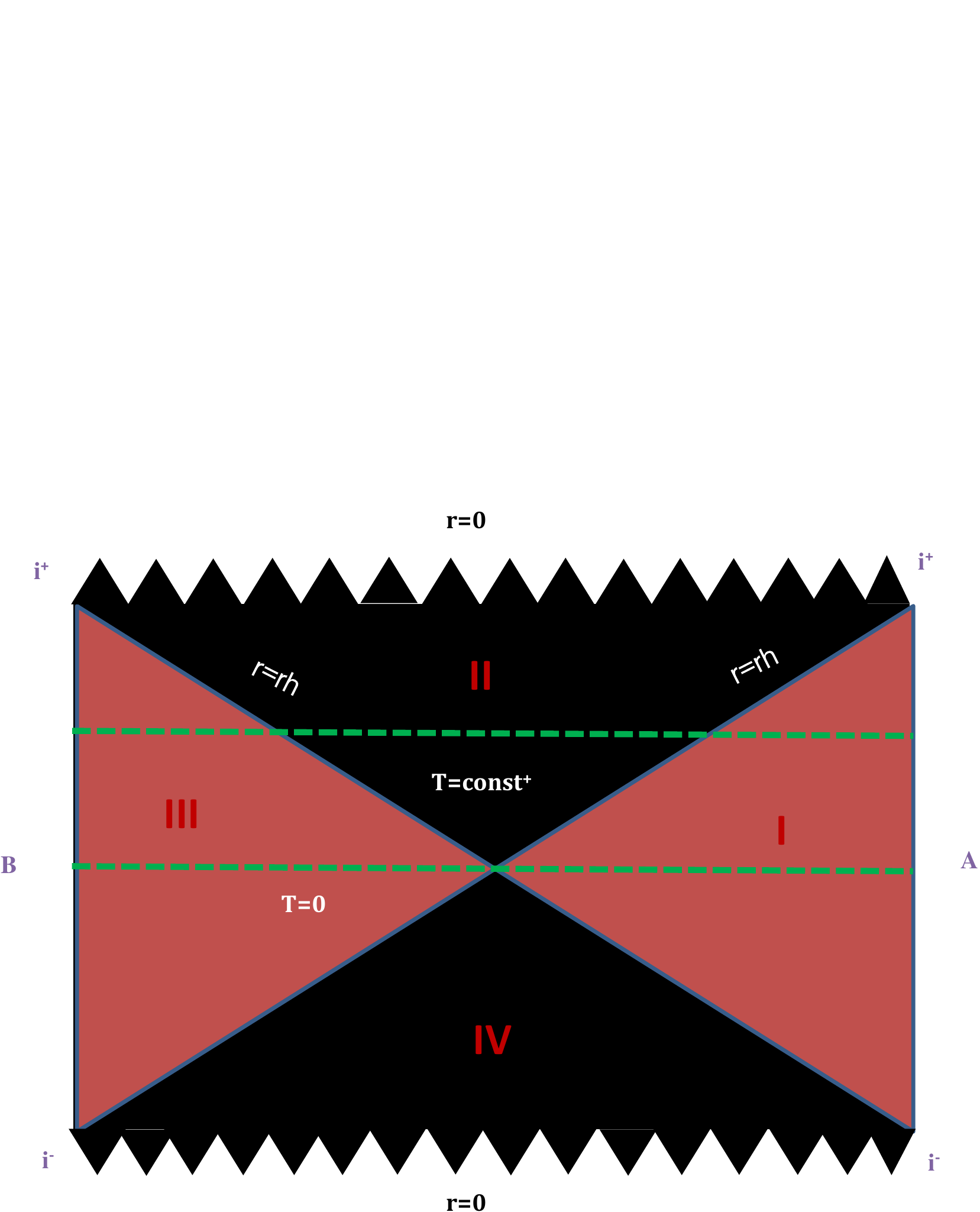}
\end{center}
\caption{The  Schwarzschild-AdS spacetime.}\label{sch1p}
\end{figure}
\subsubsection{The first law of entanglement and Einstein's equations}

  We consider a one-parameter family of CFT states $|\psi(\zeta)\rangle$. The dual spacetime of the vacuum state $|\psi(0)\rangle$ is pure AdS. The dual spacetime ${\cal M}_{\psi}$ of the perturbed state  $|\psi(\zeta)\rangle$  is a perturbation of AdS with boundary $\partial {\cal M}_{\psi}$ which is a perturbation of Minkowski spacetime.

  The accessible region $B$ is a ball shaped region of radius $R$ on the boundary  $\partial {\cal M}_{\psi}$.

  The entanglement entropy of $B$ in the CFT is equal to the von Neumann entropy of the reduced density matrix $\rho_B=tr_{\bar{B}}|\psi(\zeta)\rangle\langle\psi(\zeta)|$. We compute immediately 
    \begin{eqnarray}
  \frac{d}{d\zeta}S_B   &=&\frac{d}{d\zeta}E_B^{\rm Hyp}.
    \end{eqnarray}    
    The expectation value of the modular Hamiltonian $ H_B=-\log \rho_B|_{\zeta=0}$ is the hyperbolic energy  $E_B^{\rm Hyp}$.
    
    This is effectively the first law of thermodynamics $dE=dS$ which holds in the CFT for arbitrary perturbations $\zeta$ of the vacuum state $|\psi(0)\rangle$.
    
    The holographic extension of this law to the bulk gives (after a very long calculation) precisely Einstein's equation around AdS background.  This is the prime example of how spacetime geometry is equivalent to quantum entanglement \cite{Lashkari:2013koa,Faulkner:2013ica,VanRaamsdonk:2016exw,Casini:2011kv}.

   \subsection{Emergent time from entanglement experimentally verified!}
The quantum measurement and the associated irreversible and 
instantaneous collapse of the wave function seem to lie at the heart of the 
problem of time. The only other fundamental principle in nature which is known to be irreversible is the second 
law of thermodynamics which states that the entropy of a closed system can only increase over time. The entropy of 
the universe is thus constantly increasing with the cosmological expansion from its very small starting value at the initial big bang singularity. The smallness of the initial entropy is in itself a quite mysterious property \cite{Penrose}. On the 
other hand, the constant increase of entropy with the expansion of the universe is thought to be correlated with the observed arrow of time which is directed from past to future in the same way that the quantum measurement as well as causality are directed from past to future. 

Thus, the problem of time is reduced to the problem of the flow of time, which in turn is reduced to the problem of the 
arrow of time, and then reduced even further to the law of the entropy increase. It seems that this can be reduced 
even further to quantum entanglement which is incidentally measured by entropy. 
measurement involves the entanglement between the observer and the observed in an essential way. 


The problem of time \cite{Isham:1992ms,Sorkin:1987cd,Ashtekar:2004vs,Anderson:2010xm}, in its simplest forms, is the realization that the Wheeler-De Witt equation \cite{DeWitt:1967yk} describes a static universe without any time.  This is because in the canonical formulation of general relativity the Hamiltonian is a Dirac constraint. So, in this almost unique instance, quantum mechanics (with its absolute time) and general relativity (with its dynamical time) are merged successfully (the Wheeler-De Witt is the Schrodinger equation for the metric without any UV divergences), only at the cost of dispensing altogether with the 
notion of time. 

But time must emerge again somehow from this equation in order to make contact with the observed time evolution of the universe. A solution of this problem is given by Page and Wootters \cite{Page:1983uc} (see also \cite{Gambini:2008ke}) in which time emerges dynamically from quantum entanglement. This goes as follows.  The universe as a whole is a static system $S$ given by a static state vector $|\Psi\rangle$ as dictated by the  Wheeler-De Witt equation, viz $H|\Psi\rangle=0$, where $H$ is the Hamiltonian of the universe. This static system $S$ is divided into two entangled subsystems: a subsystem $C$ which is called a clock, and the rest of the system given by the subsystem $R$. We will assume that we can neglect the interactions between these subsystems and hence the Hamiltonian can be rewritten as

\begin{eqnarray}
H=H_C\otimes {\bf 1}_R+{\bf 1}_R\otimes H_R.
\end{eqnarray}
The external observer sees no time since his Schrodinger equation is $H|\Psi\rangle=0$. The subsystems $C$ and $R$ both obey the Schrodinger's equation with respect to an internal observer whose time parameter is given by the clock $C$. The state vector of the clock may be written as  
\begin{eqnarray}
|\Phi(t)\rangle_c=\exp(-iH_Ct/\hbar)|\Phi(0)\rangle_C.
\end{eqnarray}
Thus, we can immediately compute  (since $H=0$ and $[H_R,H_C]=0$)
\begin{eqnarray}
|\psi(t)\rangle_R=\langle\Phi_C(t)|\Psi\rangle=\exp(-iH_Rt/\hbar)\langle\Phi_C(0)|\Psi\rangle=\exp(-iH_Rt/\hbar)|\psi(0)\rangle_R.
\end{eqnarray}
In other words, the subsystem $R$ is seen by the internal observer, using the clock $C$, as evolving in time $t$. Hence, quantum entanglement between the chosen clock and the rest of the universe, yields a static universe with respect to a hypothetical external observer, and an evolving universe with respect to an internal observer (perhaps the clock itself). The internal observer detects the evolution by inspecting the quantum correlations between the clock $C$ and the subsystem $R$, whereas the external observer inspects the global properties of the entangled state of $C$ and $R$. It is clearly seen that what is called here a clock is what Wigner calls a conscious mind entangled with, and performing measurements on, the rest of the world.

The Page-Wootters solution of the problem of time is very much in the spirit of the relational interpretation of quantum mechanics 
\cite{Rovelli:1995fv} which all fall, on the surface of it, within the philosophical relationalism of space and time advocated by Leibniz against Netwon's absolutism and Kant's transcendental idealism. Thus, on this view, time is not absolute (Newton) nor it is an intuition of the cognition (Kant) but it is a relation to clocks. But as we stated above clocks are intertwined inextricably with internal observers as in all interpretations of quantum mechanics.


This elegant and simple scenario for the emergence of time from quantum entanglement was verified experimentally in the truly beautiful experiment of \cite{Moreva:2013ska}.  In this experiment the universe is constituted of only two entangled photons prepared in the initial state 
\begin{eqnarray}
|\Psi\rangle=\frac{1}{\sqrt{2}}(|H\rangle_C|V\rangle_R-|V\rangle_C|H\rangle_R).
\end{eqnarray}
This satisfies the Wheeler-De Witt equation with $H_C=H_R=i\hbar\omega(|H\rangle\langle V|-|V\rangle\langle H|)$ where $\omega$ is the time scale of the model. Thus, if the clock photon is in a horizontal polarization state, the other photon will be in a vertical polarization state, and vice versa. The two photons are sent along two separate paths through identical birefringent plates which cause rotations of the polarization of the photons. The amount of rotation incurred is proportional to the time spent by the photon in the plates. Thus, the thickness of the plates is a measure of the coordinate time. 

The clock photon has only two time readings: $t_1$ corresponding to a horizontal polarization (detector $1$ clicking) and $t_2$ corresponding to a vertical polarization (detector $2$ clicking). The internal observer who measures the clock photon in detectors $1$ and $2$ measures also the other photon's polarization in detectors $3$ (vertical) and $4$ (horizontal). These measurements are expressed by the conditional probability $P_{3|x}=p(t_x)$: the probability that the other photon has vertical polarization (detector $3$ clicking) if the clock detector $x$ clicks. This probability is exactly given by the average number of coincidences between the detectors $3$ and $x$ over the total number of photon pairs from all detectors.

The internal observer can only access the clock time $t_x$ and not the coordinate time. This is achieved experimentally by changing in a random way the thickness of the  birefringent plates in the path of the clock photon, repeating the same conditional probability measurement, and then taking average over the thickness. This allows us to obtain the probabilities  $P_{3|x}^{\tau_i}=p(t_x+\tau_i)$, with $\tau_i=\delta_i/\omega_i$, where $\delta_i$ is the optical thickness. In this so-called ``observer mode'', it is seen that the internal observer becomes entangled with the photon clock, and thus she can see an evolution of the other photon's state as a function of the clock time. 

The experimental result of the average number of coincidences as a function of time is found in excellent agreement with the theory \cite{Moreva:2013ska} which, for a global state $\rho$, is given by the conditional probability 
\begin{eqnarray}
P(d,t)=\frac{\int dT TrP_{d,t}(T)\rho}{\int dT P_t(T)\rho}. 
\end{eqnarray}
The $P_t(T)$ is the projector corresponding to a measurement of a clock time $t$ at the coordinate instant $T$, and $P_{d,t}(T)$ is the projector corresponding to the measurement $t$ of the clock photon, and to measurement $d$ of the other photon, at the coordinate instant $T$. The fact that we integrate over $T$ means that this conditional probability does not depend on the coordinate time.

On the other hand, in the so-called ``super-observer mode'', it is shown that the global entangled state of the two photons remains the same for any thickness 
of the birefringent plates which represents, as we said, the external coordinate time. 
Thus, although the external observer has access to the 
thickness of the plates (coordinate time), she is required to coherently quantum  erase \cite{SEW,DNR} the which-way information, i.e. the clock time measurements of the internal observer,  obtained in the measurement of the clock photon polarization, in order to avoid becoming herself entangled with the clock photon. The global state is found to be indeed static with respect to coordinate time.



\subsection{Other models of quantum gravity}
\subsubsection{The causal dynamical triangulation vs  Horava-Lifshitz gravity \footnote{This section is taken verbatim from \cite{Ydri:2017riq}.}}
Another powerful approach to the emergence of time and spacetime is Lorentzian causal dynamical triangulation \cite{Ambjorn:2005qt,Ambjorn:2006hu,Ambjorn:2007jv}. In this approach spacetime is built out of four-simplices (generalization of two-simplices, i.e. triangles, to four dimensions) which are equipped with a flat Minkowski metric. The causality requirement singles out globally hyperbolic manifolds which admit a global proper-time foliation structure and as a consequence Wick rotation to Euclidean is meaningful. The Hilbert-Einstein action is given in this discrete setting by the Regge action \cite{Regge:1961px}. The path integral is obtained as the sum over the set of all causal triangulations weighted with the Regge action. The parameters of the model are Newton’s gravitational constant $G$ and the cosmological constant $\Lambda$ which appear as the parameters $K_0$ and $K_4$ in the Regge action. Also the model depends on two more parameters given by the lengths of time-like and spatial-like links $a_t$ and $a_s$ respectively. We have $a_t^2=\alpha a_s^2$ where the asymmetry factor $\alpha< 0$ appears as a parameter $\Delta$ in the Regge action.

Causal dynamical triangulation (CDT) is intimately related to Horava-Lifshitz (HL) gravity \cite{Horava:2009if,Horava:2008ih,Horava:2009uw} which also, like CDT, assumes global time foliation and introduces anisotropy between space and time but in such a way as to achieve power-counting renormalizability of quantum gravity. This theory is effectively a generalization to gravity of the $d-$dimensional Lifhsitz scalar field theory given by the Lifhsitz-Landau free energy density \cite{Goldenfeld:1992qy}
\begin{eqnarray}
S=a_2\phi^2+a_4\phi^4+...+c_2(\partial_{\alpha}\phi)^2+d_2(\partial_{\beta}\phi)^2+e_2(\partial_{\beta}^2\phi)^2+...\label{LL}
\end{eqnarray}
The anisotropy is introduced by the distinction between the indices  $\beta=1,...,m$ and $\alpha=m+1,...,d$. The three phases present in this theory are: helicoidal ($|\partial_t\phi(x)|<0$), paramagnetic ($\phi(x)=0$) and ferromagnetic ($|\phi(x)|>0$). The phase diagram is depicted in figure $(a)$ of (\ref{CDT}).

The phase structure of causal dynamical triangulation is also summarized in figure $(b)$ of (\ref{CDT}). The cosmological constant $K_4$ which controls the total volume is fixed at its critical value and the phase diagram is then drawn in the plane $K_0-\Delta$ where $K_0$ is proportional to the inverse bare gravitational coupling constant $G$ while $\Delta$ is effectively the asymmetry factor $\alpha$. There are three distinct phases \cite{Ambjorn:2010hu,Gorlich:2011ga}
\begin{itemize}
\item The de Sitter spacetime phase $C$. This is the analogue of the ferromagnetic phase $(d_2>0, a_2<0$) of the Lifshitz scalar field theory or the ordered phase in noncommutative scalar field theory (see below).
\item The crumpled phase $B$ where neither space nor time have any extent and therefore there is no geometry.  This is the analogue of the paramagnetic phase ($d_2>0, a_2>0$) of the  Lifshitz scalar field theory or the disordered phase in noncommutative scalar field theory (see below). 
\item The branched polymer phase $A$ where the geometry oscillates in time.  This is the analogue of the helicoidal phase ($d_2<0$) in Lifshitz scalar field theory or the non-uniform ordered phase in noncommutative scalar field theory (see below). 
\item The transition from $A$ to $C$ is first order whereas the transition from $B$ to $C$ could be either first order or second order and as a consequence there is a possibility of a continuum limit. Similarly, the transition between the ferromagnetic ($C$) and paramagnetic ($B$) phases in the  Lifshitz scalar field theory although usually second order it could be first order. As we will see a strikingly similar situation occurs in noncommutative scalar field theory (see below). 
\item Also in both theories CDT and HL the spectral dimension at short distances is $2$ and only becomes $4$ at large distances and the anisotropy between space and time disappear in CDT in the de Sitter spacetime phase while in HL it disappears at low energies.
\end{itemize}

\begin{figure}[H]
\includegraphics[width=8.0cm,angle=0]{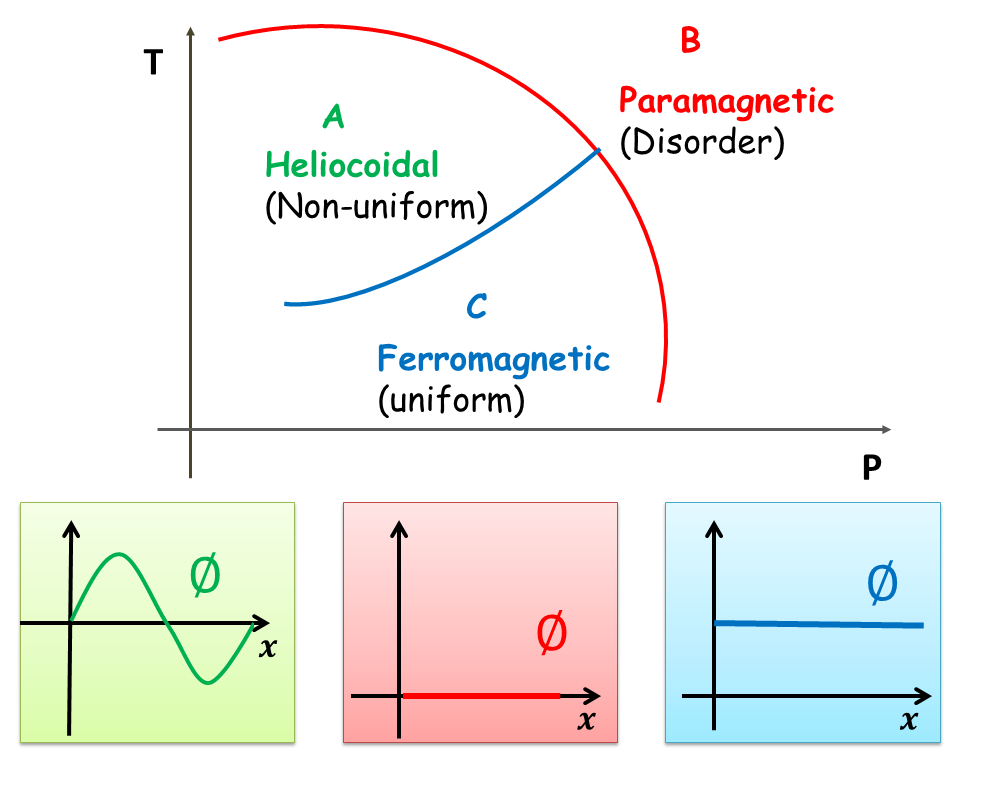}
\includegraphics[width=8.0cm,angle=0]{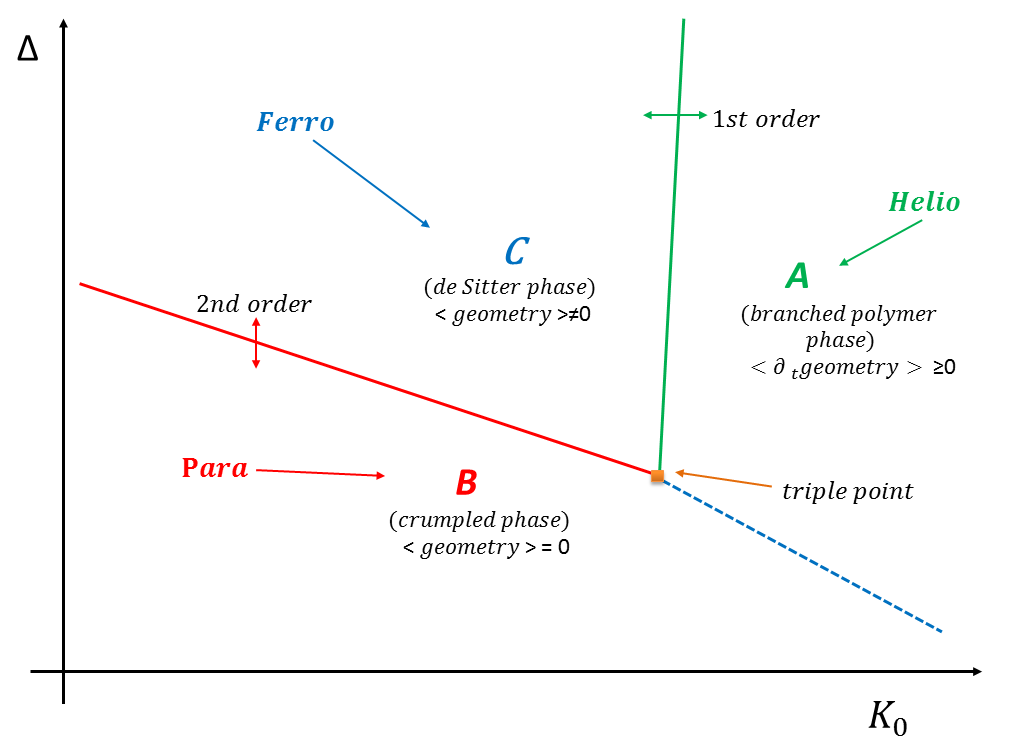}
\includegraphics[width=12.0cm,angle=0]{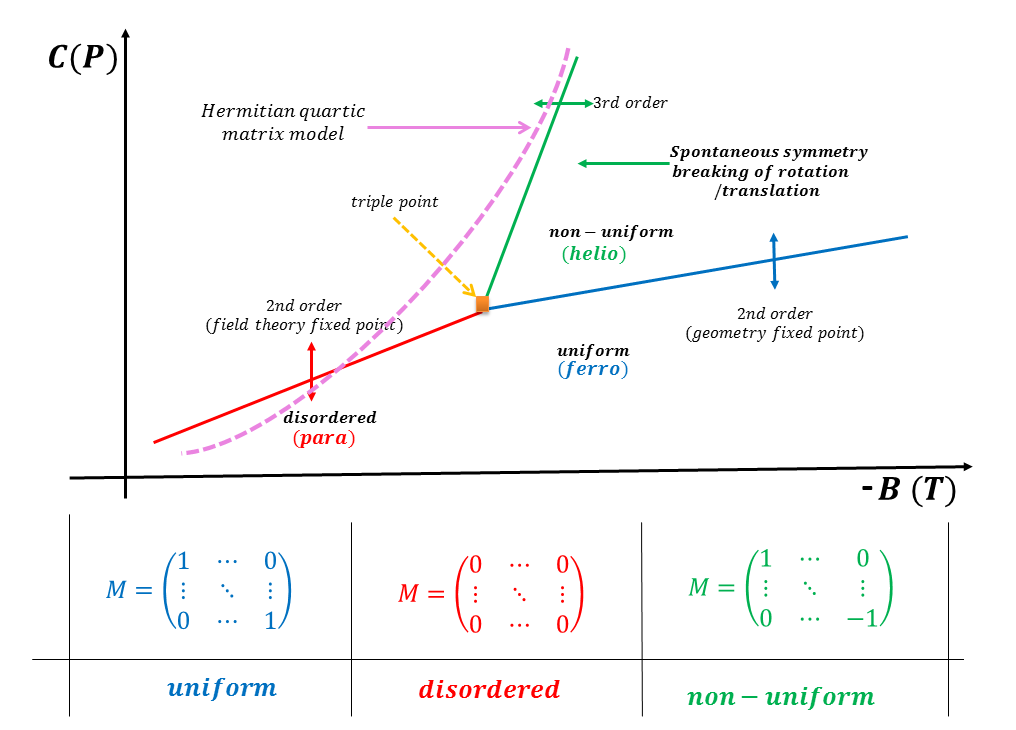}
\caption{The phase diagrams of causal dynamical triangulation \cite{Ambjorn:2010hu,Gorlich:2011ga}, Lifshitz scalar field theory \cite{Goldenfeld:1992qy} and multitrace matrix model \cite{Ydri:2017riq}.}
\label{CDT}
\end{figure}

\subsubsection{Matrix models of string theory and non-commutative geometry}
  
The combination of the matrix models of string theory (such as the IKKT matrix model \cite{Ishibashi:1996xs}, the BFSS matrix model\cite{Banks:1996vh} and the BMN matrix model \cite{Berenstein:2002jq}) and Conne's noncommutative geometry \cite{Connes:1996gi} provide the most promising (the author's view) candidate of quantum gravity as an emerging structure. The quantum entanglement of quantum mechanics is due fundamentally to the non-commutative structure of the phase space after Dirac quantization. In the matrix model approach the starting point is precisely a non-commutative structure between the spacetime coordinates. For a systematic review see \cite{ydri} and references therein. A good representation of this approach is the work of Steinacker \cite{Steinacker:2011ix}.

\section{Synthesis: quantum dualism}

Thus, the reduction/collapse of the state vector is formally very similar to the information loss problem. Indeed, they both involve going from a pure state to a mixed state which results in a  loss of quantum entanglement and decrease of information. In summary, we have 
    \begin{itemize}
  \item {\bf The measurement problem:} $\rho_{\rm e}\longrightarrow \rho_r$ (inaccessible degrees of freedom of the environment).
  \item {\bf The information loss problem:} $\rho_{\rm in}\longrightarrow \longrightarrow \rho_{\rm out}$ (inaccessible degrees of freedom of the information behind the horizon).
    \end{itemize}
    In more detail we have
    \begin{eqnarray}
      \rho_e=|\Phi_e\rangle\langle\Phi_e|\longrightarrow \rho_r=\rho_{S+D}=Tr_E|\Psi\rangle\langle\Psi|~,~|\Psi\rangle=U|\Phi_e\rangle|E_0\rangle.
    \end{eqnarray}
    \begin{eqnarray}
      \rho_{\rm in}=|\psi_{\rm in}\rangle\langle\psi_{\rm in}|\longrightarrow \rho_{\rm out}=\rho_{\rm Hawking}=Tr_S|\psi_{\rm out}\rangle\langle\psi_{\rm out}|~,~|\psi_{\rm out}\rangle=S|\psi_{\rm in}\rangle.
    \end{eqnarray}    
    The environment E acts therefore as an observer (Wigner for example) performing a measurement on the joint system S+D where S is the physical system under observation (Schrodinger's cat for example) and D is the detector which also acts as an observer (Wigner's friend). The joint system S+D+E evolves by means of a unitary matrix $U$ since the three parts S, D and E are highly entangled.

    The environment E contains therefore the inaccessible degrees of freedom of the system which lie at the root of the collapse of the wave function in the same way that the inaccessible degrees of freedom which went behind the horizon lie at the root of the information loss problem. In other words, the environment E plays the role of the interior of the black hole whereas the joint system S+D plays the role of the exterior of the black hole with an effective horizon separating them. In fact the measurement process (like a radiating black hole) can always be thought of as involving two entangled systems which are separated with a horizon where one of the systems (the observed system) contains the accessible degrees of freedom whereas the other one (the observing system) contains the inaccessible degrees of freedom which since they are inaccessible they are required to be traced out yielding as a consequence a sort of information loss or collapse of the state vector. 

    Of course the environment in this argument can be replaced with consciousness (for those who think that the observer's mind is what precipitates collapse \cite{Wigner1, Wigner2}) or quantum gravity (for those who think that spacetime curvature at the Planck scale is what precipitates an objective collapse \cite{hp}).

    By the holographic gauge/gravity duality \cite{Maldacena:1997re} the black hole is in fact dual to a certain unitary conformal field theory (CFT) and hence the information loss problem is only a coarse-graining effect since the information will eventually start to emerge with the radiation at the Page time when the entanglement entropy reaches its maximum (Page curve). A very explicit example is the case of Schwarzschild-AdS black hole which is conjectured to be dual to the thermofield double state $|\Psi\rangle$ of a pair of CFT's living on the conformal boundaries of AdS space \cite{Maldacena:2001kr} (see also \cite{VanRaamsdonk:2010pw}).

    A rigorous confirmation, using non-perturbative Monte Carlo methods, of the fact that black holes follow the predictions of the AdS/CFT correspondence is carried out in \cite{Hanada:2013rga}. This is the case of M-(atrix) theory, i.e. the case of the celebrated BFSS matrix model \cite{Banks:1996vh}, which defines the dual gauge theory of a black 0-brane as given by a bound state of $N$ D0-branes. In this case it is explicitly seen that in the limit $N\longrightarrow \infty$ both the Hawking radiation (as computed from the matrix model on the gauge side) and the measurement problem (since the theory becomes classical supergravity on the gravity side) disappear.

The measurement problem similarly to the information loss problem is not considered here to be a fundamental problem but only a coarse-graining effect. Indeed, this problem can be further related to black holes and thus made more amenable to the holographic gauge/gravity duality as follows.

We have a physical system S (which for concreteness we take to be the Schrodinger's cat C), a detector D and the environment E which both act as encapsulated observers whom we call Wigner's friend (WF) and Wigner (W) respectively.  We add a third encapsulated super observer whom we call Einstein which observes the joint system S+D+E. This is a variant of Wigner's friend experiment discussed by Susskind \cite{Susskind:2016jjb,Susskind:2014yaa}. 
    
The cat C is as usual in the superposition state of being dead and alive. 
 This is just one qubit that Wigner's friend WF (detector D as an observer) is usually taken to measure. However, the cat C is actually a collection of large $N$ of qubits and WF measures all these qubits.

In the Copenhagen interpretation the measurement of Wigner's friend collapses the state of the cat to one of its $2^N$ possible states in an irreducible way.

In the many-worlds formulation the measurement of Wigner's friend causes the qubits of the cat C to become entangled with the qubits of the memory states of Wigner's friend. This is the description which Wigner W (the environment E as a Copenhagen observer) should use  before he performs his measurement and collapses the state vector of the joint system C+WF (subjective-collapse model). But Wigner W himself as a many-worlds observer he becomes entangled with the system C+WF when he performs his measurement. 

By the stronger ER=EPR conjecture the qubits of the cat C which are entangled with the qubits of Wigner's friend are actually entangled because they are connected by quantum Einstein-Rosen bridges or quantum wormholes.

The weaker form of the ER=EPR uses the fungibility, i.e. the interchangeability, of the entanglement between its various forms \cite{Susskind:2016jjb}. If we assume that the degrees of freedom of a black hole can be represented as a collection of qubits, then by compressing the cat C and Wigner's friend WF into black holes, these black holes will be entangled, and as a consequence there will be quantum Einstein-Rosen bridges between the cat C (first black hole) and Wigner's friend WF (second black hole).  
We will also assume that messages between the cat C and Wigner's friend can be sent to meet in the wormhole.

We can also compress the cat into Hawking radiation and Wigner's friend into a black hole which allows us to the view the joint system C+WF  as an evaporating black hole entangled with its Hawking radiation. 

Next comes the measurement of Wigner W himself as a Copenhagen observer (the measurement of the environment E itself through decoherence).  
The Einstein-Rosen bridge is cut at the WF's end (because he is the one performing the measurement) and thus messages between the cat S and Wigner's friend WF can not meet in the wormhole after the completed measurement is done. The snipped ER bridge corresponds to a mixed density matrix. And the snipping will cause a firewall between the cat C (viewed as Hawking radiation) and Wigner's friend (viewed as an evaporating blakc hole), i.e. the information loss as characterized by the release of the firewall is exactly seen as the collapse of the state vector as characterized by the snipped ER bridge. 
The snipped ER bride can be made smooth again (corresponding to a unitary process) if Wigner cooperates with his friend as we will see shortly.

Indeed, with respect to Einstein (the third observer) the joint system Wigner+Wigner's friend+cat is described by a single quantum state. 
When Wigner performs his measurement on the system Wigner's friend+cat he becomes entangled with it. The system Wigner+Wigner's friend+cat is described by an entangled tripartite state in which the cat is entangled with the union of Wigner and his friend. This is a GHZ state \cite{GHZ}.

Thus, when the three systems (Wigner, Wigner's friend and the cat) are compressed into three black holes we get an Einstein-Rosen bridge connecting the three which Susskind also call the GHZ brane  \cite{Susskind:2016jjb}. This bridge allows messages between the cat and Wigner's friend to be sent to meet inside the GHZ brane in contradiction with the conclusion of Wigner, using the Copenhagen, which states that messages can not be communicated.

The crucial property of the GHZ brane or Einstein-Rosen bridge connecting the three black holes (compressed cat, compressed friend and compressed Wigner) is the fact that it corresponds to a maximally entangled GHZ tripartite state in the same way that the earlier Einstein-Rosen bridge connecting the two black holes (compressed cat and compressed friend) corresponds to a maximally entangled bipartite Bell state. In some sense then we might have called the earlier Einstein-Rosen bridge a Bell brane. The GHZ state is a state in which the union of two of its sub-systems is maximally entangled with the third one (as opposed for example to the W state in which the entanglement is less than maximal).

This property will also lie at the heart of the resolution of the apparent inconsistency between the Copenhagen and the many-worlds with regard to whether or not there can be messages sent out to meet in the wormhole. As it will turn out, these two interpretations are actually complementary or dual rather than contradictory.

The GHZ brane is thus characterized essentially by maximal GHZ entanglement which is invariant under local unitary transformations. This is a geometric structure localized behind the three horizons of the three entangled black holes which generalizes the Einstein-Rosen bridge between two entangled black holes. The GHZ brane is different from a classical tripartite wormhole which has no GHZ entanglement and thus the two structures can be differentiated.
The properties of the GHZ brane encodes the duality between the Copenhagen and many worlds interpretations.

To construct the GHZ state we start by simply replacing the cat and Wigner's friend by single qubits. The initial state of Wigner is denoted by $|0\rangle$ and his measurement of the state of his friend and the cat will cause him to become entangled with the state of his friend. If he finds his friend in the state $|0\rangle$ (the friend saw a dead cat) he will remain in the state $|0\rangle$ whereas if he finds his friend in the state $|1\rangle$ (the friend saw an alive cat) Wigner will transition to the state $|1\rangle$. The state of Wigner+Wigner's friend+cat is then given by
\begin{eqnarray}
  \frac{1}{\sqrt{2}}(|00\rangle|0\rangle+|11\rangle|1\rangle).
  \end{eqnarray}
This is the GHZ state for single qubits. It is generally true that when a measurement is performed on one member of an entangled  bipartite system a GHZ state is induced. 
The main two properties of the GHZ state are given by
\begin{itemize}
\item    By tracing over any two qubits the density matrix of the third qubit is maximally mixed. This means that the union of any two qubits is maximally entangled with the third qubit.
\item    By tracing over any one qubit we get a separable density matrix of the other two. In other words, the density matrix is a sum of projection operators on unentangled pure states and as a consequence there is no entanglement between any two parties.
\end{itemize}
This can be generalized to any three complex systems each of which is formed out of a collections of large number of qubits. We get then a product of GHZ triplets corresponding to some tensor network.

It is now almost obvious how the discrepancy found between the Copenhagen and the many worlds can be resolved. In the Copenhagen interpretation when Wigner measures the state of the system friend+cat the Einstein-Rosen bridge between Wigner's friend and the cat is snipped, because collapse has occurred and entanglement is lost, and as a consequence any messages between the two (Wigner's friend and the cat) can not be sent to meet inside the wormhole. See figure (\ref{WBH}).

The many-worlds formulation is appropriate for Einstein when describing the state of Wigner, the friend and the cat. This is given by a GHZ state and indeed because of the separability condition no messages can be sent between any two parties since there is no entanglement between any two parties. So the two interpretations are consistent. But from this perspective the many-worlds seems more general than the Copenhagen because it captures another very important effect due to the preservation of reversibility and entanglement. For Einstein there can still be messages sent between Wigner's friend and the cat if the friend cooperates with Wigner. This is indeed true because of the first property above of the GHZ state which asserts that the union between any two parties is entangled with the third party and as a consequence messages can be sent between the union of any two parties and the third one. Hence the two perspectives are consistent and it is in this sense that the many-worlds is dual to the Copenhagen.

The duality between Copenhagen and many-worlds guarantee that our description is strictly unitary and as a consequence there can be no measurement problem more than as a coarse-graining effect.

Indeed, the measurement problem is due mainly to the quantum dualism inherent in the Copenhagen local view of the single-world reality which fortunately can be mapped to a strictly physicalist view of the many-worlds reality in the same way that the information loss problem is only due to the existence of an event horizon for the local Schwarzschild observer at infinity. The environment plays for the Copenhagen observer the same role as the interior of the black hole for  Schwarzschild observer.

Equally, and perhaps more importantly, is the possibility of fungibility  of entanglement which allows us the conversion (through unitary transformations) of physical systems, observers and environments into black holes and/or Hawking radiation which in principle can be made dual to a unitary conformal field theory via the holographic gauge/gravity correspondence. The  measurement problem can then be solved through the same means as the information loss problem at least in principle.

 

\begin{figure}[htbp]
\begin{center}
  \includegraphics[width=8.0cm,angle=0]{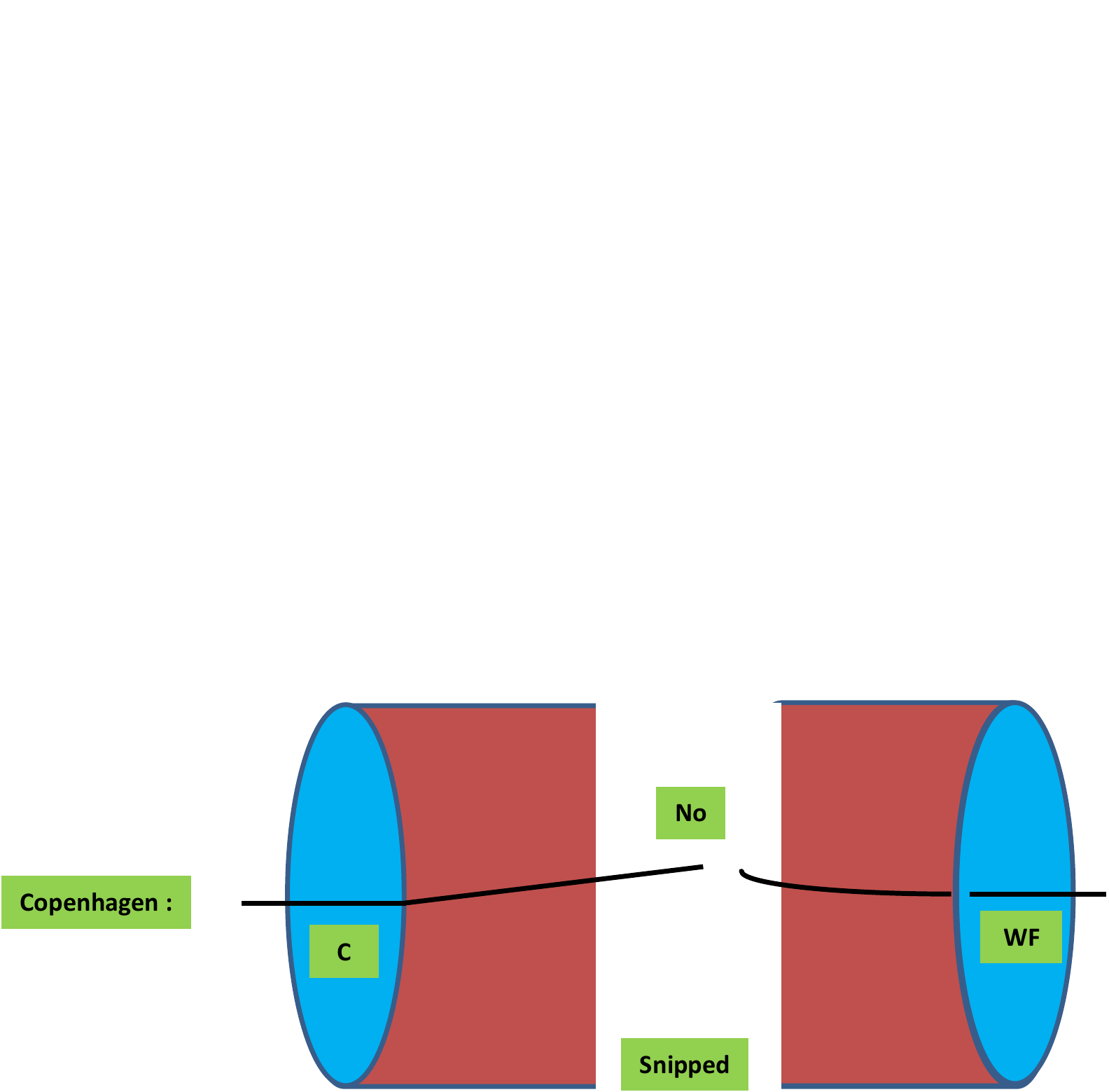}
  \includegraphics[width=10.0cm,angle=0]{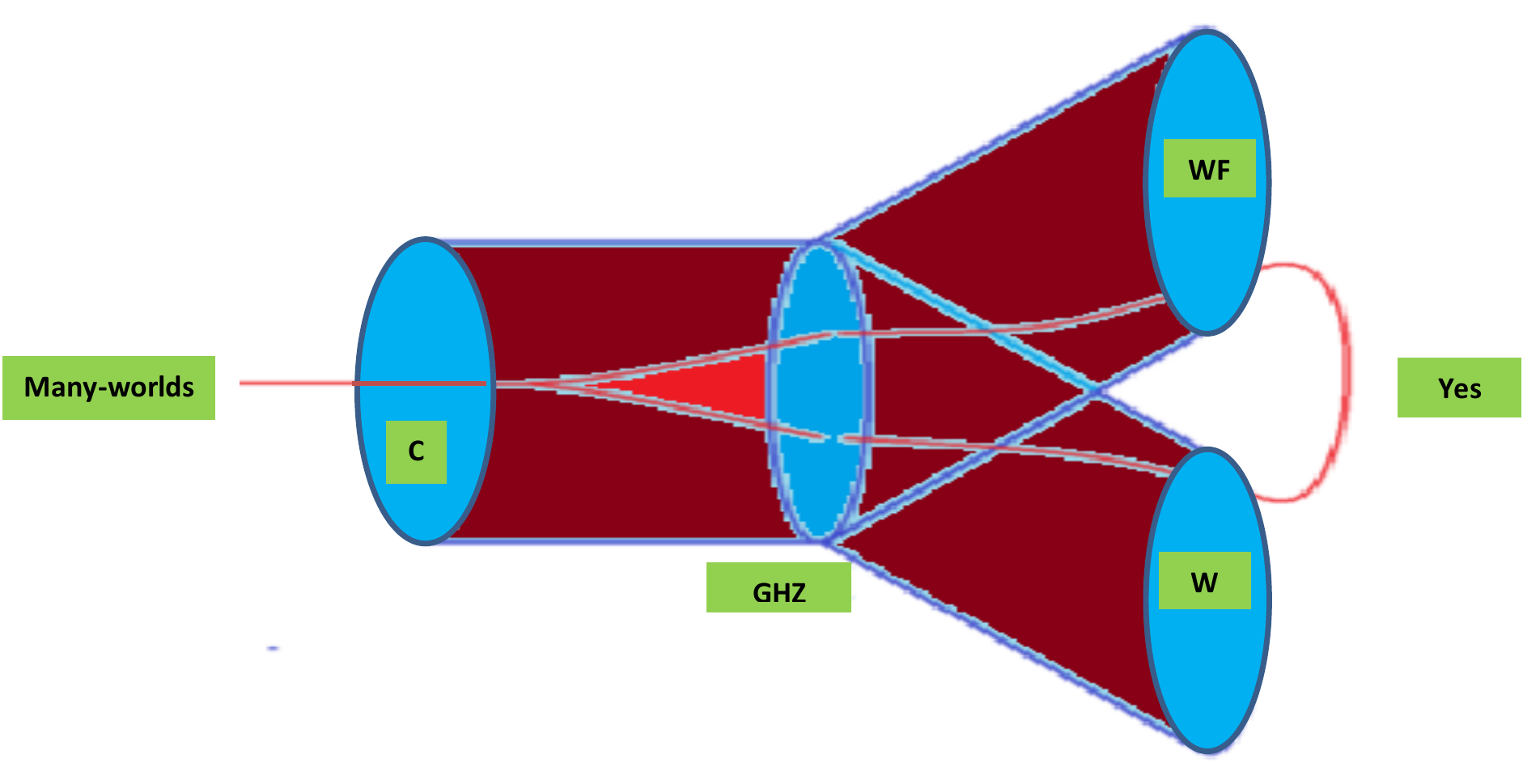}
\end{center}
\caption{The duality between Copenhagen and many-worlds interpretations.}\label{WBH}
\end{figure}

    \section{Summary}
\begin{itemize}
\item {\bf Measurement problem and interpretations:}
        \begin{itemize}
        \item Copenhagen: unitarity evolution + collapse of the state vector. Measurement: The collapse takes pure state to a mixed state  through decoherence or consciousness or gravity.
        \item Many-worlds: unitarity evolution + branching. Measurement: Branching into coherent worlds (parallel worlds through decoherence).
        \item Non-unitary collapse (irreversible) is an approximation of unitary branching (reversible). It works because the branches are completely decohered.
          \item A stronger view is that the Copenhagen interpretation provides the local view of reality whereas many-worlds provide the manifold view (more on this below).
          \item Bell's theorem: The quantum entangled state does not exist before measurement. 
          \item Bell's Theorem+EPR: {Causal realism or locality} must go.
          \item Decoherence:
            The state of the system + environment is a pure state. However, due to entanglement, the state of the system is mixed given by tracing over the degrees of freedom of the environment which gives a reduced density matrix.  Hence, {entanglement seems to give rise to collapse}. The density matrix undergoes therefore a decrease of information called {decoherence} through interaction with the environment which is seen as collapse.
            
        \end{itemize}
    
\item {\bf Black hole information loss:} A correlated entangled pure state near the horizon gives rise to a thermal mixed state outside the horizon.
\item {\bf Laws of nature:} Black hole dynamics contradicts one or more of the following laws of nature: unitarity (information conservation), equivalence principle (smoothness of the spacetime manifold), quantum zerox principle (linearity of quantum mechanics), Bekenstein-Hawking entropy (number of states equal the exponential of the surface area of the horizon divided by $4G$), and effective field theory (quantum field theory in weak gravitational backgrounds is valid).
\item {\bf Page curve:} The computation of the Page curve starting from first principles will provide, in some precise sense, the mathematical solution of the black hole information loss problem (information starts to get out at the Page time when the entanglement entropy between the interior and the exterior becomes maximal).

\item {\bf Black hole complementarity principle:} The information is both reflected at the event horizon (with respect to the external observer) and passed through the horizon towards the singularity (with respect to the infalling observer).    
     \item {\bf Firewall:} A Hawking particle $B$ must be entangled with the early part $R_B$ of the Hawking radiation and with the particle $A$ which went behind the horizon (complementarity). This is however forbidden by monogamy. The AMPS resolution is via a loss of entanglement by breaking the entanglement between $B$ and $A$. Since $A$ is a high energy mode the breaking will release a firewall and the interior of a black hole is thus not smooth.
     \item {\bf ER=EPR:} The entanglement between $A$ (behind horizon) and $B$ (outside horizon) is possible via an ER bridge connecting them. Thus it resolves the information loss problem by stating that the Hawking radiation is actually connected by ER bridges to itself and also to the black hole horizon and interior.

  \item {\bf Quantum dualism:}

    \begin{itemize}
    \item
      The {collapse is dual to branching} and {Copenhagen is dual to many-worlds}. The many-worlds is an external view (like the curved manifold structure in general relativity) in which the mathematics has an objective reality whereas the Copenhagen is an internal view (the perceived flatness around every point in a manifold) in which the mathematics has certainly also an objective reality but the representation of reality is only local, i.e. vis a vis a local conscious/zombie observer \footnote{In this paper the word "zombie" refers to philosophical zombie which shares with conscious human beings structure and even functional consciousness but lack subjective qualitative experience (qualia).}
    \item Quantum mechanics is genuinely dualistic (Copenhagen and von Neumann-Wigner interpretations) and in the same time physicalist (many-worlds) and the two views are complementary not contradictory, i.e. one can always dualize one into the other consistently.

\item This quantum dualism is rooted on the assumptions that:

\begin{itemize}
 \item The physical world is quantum mechanical.
   \item The von Neumann-Wigner interpretation (quantum dualism, i.e. the independent existence of mind and consciousness).
        \item Copenhagen/many-worlds duality (the consciousness of the observer causing
 collapse  can be recast as a unitary description by a super-observer).
\end{itemize}
\item An efficacious consciousness in a single-world as described by Copenhagen quantum mechanics is thus effectively equivalent to a many-worlds where consciousness plays no more than the role it plays in classical mechanics.
\end{itemize}
  \item  {\bf The unification of the measurement and the information loss problems:}
\begin{itemize}
\item    The information loss problem in quantum black holes and the measurement
  problem in quantum mechanics share a unified mathematical structure, namely that an initial pure state is evolved to a final mixed state where the inaccessible environment in quantum mechanics plays exactly the same role as the interior of the event horizon in black holes. 

\item The information loss problem is only a coarse-graining effect in any model of quantum gravity based on the holographic gauge/gravity duality because simply unitarity is guaranteed in an obvious way.

\item Similarly, the duality between Copenhagen and many-worlds guarantee that our description is strictly unitary and as a consequence there can be no measurement problem more than as a coarse-graining effect. Thus, the measurement problem is mainly due to the quantum dualism inherent in the Copenhagen view of the single-world which fortunately can be mapped to a strictly physicalist account of the many-worlds view.

\item Together with the fungibility  of entanglement, which allows us to compress physical systems, observers and environments into black holes and/or Hawking radiations,  the measurement problem can be recast as an information loss problem. Thus in principle it can be solved
through the same means, i.e. by employing  the holographic gauge/gravity duality which maps black holes into conformal field theory.

\item Indeed, the gauge/gravity duality provides a framework for a novel interpretation of quantum mechanics in which it is seen that the large $N$ limit of say the M-(atrix) quantum mechanics of BFSS
becomes given by classical supergravity around a classical black hole. Thus, in the large $N$
limit the role of the conscious observer decouples from the unitary evolution and
there are no measurement problem and collapse of the wave function. The effect of the observer
can then be seen as given by a $1/N$ expansion and as such it is also intimately related to the
evaporation of the black hole.
\end{itemize}
\item {\bf Other interpretations of quantum mechanics and theories of quantum gravity:} Although we have only concentrated mainly on the Copenhagen and many-worlds interpretations and on string theory as candidate for quantum gravity there are many more ingenious interpretations and theories of quantum gravity that it is very hard for anybody to know them all. 
 
\end{itemize}

\paragraph{Acknowledgment:} I would like to thank the organizers of  "Jijel 1st meeting on theoretical physics -Quantum Mechanics, Gravitation and Particle Physics-" held in Jijel between 29-31 October 2018, and the organizers of  "Study days on quantum physics and its applications" held in El-oued  between 11-15 March 2018, where this article was presented as a series of lectures, for their invitation, their warm hospitality and for the opportunity to think, write and talk about these topics of great importance to fundamental quantum physics.


\begin{thebibliography}{10}
\bibitem{DM82} 
  D.~Mermin,
  ``Is the moon there when nobody looks? Reality and the quantum theory,''
  Physics Today. 38: 38–47.
\end{thebibliography}
 \end{document}